\pdfoutput=1

\documentclass [PhD] {uclathes}

\usepackage{amsmath}
\usepackage{graphicx}
\usepackage{amssymb}
\usepackage{epstopdf}
\usepackage{wasysym}
\usepackage{oldlfont}
\usepackage{subfigure}
\usepackage{textcomp}

\usepackage{ifthen}
\newboolean{JpgFigs}
\setboolean{JpgFigs}{true}

\def \rhovec { \mbox{\boldmath $\rho$} }

\def \ellvec { \mbox{\boldmath $\ell$} }
\def \e { \mbox{$\mathrm{e}$} }
\def \xivec { \mbox{\boldmath $\xi$} }

\newcommand{\supercite}[1]{\textsuperscript{\cite{#1}}} 

\newcommand{\fluxavg}[1]{\left\langle#1\right\rangle_{\psi}}
\newcommand{\annavg}[1]{\left\langle#1\right\rangle_{\Delta V}}
\newcommand{\patch}[1]{{\cal P}[#1]}
\newcommand{\gyroavg}[1]{\left\langle#1\right\rangle_{\bf R}}
\newcommand{\angleavg}[1]{\left\langle#1\right\rangle_{\bf r}}
\newcommand{\smooth}[1]{{\cal S}[#1]}

\newcommand{\field}[1]{\mathbb{#1}}

\usepackage{natbib}

\def \varepsilonE {\mbox{$\varepsilon_{\mbox{\tiny{$E$}}}$}}
\def \Sthree { \mbox{$\boldsymbol{S}_3$}}
\def \SthreeE { \mbox{$\boldsymbol{S}^{(E)}_3$}}
\def \Stwo { \mbox{$S_2$}}
\def \StwoE { \mbox{$S_{2E}$}}
\def \Source {\mbox{$\mathcal{F}_{\mbox{\tiny{}}}$}}
\def \SourceCHM {\mbox{$\mathcal{F}_{\mbox{\tiny{CHM}}}$}}
\def \SourceE {\mbox{$\mathcal{F}_{\mbox{\tiny{$E$}}}$}}
\def \ellF {\mbox{$\ell_i$}}
\def \hE {\mbox{$h_{\mbox{\tiny{$E$}}}$}}
\def \hPhi {\mbox{$h_{\mbox{\scriptsize{$\left\langle\phi\right\rangle$}}}$}}

\newcommand{\gyroavgP}[1]{\left\langle#1\right\rangle_{{\bf R}^{\prime}}}
\newcommand{\angleavgP}[1]{\left\langle#1\right\rangle_{{\bf r}^{\prime}}}
\newcommand{\angleavgRR}[1]{\left\langle#1\right\rangle_{({\bf r},{\bf r}^{\prime})}}
\newcommand{\ensbl}[1]{\overline{#1}}
\newcommand{\CollisionOp}[1]{\gyroavg{C[#1]}}
\newcommand{\CollisionOpP}[1]{\gyroavgP{C^{\prime}[#1]}}
\newcommand{\CollisionOpZ}[1]{C[#1]}

\providecommand\bnabla{\boldsymbol{\nabla}}

\newcommand\etc{etc.\ }
\newcommand\eg{e.g.\ }
\newcommand\ie{i.e.\ }

\title          {The theory of gyrokinetic turbulence: \\ A multiple-scales approach}
\author         {Gabriel Galad Plunk}
\department     {Physics}
\degreeyear     {2009}

\chair          {Steven Cowley}
\member         {Troy Carter}
\member         {George Morales}
\member         {James McWilliams}

\dedication     {\textsl{For Nicole, with love}}

\acknowledgments {
First, I thank my advisor, Professor Steve Cowley, who I feel very privileged to have been mentored by.  He is truly a physicists' physicist.  In my experiences working with him, I continually have a renewed appreciation for his creativity, rigor, strength of intuition and the depth of insight.  He is a great thinker and a kind and generous advisor.

Thanks, with love, to my mother who fostered a spirit of independence and creative thinking.  Thanks to my stepfather Ken Rowe for telling me about the speed of light and encouraging the analytical and generally scientific direction of my thinking at a young age.  Thanks to my childhood mentor and friend Kevin Carmody for sharing his passion for mathematics and encouraging my own passion for computer programming.

My high school science teachers were excellent and deserve much thanks and credit for my decision to continue in the sciences during college and beyond.  Thanks to Mrs. Canham for being so supportive and open to odd questions based upon a thirteen year old's understanding of quantum mechanics.  Thanks to Mrs. Parsons for an excellent introduction to biology.  Thanks to the remarkable Mr. Lilga who was a difficult teacher but taught me more chemistry than I thought I could learn.  Thanks to Mr. Bascom, my advanced placement physics teacher who really did a fantastic job enabling us to choose any path in physics that we desired.

I had some especially influential and noteworthy college teachers during my time at Cornell.  Thanks to Nick Jones for dedication and excellence as a TA.  And also thanks to the excellent TA Ben D. Pecjak.  Thanks to Professor Philip Argyres for an inspiring college introduction to quantum mechanics.  His lectures were seeds that the germinated in the mind slowly, so that understanding came continually as gifts during the weeks and months that followed them.  Thanks to Professor Matthias Neubert for Euler angles -- and for, in general, immaculately organized and cogent lectures which inspired a personal physics renaissance during my junior year.  Thanks to Professor N. David Merman for teaching me quantum information theory and for his kindness in helping me with independent research during my senior year.

Special thanks are due also to Chris Fuchs, my research advisor at Bell Labs during the summer of 2002.  He gave me my first taste of physics research and opened my eyes to the many conceptual and philosophical mysteries in the foundations of quantum theory.

Thanks to fellow physics undergrads Jake Stevenson, Ilya Berdnikov, Adam Frankel and Jaka Skrlj.  Thanks to Dan Sullivan for friendship and inspiration.

Thanks to David Schwab who had an immeasurable influence on my physics education and who is my best friend.  I look forward to future creative collaborations with him.

Thanks to all the Cafes and other favorite places where I've done physics:  From my Cornell days, thanks go to Collegetown Bagels, Stella's Cafe, Dino's, Jasmine's, Juna's Cafe, The Lost Dog Cafe in Binghamton, Rongovian Embassy To the USA in Trumansburg, The ABC Cafe and especially Wegman's.  In Los Angeles, thanks go to Anastasia's Asylum, Jerry's Famous Deli, Novel Cafe in Venice Beach (and then in Westwood too), Peet's in Westwood, The Cow's End in Venice, The 18th Street Cafe and The Bagel Nosh.

Thanks to Eric Wang for his loyal friendship and unbounded generosity.  He is very talented and radiates positivity -- his path in life is surely in the direction of great success.  Also I thank him for the great memories in Los Angeles and the greater Southern California area.

Thanks to Professor Bill Dorland for his interest, support and advice at various times during the journey.    Thanks to Kyle Gustafson for his friendship and general help during my transition from UCLA to UMD.  Thanks to Alex Schekochihin.  His creativity and quality of his scholarship make him an inspiring example to follow and his passion is a driving force for many people, me included.  Thanks to Uriel Frisch for writing his book \cite{frisch} which taught me about fluid turbulence.  Thanks to Professor Troy Carter for being extraordinarily supportive and kind during a generally difficult final period of organizing and scheduling the defense.  Also thanks to Professor Jim McWilliams for his invaluable role in my thesis committee.  Thanks to Professor George Morales for his careful reading of my thesis and constructive feedback.

Last (that is, in a very prominent and honored place), I would like to give a loving thank you to Lisa Chemery.  Around our lives she has built a great home in the physical, emotional and spiritual sense.  Through this, she has given me the love, stability and sense of peace which I have really relied upon while I carried out this work.

The material of chapter \ref{chap-2-sec} is based on work done in collaboration with Eric Wang and Steve Cowley.  I would like to thank Michael Barnes for his careful reading of our notes and his helpful suggestions.

The material of chapter \ref{secondary-chap} is based on the publication \cite{plunk}.  The text has some slight modifications but the figures are unchanged.  Therefore, as instructed by the publishers, acknowledgment of permission for reprinting the copyrighted material is given in text at the beginning of the chapter.  For this work specifically, I would like to thank Bill Dorland and Ron Waltz for generous and helpful advice, and Eric Wang for many stimulating discussions.

Chapter \ref{phase-space-turbulence-chapter} is a version of the paper ``Two dimensional gyrokinetic turbulence'' to be submitted for publication in the Journal of Fluid Mechanics.  The authors are Gabriel G. Plunk (GGP), Steven C. Cowley (SCC), Alexander A. Schekochihin (AAS) and Tomo Tatsuno (TT).  The following acknowledgments are made specifically for this work:  GGP would like to acknowledge scholarship support from the Wolfgang Pauli Institute and support from the US DOE sponsored Center for Multiscale Plasma Dynamics, Grant No. DE-FC02-04ER54785.  GGP and TT thank the Leverhulme Trust International Network for Magnetised Plasma Turbulence (Grant F/07 058/AP) for travel support.  TT was supported by the US DOE Center for Multiscale Plasma Dynamics.  AAS was supported by an STFC Advanced Fellowship and by the STFC Grant ST/F002505/1.

The work of this thesis was supported by the US DOE sponsored Center for Multiscale Plasma Dynamics, Grant No. DE-FC02-04ER54785. 
}

\vitaitem   {1981}
                {Born, Capitola, California, USA.}
\vitaitem   {1995--1999}
                {Wootton High School, Rockville, MD.}
\vitaitem   {2002}
                {Summer Research Assistant in Quantum Information Theory, Mentor: Chris Fuchs, Bell Labs, Murray Hill, New Jersey.}
\vitaitem   {2002--2003}
                {Independent Research, Advisor: N. David Mermin, Cornell University, Ithaca, New York.}
\vitaitem   {2003}
                {B.A.~(Physics) and B.A.~(Mathematics),
                Cornell University, Ithaca, New York.}
\vitaitem   {2003--2007}
                {Teaching Assistant, Physics Department, UCLA.}
\vitaitem   {2005--2009}
                {Graduate Student Researcher, Advisor: Steven Cowley, University of California, Los Angeles.}
\vitaitem   {2005}
                {M.S.~(Physics), UCLA, University of California, Los Angeles.}

\publication{
``Gyrokinetic secondary instability theory for electron and ion temperature gradient driven turbulence,''
Gabriel Plunk, Phys. Plasmas 14, 112308 (2007)

``Gyrokinetic turbulence: a nonlinear route to dissipation through phase space,''
A. A. Schekochihin, S. C. Cowley, W. Dorland, G. W. Hammett, G. G. Howes, G. G. Plunk, E. Quataert, and T. Tatsuno, Plasma Phys. Control. Fusion 50 124024 (2008)

}

\abstract{Gyrokinetics is a rich and rewarding playground to study some of the mysteries of modern physics -- such as turbulence, universality, self-organization and dynamic criticality -- which are found in physical systems that are driven far from thermodynamic equilibrium.  One such system is of particular importance, as it is central in the development of fusion energy -- this system is the turbulent plasma found in magnetically confined fusion device.  In this thesis I present work, motivated by the quest for fusion energy, which seeks to uncover some of the inner workings of turbulence in magnetized plasmas.  I present three projects, based on the work of me and my collaborators, which take a tour of different aspects and approaches to the gyrokinetic turbulence problem.  

I begin with the fundamental theory of gyrokinetics, and a novel formulation of its extension to the equations for mean-scale transport -- the equations which must be solved to determine the performance of magnetically confined fusion devices.  The results of this work include (1) the equations of evolution for the mean-scale (equilibrium) density, temperature and magnetic field of the plasma, (2) a detailed Poynting's theorem for the energy balance and (3) the entropy balance equations.

The second project presents gyrokinetic secondary instability theory as a mechanism to bring about saturation of the basic instabilities that drive gyrokinetic turbulence.  Emphasis is put on the ability for this analytic theory to predict basic properties of the nonlinear state, which can be applied to a mixing length phenomenology of transport.  The results of this work include (1) an integral equation for the calculation of the growth rate of the fully gyrokinetic secondary instability with finite Larmor radius (FLR) affects included exactly, (2) the demonstration of the robustness of the secondary instability at fine scales ($k\rho_i$ for ion temperature gradient (ITG) turbulence and $k\rho_e \ll 1$ for electron temperature gradient (ETG)) which rules out the possibility that ultra-fine streamers could produce significant transport, (3) a demonstration that the variation in the phasing of the primary mode (which depend on the values of the equilibrium scale lengths of the system) effects the strength of the secondary instability, distinguishing the gyrokinetic model from a previous gyrofluid model, (4) parameter scans for the mean-scale gradient lengths which suggest a possible role of secondary instabilities in the Dimits shift and the formation of electron internal transport barriers (ITB) in tokamaks, (5) a formulation of the theory for fully gyrokinetic ions and electrons in order to explore the transition between ETG and ITG scales and (6) demonstrate the existence of a mechanism for the saturation of long-wavelength ETG modes in this ETG-ITG transition range (modes which have been demonstrated in simulations not to saturate when employing the ETG Boltzmann-ion gyrokinetic system).

The final project is an application of the methods from inertial range understanding of fluid turbulence, to describe the stationary state of fully developed two-dimensional gyrokinetic turbulence.  This work explores the relatively new idea of a phase-space cascade, whereby fine scales are nonlinearly generated in both position space and velocity space, and ultimately smoothed by collisional entropy production.  This process constitutes the thermodynamic balance which occurs in the true steady state of a turbulent plasma, including those found in fusion devices.  The results of this work include (1) exact third order relations (in analogy to Kolmogorov's four-fifths law), (2) phenomenological scaling theories for the forward and inverse cascades, (3) a detailed description of the relationship of the two-dimensional gyrokinetic cascade to the Charney-Hasegawa-Mima and two-dimensional Navier-Stokes cascades, (4) a Hankel transform formalism for treating velocity scales in the distribution function and (4) power law predictions for the phase-space free energy spectra.}

\begin {document}

\makeintropages

%
%

\chapter{Introduction}

The work of this thesis addresses related but distinct problems in the theory of gyrokinetic (GK) turbulence, or {\em gyrokinetics}.  This introduction will motivate the problems addressed by this thesis and outline the main results of the work.

\section{Gyrokinetics}

When inter-particle collisions in a plasma are too weak to maintain local thermodynamic equilibrium (over time-scales of interest) the velocity distribution of the particles becomes an important dynamical feature, and a kinetic description (\ie the Fokker-Planck equation) is necessary to capture this.  In kinetics, a plasma is described by a scalar field over 6 dimensions in phase space.  This is clearly more complicated than a fluid description, over 3 spatial dimensions.  The complexity of plasmas is furthered by its famously rich ``zoology'' of waves and instabilities, existing over an enormous dynamical and spatial range of scales.

In magnetized plasmas, this complexity may be reduced substantially when the dynamics of interest occur over time-scales much longer than the cyclotron period, the period of rotation of particles about the mean magnetic field.  This is the starting point for the development of gyrokinetic theory.  The complexity of the problem is reduced in two ways.  First, phase space is reduced by one dimension (by eliminating the gyro-angle of particles about the magnetic field).  Second, and most important, the dynamical range of the problem is reduced by averaging away behavior at the cyclotron time-scale.  This yields a vast improvement in the efficiency of numerical simulations.  One estimate gives a total speedup of $10^{11}$ achieved by the nonlinear gyrokinetic equation (when including the elimination of the plasma frequency scale, Debye and electron Larmor radius scales, and the ion cyclotron scale).\supercite{hammett-talk}

The historical milestones in the development of gyrokinetics are as follows.  The evolution begins with the work of \cite{taylor-hastie} and \cite{rutherford}.  The discovery of a generalized adiabatic invariant in the presence of low frequency fluctuations allowed the generalization of guiding center theory in the form of gryo-orbit-averaged kinetic equation, which became known as the linear gyrokinetic equation.  The linear theory was refined by introduction of the Catto-transformation and the inclusion of collisional effects in the paper \cite{catto1977}.  The work by \cite{antonsen} developed the application of gyrokinetics to the flux-surface equilibrium field geometry of axi-symmetric magnetically confined plasmas.  Then a critical breakthrough came with the nonlinear formulation of gyrokinetics by \cite{frieman}.  The hamiltonian formulation of gyrokinetics has also brought deeper understanding, and its history, as well as a general review of gyrokinetics, is reviewed in \cite{brizard-hahm}.

\section{The Tokamak Transport Problem}

The motivation and context for the development of the gyrokinetic equations is in magnetic fusion.  Although the applicability of gyrokinetics extends to astrophysical plasmas, the application on which this thesis will focus will be fusion.  We begin this introduction with a statement of the problem of Tokamak transport.

\subsection{Confinement and Fusion}

Following the presentation in \cite{wesson} and \cite{friedberg}, we now take a look at some simple considerations of confinement and fusion.  The fusion reaction rate per unit volume is give by the product of the densities of the the two fusing species multiplied by an integral over the reaction cross section which accounts for the relative contribution from different energy levels of the bulk Maxwellian:

\begin{align}
\Re &= \left(\frac{8}{\pi}\right)^{1/2} n_1n_2\left(\frac{\mu}{T}\right)^{3/2}\frac{1}{m_1^2} \int \sigma(\varepsilon)\varepsilon\exp(-\frac{\mu\varepsilon}{m_1 T})d\varepsilon \nonumber\\
&= n_1n_2 \left\langle\sigma v \right\rangle
\end{align}

\noindent where $\mu = m_1m_2/(m_1+m_2)$ is the reduced mass of the fusing nuclei and $\sigma$ is the collisional cross section.  To determine the conditions necessary for a fusion reactor, let's consider how the heating by fusion power must balance with the thermal losses.  The fusion reaction between the tritium and deuterium nuclei produces an alpha particle and a neutron with 3.5 MeV and 14.1 MeV of kinetic energy respectively.  The high-energy neutrons are not confined and pass from the plasma volume.  The alpha particles are sufficiently confined to impart their energy back to the plasma.  We may calculate the total power of this heating by multiplying the fusion rate $\Re$ by the energy of $\alpha$ particles, ${\cal E}_{\alpha} = 3.5 MeV$, and integrate over the plasma volume:

\begin{align}
P_{\alpha} &=  \int d^3{\bf r}{\cal E}_{\alpha} \Re \nonumber\\
&= {\cal E}_{\alpha}\int d^3{\bf r} n_1 n_2 \left\langle\sigma v \right\rangle\nonumber\\
&= \frac{n^2}{4}{\cal E}_{\alpha}V \overline{\left\langle\sigma v \right\rangle}\label{p-alpha}
\end{align}

\noindent where $V$ is the plasma volume and we assume a constant density $n_1 = n_2 = n/2$ between lines two and three.  The over-bar indicates volume-averaging.  Now consider the overall energy balance equation in steady state.  The total plasma energy is $W = 3\int d^3{\bf r}\; nT$ and in steady-state, its decay due to losses is balanced by the externally applied heating $P_H$ and the heating by alpha particles:

\begin{align}
P_{\alpha} + P_H &= \frac{W}{\tau_E} \nonumber \\
&= \frac{3n}{\tau_E}\int d^3{\bf r}\;T \nonumber \\
&= \frac{3n\bar{T}V}{\tau_E}
\end{align}

\noindent This introduces the important measure, $\tau_E \equiv (d\ln W/dt)_{\mbox{\scriptsize{losses}}}^{-1}$, which is called the confinement time.  The confinement time is the e-folding time of the total plasma energy due to thermal conduction.  Ignition is achieved when the $\alpha$ particle heating matches or exceeds thermal losses.  When this occurs,  the external heating $P_H$ is not needed to sustain the plasma energy.  By setting $P_H = 0$, and using the expression \ref{p-alpha} for $\alpha$-heating power, we obtain an expression for a minimal confinement time needed for ignition:

\begin{equation}
n\tau_E = \frac{12\bar{T}}{\overline{\left\langle\sigma v \right\rangle}{\cal E}_{\alpha}}
\end{equation}

\noindent To get a sense of the performance requirements of a fusion reactor, an approximate expression may be obtained for a constant temperature (flat profile) between $10$ and $20$ keV (from equation 1.5.5 in \cite{wesson}):

\begin{equation}
nT\tau_E \approx 3 \times 10^{21} m^{-3}keV\;s \label{minimal-ignition-time}
\end{equation}

\noindent The typical densities found in present-day tokamaks vary from $10^{19}$-$10^{20}\; \mbox{m}^{-3}$.  Assuming the temperature range $10$-$20$ keV, this gives an ignition confinement time of approximately $\tau_E \sim 1.5 - 30$ s.  

This timescale is very large in comparison to the micro-scales in the plasma such as inter-particle collisions, particle transit and bounce frequencies, linear and nonlinear timescales.  This is the basis for the assumption of scale separation and suggests that a suitably defined time-average may be used as the ensemble average in the formal statistical treatment of fluctuations.  We will explore this idea in detail in chapter \ref{chap-2-sec}.

The preceding simple calculation of the ignition confinement time sets a rough goal for the required performance of a tokamak-based reactor.  There are several processes which enter in the determination of the confinement measure $\tau_E$.  The basic mechanisms of confinement degradation, that are of chief concern, are collisional (classical) transport and turbulent (anomalous) transport.  The former process is well-understood but unfortunately subdominant to the latter process, which is a subject of intense research, and a focus of this thesis.

\subsection{Classical transport}

The transport due to collisions in a magnetized plasma in local thermodynamic equilibrium is referred to as classical transport.  The theory of this process is significantly more detailed for a toroidal plasma, so has been given the name neoclassical transport for that case.  The classical and neoclassical theories are not the main subject of this thesis.  However, there are some scenarios where anomalous transport is suppressed, and the neoclassical levels are observed (for example, transport barriers -- see \cite{stallard}).  Thus classical transport is discussed here briefly as a baseline mechanism for transport and to outline some general considerations of scaling in experiments.

The diffusion of particles in the classical scenario occurs via random collisions.  For this process, the scale of interest in the presence of a uniform background magnetic field is the Larmor radius.  The random change in the particle velocity implies a change in the the position of the center of the particle orbit, the gyro-center.  Thus the step length for this process is the Larmor radius $\rho$ and the time between steps is set by the inter-species collisional frequency (like-particle collisions do not produce particle diffusion).  This gives a classical particle diffusivity of $D^{(C)} \sim \nu_{ie}\rho_i^2 \sim \nu_{ei}\rho_e^2$.  On the other hand, the thermal diffusivity is caused by like-particle collisions and is given $\chi^{(C)} \sim \chi_i^{(C)} \sim \nu_i \rho_i^2$.  The associated confinement time is $\tau_E \sim a^2/\chi$, where $a$ is the minor radius of the plasma.  

As an example of a tokamak which is approaching reactor conditions, consider the Joint European Torus (JET), which is capable of achieving peak density and temperatures $T_i \approx 28 keV$ and $n \approx 4 \times 10^{19}m^{-3}$ (see \cite{gormezano} and also section 13.7 of \cite{friedberg}).  Using these numbers and the basic parameters of the machine ($B = 3.6 T$, $a = 1.25 m$), we may calculate the ion collisional frequency, $\nu_{i} =  3.5 s^{-1}$, and ion Larmor radius, $\rho = 6.7 mm$.  The corresponding classical thermal diffusivity gives a confinement time

\begin{equation}
\tau_E \sim 2.8 hrs.\label{classical-confinement-time}
\end{equation}

\noindent Obviously, this greatly exceeds the minimal ignition confinement time (which, using equation \ref{minimal-ignition-time}, is $\tau_E \sim 2.7 s$).  However, it also greatly exceeds the actual confinement time observed, $\tau_E \sim 1 s$.

Neoclassical transport theory can account for a small part of this discrepancy.  Toroidal geometry introduces, among other features, additional timescales and spatial scales associated with the toroidal orbits of the particles.  (We will not review the details here.)  The net result is a substantial enhancement of collisional diffusion.  With the approximations of high-aspect-ratio and circular cross-section, the neoclassical thermal diffusivity was derived by \cite{rosenbluth-hinton-hazeltine} in terms of the classical diffusivity to be $\chi^{(NC)} \sim 0.68 q^2 (R/a)^{3/2} \chi^{(C)}$\footnote{See also equation (14.126) of \cite{friedberg}}, where $(R/a)$ is the aspect ratio and $q$ is the safety factor which characterizes the trajectory of the magnetic field lines as they trace out the toroidal surface.  For the high-performance JET discharge (taking $q \sim 4$ and $(R/a) \sim 2.4$), these factors give an enhancement of about $40$ over classical diffusion so that the confinement time is reduced to 

\begin{equation}
\tau_E \sim 4 min.\label{neo-ignition-time}
\end{equation}

\noindent Clearly, neoclassical transport not nearly enough to account for the experimental observed confinement.  The remainder of the thermal conduction is accounted for by the so-called anomalous transport as we shall now describe.

\subsection{Anomalous transport}

Micro-instabilities drive turbulent plasma flows, on scales ranging from the ion to the electron Larmor radius, thereby mixing plasma and enhancing transport, well beyond base levels set by classical diffusion.  This ``anomalous transport'' has the affect of degrading heat confinement, so that fusion scientists widely view turbulence as an ailment of fusion devices.

Let's examine turbulent transport in a little more detail.  The anomalous contribution to transport of density and temperature can described by the equations

\begin{align}
&\left(\frac{dn}{dt}\right)_a = -\partial_x \Gamma_x \nonumber\\
&\frac{3}{2}n\left(\frac{dT}{dt}\right)_a = -\partial_x Q_x
\end{align}

\noindent where $x$ is the radial direction -- the direction in which the equilibrium temperature and pressure vary.  The complexity and difficulty of the problem enters in that these fluxes must be calculated from a statistical description of the turbulent steady state.  A standard expression for the anomalous particle flux (the formal theory will be worked out in chapter \ref{chap-2-sec}) is:

\begin{subequations}
\begin{align}\label{transport-statement}
\Gamma_x &= \int d^3{\bf v} \; \overline{{\bf v}_{\chi} h} \\
&= \int d^3{\bf v}\int d^3{\bf k} \; \overline{{\bf \hat{v}}_{\chi}({\bf k}) \hat{h}(-{\bf k})} \label{flux-spectrum}
\end{align}
\end{subequations}

\noindent where the over-bar is a suitably defined averaging operator (to be specified later), $h$ is the fluctuating part of the (gyrokinetic) distribution function, and ${\bf v}_{\chi}$ is the turbulent drift-velocity.  (Note that we have used homogeneity to obtain the expression \ref{flux-spectrum}.)  We may state the transport problem thus either as the calculation of the steady state averaged value of ${\bf v}_{\chi} h$ or the average spectrum ${\bf \hat{v}}_{\chi} \hat{h}$.  In other words, the transport problem is that of determining the stationary spectrum of fluctuations in fully developed plasma turbulence.

Thus a central problem in the general theory of `turbulence'--that is, obtaining a statistical description of the turbulent state which is capable of predicting mean-scale quantities such as drag coefficient conductivity, diffusivity, etc--is the essence of the anomalous transport problem.  From a theoretical perspective, this is obviously a difficult, unsolved problem and although an exact solution has not been achieved, there are approaches to estimating the transport fluxes.  The simplest approach is the random walk {\bf mixing length} estimate.  This estimate assumes that the transport spectrum is dominated by a single characteristic scale and assumes a diffusive random-walk process such that the fluxes are proportional to a diffusivity $D$:

\begin{subequations}
\begin{align}
&\Gamma_x \sim -D\; \partial_x n  \nonumber\\
&Q_x \sim -n\chi\; \partial_x T
\end{align}
\end{subequations}

\noindent and

\begin{equation}
D \sim \chi \sim \frac{\Delta x^2}{\Delta t}\label{mixing-diffusivity}
\end{equation}

\noindent where $\Delta x$ is the characteristic step length and $\Delta t$ is the typical time between steps.  If one assumes that the step length $\Delta x$ is the ion Larmor radius (\ie that the transport spectrum is dominated by $k \sim \rho^{-1}$, as is supported by nonlinear simulations) and the step time is the inverse of the linear diamagnetic drift frequency ($\omega_{\star} \sim (k\rho)\;v_{th}/L_T \sim v_{th}/L_T$), one obtains gyro-Bohm diffusivity

\begin{equation}
D_{GB} = \rho_i^2 v_{th}/L_T\label{gyro-bohm-intro}
\end{equation}

\noindent where $L_T^{-1} = -d\ln T/dx$.  Assuming the high performance JET parameters from above (so $v_{th} \sim 1.2\times 10^6 m/s$ and $L_T \sim 1 m$), we obtain a resulting confinement time $\tau_E \sim a^2/D_{GB} \approx 3 s$.  Clearly, this is much closer than the classical or neoclassical estimates (\ref{classical-confinement-time} and \ref{neo-ignition-time}) to the observed confinement time $\tau_E \sim 1s$.  

The gyro-Bohm estimate for the turbulent diffusivity reveals the scaling of the transport fluxes (under the assumption of scale-separation) but does not give a specific quantitative prediction or include details of the nonlinear dynamics.  However, the validity of gyro-Bohm scaling has been tested via simulations\supercite{waltz-rho-star, candy-dorland} and it is the accepted scaling in the limit of strong scale separation between the Larmor radius scale and the equilibrium variation scale -- \ie the limit of small $\rho_{\star} \equiv \rho/L_T$.

As a general note on scaling, let's compare the neoclassical confinement time to the gyro-Bohm confinement time.  In particular, the temperature dependence of the collisional frequency goes as $T^{-3/2}$ so that the classical and neoclassical confinement times will increase as $T^{3/2}$ and increase with the system size as roughly $a^2$ (assuming a constant aspect ratio).  On the other hand the gyro-Bohm diffusivity increases as $T^{3/2}$ due to increased Larmor radius and linear diamagnetic frequency.  This means gyro-Bohm confinement worsens at higher temperature (of course, the positive scaling of confinement in system size is the dominant effect in scaling to reactor designs).  Thus the hotter temperatures expected in reactor conditions should make classical and neoclassical contributions to transport even less important, compared with the anomalous contribution.

\subsection{Weak turbulence}

Quasi-linear theory gives an estimate for the anomalous transport fluxes under the assumption of weak turbulence.\supercite{tsytovich, antonsen-coppi, connor-pogutse}  This estimate is more sophisticated than the dimensional analysis of random walk estimate (given above) in that it includes linear mode solutions for a specific set of plasma parameters and equilibrium geometry.  Thus it is capable of capturing the parametric dependence of specific features such as the anomalous particle pinch.\supercite{angioni}  It also has the advantage of being built upon rigorous mathematical footing via a perturbative expansion in the fluctuation amplitude.  However, it is limited since a linear mode spectrum may not adequately describe the fully developed state.  The quasilinear theory must also provide a guess for amplitudes of the modes -- the ``saturation amplitude.''  To sketch an example, let's return to the expression \ref{flux-spectrum} and approximate $\hat{h}$ quasilinearly.  From the gyrokinetic equation (Eqn.~\ref{the-gyro-equation}) we may write (neglecting the nonlinearity and the equilibrium curvature and $\nabla B$ drifts)

\begin{equation}
-\dot{\imath} \omega({\bf k}) \hat{h}({\bf k}) \sim {\bf k}_{\star}\cdot{\bf \hat{v}}_{\chi}({\bf k}) F_0
\end{equation}

\noindent where ${\bf k}_{\star} = \bnabla\ln F_0 \sim {\bf \hat{x}}L_T^{-1}$ -- assuming that the density gradient is weaker than the temperature gradient, \ie $L_T^{-1} \gg L_n^{-1}$.  Also, .  (Note, we have only included the ``irreversible'' contribution to $\hat{h}$ in going from the first to second line.\supercite{connor-pogutse})  Substituting this into the expression \ref{flux-spectrum} we get

\begin{align}
\Gamma_x &\sim \int d^3{\bf v}\int d^3{\bf k} \; \overline{{\bf \hat{x}}\cdot{\bf \hat{v}}_{\chi}({\bf k}) \frac{\dot{\imath}}{\omega}{\bf k}_{\star}\cdot{\bf \hat{v}}_{\chi}(-{\bf k}) F_0 } \nonumber\\
&= \int d^3{\bf k} \; \frac{n_0}{L_T} \overline{|{\bf \hat{x}}\cdot{\bf \hat{v}}_{\chi}({\bf k})|^2}\frac{\gamma}{|\omega|^2}
\end{align}

\noindent where $\gamma = \mbox{Im}[\omega]$ and the second line is obtained by noting that the $\Gamma_x$ is real so we may take the real part of the expression.  We may eliminate ${\bf v}_{\chi}$ in favor of the fluid displacement variable, $\xi$, which is the time integral of the drift velocity: $d\xivec/dt = {\bf v}_{\chi}$.  Thus we may write $|{\bf \hat{x}}\cdot{\bf \hat{v}}_{\chi}({\bf k})|^2 \sim |\omega|^2|\hat{\xi}_x({\bf k})|^2$ leading to the expression

\begin{equation}
\Gamma_x \sim \int d^3{\bf k} \; \gamma\frac{n_0}{L_T}\overline{|\hat{\xi}_x|^2}
\end{equation}

\noindent Now if we assume that the dominant part of the spectrum comes from $k \sim \rho^{-1}$ and furthermore assume that $\xi \sim \rho$, we obtain the quasi-linear estimate for the transport flux.\footnote{The additional approximation $\gamma \sim v_{th}/L_T$ applied to equation \ref{ql-flux} gives gyro-Bohm diffusivity argued above by random walk.  Thus the random walk mixing length estimate is sometimes referred to as a quasi-linear estimate.}

\begin{equation}
\Gamma_x \sim \left(\gamma\frac{n_0}{L_T}\rho^2\right)_{k_{\perp} = \rho^{-1}}\label{ql-flux}
\end{equation}

\begin{figure}
\subfigure[{\bf Isotropic turbulent mixing}: gyro-Bohm diffusivity assumes $k_x \sim k_y \sim \rho^{-1}$ and $\xi_x \sim \pi/k_x$]{
\ifthenelse{\boolean{JpgFigs}}
{\includegraphics[width=.8\columnwidth]{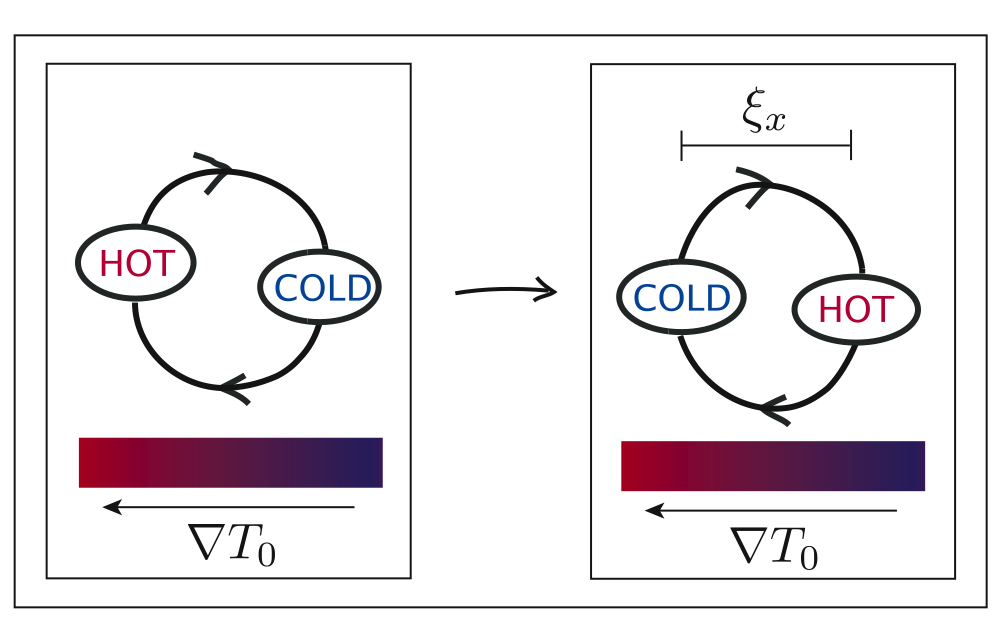}}
{\includegraphics[width=.8\columnwidth]{Figures/mixing-schematic-isotropic.pdf}}
\label{mixing-scheme-fig-a}
}
\subfigure[{\bf Mixing by radially elongated streamer}: Enhancement of transport by anisotropic turbulent mixing.]{
\ifthenelse{\boolean{JpgFigs}}
{\includegraphics[width=.8\columnwidth]{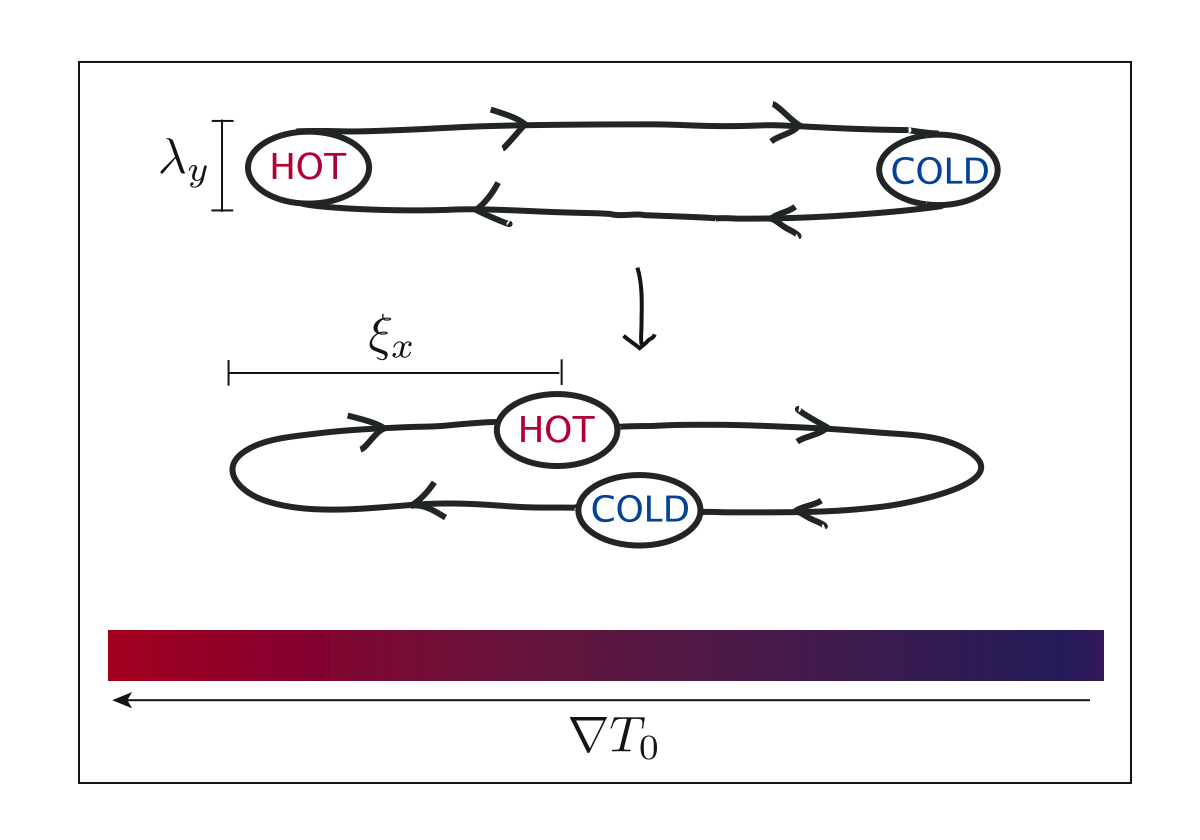}}
{\includegraphics[width=.8\columnwidth]{Figures/mixing-schematic-streamers.pdf}}
\label{mixing-scheme-fig-b}
}
\caption{{\bf Transport by turbulent mixing}: The radial mixing length is set by the strength of the flow (the saturation amplitude) but limited by the dominant wavenumber in the transport spectrum $\xi_x \lesssim \pi k_x^{-1}$.}
\end{figure}

One limitation of the assumption $\xi \sim k_{\perp}^{-1}$ is that the typical fluid displacement may in fact be less than $k_{\perp}^{-1}$ if the saturation amplitude is low.  Thus $\xi \sim \rho$ may be considered an upper bound imposed by the geometry of the flow for the case where the spectrum is dominated by $k_x \sim \rho^{-1}$.  This case is represented in figure \ref{mixing-scheme-fig-a}.  If, on the other hand, there are anisotropic radially elongated streamers, as can be the case with electron temperature gradient (ETG) driven turbulence, the dominant wavenumber may correspond to $k_y \sim \rho^{-1}$ but $k_x \ll k_y$ so that we may have $\xi_x \gg \rho$ -- see figure \ref{mixing-scheme-fig-b}.  It is clear that the proper formulation of mixing length phenomenology requires insight into the fully developed turbulent state -- a state which may or may not be successfully treated with quasi-linear theory.  

\subsection{Strong turbulence}

In the strong-turbulence limit, the nonlinearity is treated non-perturbatively and the spectrum of turbulent excitations cannot {\em a priori} be assumed to retain linear mode features.  This problem is treated most commonly by direct numerical simulation.  Although there is great progress in the ability to simulate turbulent transport, it remains a problem that pushes the limits of computational technology and progress in theoretical understanding remains critically important.  Strong turbulent transport may be divided into two categories: {\bf Inertial range} turbulence and {\bf large scale} or {\bf energy-containing range} turbulence (see for instance \cite{canuto-goldman-0}).

The inertial range understanding of turbulence, pioneered by Kolmogorov\supercite{kolmogorov41a, kolmogorov41b, kolmogorov41c}, is characterized by the nonlinear cascade, whereby dynamically invariant quantities are transfered locally in $k$-space, to different scales, producing a self-similar spectrum which is independent of the specific way that the system is driven (universality).  It is a common point of view that the inertial range understanding of fluid turbulence has a limited applicability to the tokamak transport problem.  Indeed, simulations suggest that the majority of transport occurs at linearly unstable scales where the majority of free energy injection occurs -- such scales, by definition, are outside of the inertial range.\footnote{It is also possible to sustain turbulence without unstable linear modes via a nonlinear process which draws from the background gradient in free energy.\supercite{drake-zeiler}  Thus, linear instability is not a definitive criteria that separates inertial and non-inertial ranges.}  Also there is evidence of self-organization in plasma dynamics at scales that are not well-separated from the system scale\supercite{itoh-itoh-mesoscale} so that the traditional assumptions of homogeneity will not apply to these cases.  For these reasons, the application of inertial range concepts to gyrokinetic turbulence has been largely unexplored, although much is known about fluid models such as the Charney--Hasegawa--Mima (CHM) turbulence.  The final chapter, chapter \ref{phase-space-turbulence-chapter}, of this thesis is devoted to an inertial range theory of gyrokinetics and we will argue that this theory has a place in the study of anomalous transport.  For instance, the inertial range understanding is important in determining the proper resolution needed in numerical simulations.  That is, the fine-scale inertial range transfer of energy to the collisional scale is the process by which the true steady state is achieved in a turbulent plasma; and the collisional scale determines the smallest scale which must be resolved in a simulation.  This will be discussed more in chapter \ref{phase-space-turbulence-chapter}.  At this point, we emphasize that although an inertial range theory can not come close to fully describing tokamak turbulence, we believe it is a key part of the picture and that, in general, its applicability is an open question.

The range where energy injection is non-negligible is the ``energy-containing'' range because the fully developed steady-state tends to have the majority of energy contained here.  A statistical description of the turbulence spectrum in this range depends on the particular features of instability drives and other features of equilibrium plasma which can strongly affect the nonlinear physics.  It therefore does not exhibit the universality that characterizes the inertial range.  For this reason, there does not exist as much of a unifying framework to describe phenomena of the energy-containing range.  Theoretical progress in this subject is driven by (1) the intuition built by experiment and direct numerical simulations, and (2) theoretical models of nonlinear processes inspired, at least in part, by these observations.  The material in chapter \ref{secondary-chap}, which is mostly concerned with the energy containing range, falls into the second category.

As a final note about strong turbulence, note that although the random walk mixing length diffusivity estimate \ref{gyro-bohm-intro} may be obtained within quasilinear framework, mixing length phenomenology in general does not assume weak-turbulence so it can be applied to strong turbulence regimes as well.  This will be discussed in chapter \ref{secondary-chap}.







\section{Results}

We now turn to the results of this thesis.  There are three chapters in the body of the thesis, corresponding to the three separate but related projects undertaken by the author and his collaborators.  These projects all investigate theoretical aspects of gyrokinetic turbulence and are closely related but ultimately, stand alone individually.  The first project, presented in chapter \ref{chap-2-sec} and the appendix \ref{appendix-sec}, is a yet unpublished study of the equations of gyrokinetics extended to describe the mean-scale evolution of a turbulent plasma.  The final two chapters are adapted, with little modification, from papers by the author and his collaborators -- the first published in {\em Physics of Plasmas} \supercite{plunk} and the second submitted to the {\em Journal of Fluid Mechanics}\supercite{plunk-2}.  These two works are a closer look at the details of fully developed gyrokinetic turbulence and are described below.

\subsection{Chapter \ref{chap-2-sec}: Gyrokinetics as a transport theory}

The theory of transport in magnetically confined fusion plasmas began with the formulation of classical transport theory\supercite{braginskii} and then neoclassical transport theory (a review is given by two of the pioneers in \cite{hinton}).  These theories followed the example set by the theory of collisional transport in neutral gases pioneered by Chapman and Enskog.  While being great achievements, neither of the classical transport theories -- which considered transport by diffusive collisional processes -- could, in general, come close to predicting the level of transport in tokamaks.

As we have been discussing, it has become the popular belief that the equation of magnetized turbulence, namely the gyrokinetic equation, must be solved to predict ``anomalous'' transport levels.  On the other hand, it is observed that under some conditions, neoclassical transport does correctly predict the base level of transport -- \ie ion thermal transport in transport barrier regions.  Thus, the most complete descriptions unify neoclassical and anomalous transport.  We provide such a description in this chapter, chapter \ref{chap-2-sec}.


\subsubsection{Scale separation, locality and gyro-Bohm scaling}

Our treatment of gyrokinetics is standard in many regards.  In particular, the ordering of terms and the final gyrokinetic equation obtained is consistent with standard non-linear delta-f gyrokinetics \cite{frieman}.  The main feature which distinguishes our approach from others is (1) the assumption of separation of scales between equilibrium scales and the fluctuation scales and (2) the method (intermediate-scale spatial averaging and time averaging) by which we separate fine scale fluctuations and mean-scale quantities.  This permits the simultaneous derivation of (neo)classical and anomalous transport theory.

We can view separation of scale as an assumption of local homogeneity.  Consider a sub-system which extends over a volume much smaller than the equilibrium scale but much larger than the turbulent scale.  Although there are turbulent fluctuations locally, we assume that these fluctuations do not accumulate coherently to produce variation on the size of the sub-system--for instance, the root-mean-square of the fluctuation can only change significantly on the mean scale, so is approximately constant (homogeneous) in the intermediate-scale sub-system.

\begin{figure}
\ifthenelse{\boolean{JpgFigs}}
{\includegraphics[width=\columnwidth]{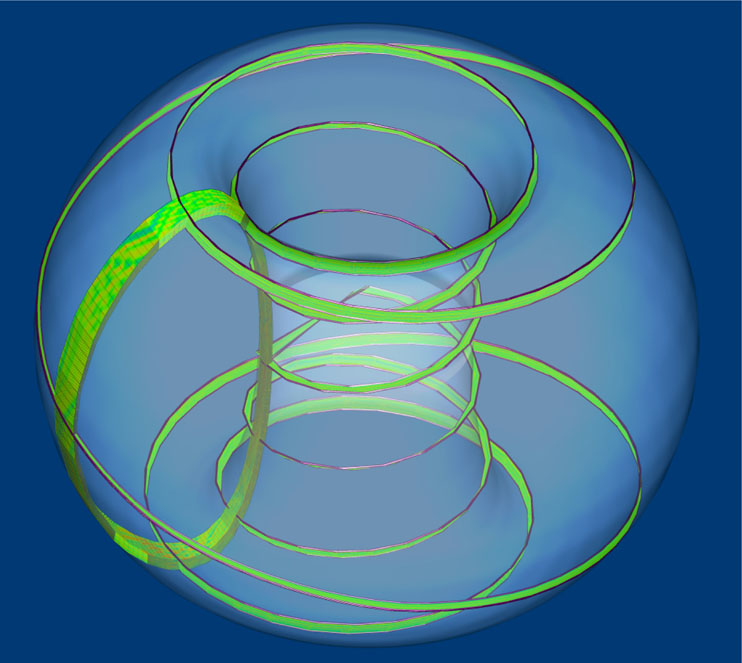}}
{\includegraphics[width=\columnwidth]{Figures/flux-tube-mast-08.pdf}}
\caption{Flux tube simulation domain: A thin flux tube volume surrounding an equilibrium field line is rendered in green.  Also, a small segment of the full annulus volume is included on the left side of the torus.  A larger segment of an annulus volume is pictured in figure \ref{annulus-volume-fig}.  (courtesy of G. Stantchev)}
\label{sim-domain-fig}
\end{figure}

At the level of direct numerical simulation, scale separation implies that small sub-domains of the full system may be simulated separately.  Such a domain may be chosen to be a full annulus segment comprised of the full volume contained between two toroidal surfaces (see section \ref{annulus-vol-sec}), or even smaller domains called flux tubes, which cover a small volume following a magnetic field line.  These domains are represented in figure \ref{sim-domain-fig}.  A reduced domain yields a large improvement in computational efficiency and has made possible the direct simulation of the macro-evolution of the plasma.  This has been achieved in the work by \cite{barnes-thesis} which, has used the formulation of transport (based upon the separation of scales) which will be presented in this chapter.

\subsubsection{Chapter \ref{chap-2-sec} overview}

The presentation given in section \ref{chap-2-sec} gives a detailed description of the assumptions and methods leading to gyrokinetic theory and the mean-scale transport theory -- the solutions to which constitutes the ultimate goal of the study of turbulence in fusion plasmas.  This material will constitute the fundamental framework upon which the subsequent chapters are built.  The notation will largely carry through, with the exception of specific normalizations which will be introduced explicitly before they are used.

The chapter will begin with introductory material describing the ordering assumptions, notations and then proceed to the derivation of the gyrokinetic equation.  The ordered expansion of the Fokker-Planck equation is extended to one order beyond the gyrokinetic equation to obtain the transport equations for density and temperature.  Finally a detailed look at Poynting's theorem and entropy balance will give a perspective of the overall energy balance and mean-scale thermodynamics.

An extension of this work to the case of toroidal magnetic field geometry is given in the appendix \ref{appendix-sec}.  The material in chapter \ref{chap-2-sec} and appendix \ref{appendix-sec} constitute a body of work that is still developing.  In light of its special applicability to the ITER design, it may be adapted to an ITER-specific theory.\supercite{cowley-vienna-talk}  However, in its present form it does have several complete results, some of which have already been incorporated into the work by \cite{barnes-thesis} as mentioned above.

\subsection{Chapter \ref{secondary-chap}: Primary and secondary mode theory}

To understand micro-turbulent transport, one typically starts with the linearly unstable modes which drive the turbulence.  These ``primary'' modes in general exploit a gradient in the free energy of the background plasma and magnetic field.  As these modes grow, they begin to interact nonlinearly, causing the transition to a turbulent state -- a state where transport by the mixing of plasma would bring about the relaxation of the driving gradient, if an external drive were absent.  In the case of tokamak turbulence, however, the steady state operation corresponds to a balance between the overall loss by transport and injection of heat and particles.  This fully developed state remains highly turbulent throughout operation of the device.

This chapter will discuss some details of primary mode analysis, but is mostly focused on ``secondary instability'' theory.  A secondary instability is an exponentially growing mode induced by the presence of a linearly unstable mode which has itself grown to sufficient-amplitude.  Secondary modes or `secondaries' are a key part of the nonlinear physics.  Their study in the context of plasma turbulence can bring understanding of the transition to turbulence.  Also, and perhaps surprisingly, secondary instability theory may also be used to determine features of the fully developed state.

In fully developed plasma turbulence, it cannot be assumed that the chaotic state will bear resemblance to the unstable linear modes from which the state originated.  However, it has been observed that radial streamers (structures existing in the fully developed turbulent state) bear a close resemblance to unstable linear modes.\supercite{dorland-conversation, jenko-conversation}  In light of this observation, it is not surprising that the theory of secondary instabilities has been used successfully to predict saturation amplitudes in fully developed turbulence.\supercite{dorland3}  These saturation amplitudes, as argued above, place an upper bound on radial fluid displacement and therefore enter directly in the mixing length phenomenology of transport.

\subsubsection{Chapter \ref{secondary-chap} overview}

Chapter \ref{secondary-chap} is largely taken without modification from the work \cite{plunk}.  As mentioned, we have added some more detail on the solution of the linear dispersion relation.  The main subject of this chapter, however, is a fully gyrokinetic secondary instability theory.  

Our investigation into the fully gyrokinetic secondary was first motivated, in part, by the possibility of an ultra-thin streamer, with $k_y\rho_e$ significantly greater than 1.  Using the mixing length ideas, such a streamer could enhance transport via its very rapid turnover and large step length.  Also, these structures could evade detection, as (1) the most advanced experimental techniques for measuring electron temperature/density fluctuations\supercite{white} are able to resolve $k \sim 1 cm^{-1}$ (which is much larger than the electron Larmor radius for tokamak plasmas) and (2) the computational requirements of a simulation would be quite large to resolve the temporal and spatial scale range of the nonlinear state in which these structures exist.

The first result coming from this work is that the very fine-scale ($k \rho_e > 1$) ETG secondary instability is robustly unstable, over the full range of parameters investigated.  This indicates that ultra-thin streamers are unlikely to exist.  Then, some unexpected results are reported.  By scanning across several parameters, we find features which suggest the role of secondary instabilities in affecting such macroscopic phenomena as profile stiffness and the electron transport barrier.  Specifically, it is found that the secondary instability for the two dimensional ETG mode (``toroidal'' mode) exhibits a significant sensitivity to the the critical gradient parameter $(R/L_{T})_{\mbox{\scriptsize{crit}}}$ (which determines the instability of the primary mode).  This finding is used, again with the aid of mixing length phenomenology, to offer an explaination of observed transport reduction near marginal stability of the primary mode (the essential idea behind the ``Dimits shift'') and the fact that ETG turbulence exhibits less profile stiffness\footnote{From experiments and simulations of tokamaks, there is a large amount of evidence that the equilibrium profile is maintained at a near-marginal state by the underlying transport.  This concept is known as profile stiffness because transport fluxes increase sharply if the equilibrium is pushed beyond a critical level -- which means the profile exhibits a stiffness or resistance against being externally tuned.} than ITG turbulence.

Using gyrokinetic secondary instability theory, we also are able to explore the intermediate scale range between the electron Larmor radius and the ion Larmor radius.  In this range, which is an example of what we later call the nonlinear phase-mixing range, the ion gyrokinetic equation is strongly affected by the gyro-averaging which cannot be approximated by a fluid treatment.  We calculate the primary and secondary instability using a full gyrokinetic description for both ions and electrons and are able to locate the wavenumber of minimal secondary growth rate.  It is argued that the trough in the secondary growth rate curve may, under suitable conditions, correspond to the peak of the turbulent spectrum, which is one of the most important features in applying mixing length phenomenology to predicting the diffusivity.

To complete the study, we also investigate the secondary instability theory for a three dimensional slab configuration.  We confirm the findings of previous works which indicated that the nonlinear behavior of ITG and ETG do not differ as strongly as they do for the toroidal case.

\subsection{Chapter \ref{phase-space-turbulence-chapter}: The phase space cascade in two dimensional gyrokinetics}

While chapter \ref{secondary-chap} is concerned with the energy containing range, chapter \ref{phase-space-turbulence-chapter} will shift to a study of inertial range gyrokinetics.  This is a very different approach to describing fully developed turbulent state, and although the inertial range represents a relatively small contribution to the ``bottom-line'' of fusion theory -- \eg turbulent transport --  it is surely host to important nonlinear phenomena at play in a general turbulent plasma state.

Gyrokinetics assumes anisotropy in the fluctuations such that structures are elongated in the direction of the equilibrium magnetic field, \ie that $k_{\parallel} \ll k_{\perp}$.  In a subsidiary limit, when the effect of $k_{\parallel}$ is subdominant to the nonlinearity, we may neglect $k_{\parallel}$ altogether to obtain two-dimensional gyrokinetics.  This two dimensional assumption is the basis of the successful Charney--Hasegawa--Mima (CHM) fluid model\supercite{charney, hasegawa} which is closely related to the equation of incompressible two-dimensional fluid turbulence.  There are at least two scenarios, relevant to tokamak turbulence, in which two dimensional gyrokinetics may be an appropriate description.

First consider the case of the cascade of ion-scale turbulence to scales much finer than the ion Larmor radius, $k\rho_i \gg 1$.\footnote{Alternatively, we may consider the cascade from the electron-scale turbulence to even finer scales.  This case relates closely to the ion-scale cascade, as is described in detail in section \ref{eqns-sec}.}  First, for $k\rho \gg 1$ the dynamics are dominated by a nonlinear phase-mixing process which precludes a fluid approximation.  Also, in this limit the linear modes may be stable (a necessary but not sufficient condition for inertial range treatment) and the strength of the nonlinearity grows with $k_{\perp}$ and will dominate over the parallel compressibility term if we assume $k_{\parallel}$ is bounded.  For instance, one conventional assumption is $k_{\parallel} \sim 1/qR$.  That is, the parallel wavelength is assumed to be roughly the distance between the inboard and outboard sides of the toroidal magnetic surface, \ie the distance between the good and bad curvature regions.  Another hypothesis, given by \cite{schek-ppcf}, is that the parallel wavenumber is determined by the distance that particles stream along the field lines during a nonlinear turnover time -- this implies that the parallel term remains balanced with the nonlinearity and cannot be formally neglected.  However, it is pointed out in \cite{schek-ppcf} that this term becomes much less efficient as a phase-mixing mechanism and does not break the conservation of the forward cascading invariant.  Thus this hypothesis of balance is not incompatible with the the exact two-dimensional theory.

Another scenario where the high-$k$ two dimensional theory of gyrokinetics may be useful is in describing the inverse cascade from turbulence driven at the electron Larmor radius to scales as large as the ion Larmor radius.\footnote{As with neutral fluid turbulence (and CHM turbulence), the two dimensional scenario in gyrokinetic turbulence two dynamically invariant quantities.  The spectral scaling of the two invariants has a fixed relationship, as with the case of fluid turbulence, which forces a dual cascade.}  (This may be relevant in cases where the ion scale turbulent drive is quenched by shear flow in, for example, a transport barrier -- although further investigation is necessary to establish this.)  This scenario is also inextricably kinetic, as the nonlinear phase-mixing in the ion gyrokinetic equation is strong at $k\rho_i > 1$.

\subsubsection{The inertial range cascade in tokamak turbulence}

Notably, there are parameter regimes for tokamaks which exhibit linear instability across nearly the full range of dynamical interest.  For instance, the ITG instability may blend continuously into the trapped electron mode (TEM) which is continued as the ETG instability.  Here, it is not clear that the inertial theory is directly applicable.  However, the rigorous justification for inertial range treatment requires a demonstration that the non-conservative terms (\ie the linear drive terms) become subdominant to the other terms in the dynamical equation.  This is possible even when a linear instability is present -- the linear instability would simply be ineffective at injecting energy.

Even when the case of interest does not exhibit an inertial range, the classical inertial range cascade should be understood as one of the possible underlying nonlinear processes.  Also, because an inertial range cascade is a universal feature which is insensitive to the specific forcing mechanism, the verification of theoretical features of the cascade may be used in diagnosing and benchmarking numerical simulations and determine resolution requirements.

As a final note of motivation, the fine-scale contribution to transport fluxes may be obtained from an inertial range theory.  Although the majority of transport is believed to originate from the linearly unstable scales $k\rho_i \lesssim 1$, the turbulent transport at fine scales may represent a non-negligible portion of transport.  A forward-thinking theory of plasma turbulence should include such contributions as they will ultimately be part of quantitative predictions.

\subsubsection{Chapter \ref{phase-space-turbulence-chapter} overview}

These preceding considerations suggest the need for a high-$k$ inertial range theory of gyrokinetics in the two-dimensional limit.  We present such a theory in the final work of this thesis.  We offer that the case of two dimensional gyrokinetic turbulence is a simple paradigm for kinetic plasma turbulence.  

We study the inertial range dual cascade, assuming a localised random forcing.  This cascade occurs in phase-space (two dimensions in position-space plus one dimension in velocity-space) via the nonlinear phase-mixing process, at scales smaller than the Larmor radius.  In this ``nonlinear phase-mixing range,'' we show that the turbulence is self-similar and exhibits power law spectra in position {\em and} velocity-space.  The velocity-space spectrum is treated via a Hankel-transform which fits naturally into the mathematical framework of gyrokinetics.  We derive the exact relations for third order structure functions, in analogy to Kolmogorov's four-fifths law.  

The two dimensional gyrokinetic system bears some resemblance to the equations of incompressible fluid turbulence and, notably, may be rigorously reduced (in the appropriate long wavelength limit) to the familiar Hasegawa-Mima equation or the vorticity equation for two dimensional Euler turbulence.  We investigate the relationship between these theories.  First we review the derivation of the CHM/vorticity equation from gyrokinetics.  Then we derive a relationship between the fluid and gyrokinetic invariants and, after a phenomenological derivation of the inertial range scaling laws, present a picture of the full range of cascades from the fluid range to the fully kinetic range.                         
\chapter{Gyrokinetics as a transport theory}\label{chap-2-sec}

\section{Introduction}

This chapter will introduce the basic equations of gyrokinetics and gyrokinetic transport.  On a basic level, the derivation is a rigorous asymptotic expansion of the Fokker-Planck equation centered around the assumption of scale separation between equilibrium and fluctuation quantities.  The ultimate goals are (1) to obtain the gyrokinetic equation to describe turbulence at the ion Larmor radius scale, (2) obtain the classical counterpart to describe perturbations that vary smoothly in space and time and (3) obtain the evolution equations for the mean-scale quantities (temperature, density, magnetic field) which are influenced by both turbulent fluctuations and collisional effects associated with the inhomogeneity of the equilibrium (classical and neoclassical transport theory).

For the sake of cohesiveness with the other chapters in this thesis, we will avoid discussion of axi-symmetric geometry, flux coordinates and neoclassical theory in this chapter.  Although the full-geometry toroidal case has been treated (and is included in the Appendix), the remainder of the chapters in this thesis will take a local slab (constant curvature) approximation and we find it to be more simple and illuminating in this chapter to formulate gyrokinetics and transport in a local, geometry-free manner.

\section{Scale separation}

The key feature which distinguishes the following approach to gyrokinetics and transport is the assumption of multiple disparate scales in both time and space.  Although there are several approaches to deriving the gyorkinetic system of equations and multiple forms of the final equations (e.g. conservative, non-conservative, gyro-center density, real space density) they share a common set of ordering assumptions.  In the following approach we take scale separation as the fundamental assumption and the ordering assumptions of spatial and time derivatives follow from this.

From this approach, classical (or neoclassical) theory is recovered from mean-scale behavior (governed by the kinetic equations under a suitably defined smoothing operation) while turbulent theory manifests itself in the fine-scale behavior of the perturbative fields.  In other words, by starting with the assumption that the physical fields have dependence concurrently on multiple scales, we can recover both theories on the same footing.  Other works having achieved this synthesis of anomalous and classical theories are \cite{shaing, balescu, sugama-transport, sugama-transport-2}.

It should be pointed out that this approach does have some weakness.  In particular it is not designed to treat meso-scale phenomena\supercite{itoh-itoh-mesoscale} where variation on a scale intermediate to the micro- and macro-scales is present.  This is possible with transport barriers (where the equilibrium background varies on scales approaching the turbulence scale).  However we reiterate that scale separation is widely observed in tokamak experiments and should become an excellent approximation for ITER where $\rho^{\star}$ will be quite small.  Thus while there are surely many interesting and undiscovered meso-scale phenomena, it is the view of the authors that the case of scale separation is crucially important to understanding tokamak turbulence and transport.

In this section we will introduce the two characteristic spatial scales (whose ratio defines the fundamental small parameter of gyro-kinetic theory) as well as the three characteristic time scales.  We then discuss the {\bf method of multiple scales} as a foundation to the derivations of this paper.  Finally we detail the gyrokinetic ordering scheme.

\subsection{Characteristic scales}

We assume the existence of two important spatial scales, the macroscopic spatial scale length $L$ and the microscopic spatial scale length $\rho_i$.  The macroscopic scale length is the distance over which equilibrium quantities (such as density and pressure) vary and is taken to be the size of the plasma--the minor radius, the major radius, etc., will all be taken to be order $L$.  The microscopic scale length is taken to be the ion Larmor radius (we will omit the subscript unless there is need to distinguish it from the electron Larmor radius).  These scale lengths define the fundamental expansion parameter:

\begin {equation}
\epsilon = \frac{\rho}{L} \ll 1
\end{equation}

\noindent (Which is what is commonly referred to as $\rho^{\star}$.)  For example in ITER these lengths will be approximately: $L \sim  n/|\nabla n| \sim 2m$ (the minor radius of ITER) and $\rho\sim 2 mm$\supercite{iter-basis}.  This gives a strong expansion parameter: $\epsilon \sim 10^{-3}$.  There are three basic time scales of interest defined by characteristic frequencies.  The first is the {\em fast} ion cyclotron frequency $\Omega_{0i}$ (we will omit the subscript here also).

\begin{equation}
\Omega_0 = \frac{qB_0}{m_i}
\end{equation}

\noindent In ITER $\Omega_0 \sim 2\times 10^8 rad/s$.  Next is the {\em medium} turbulent frequency $\omega$.

\begin{equation}
\omega = \frac{v_{thi}}{L}
\end{equation}

\noindent where $v_{thi}$ is the ion thermal speed.  On ITER $\omega \sim \times 10^5 rad/s$.  The third time scale is the {\em slow} transport time $\tau$.  On ITER $\tau \sim 3.7 s$.  These frequency scales have a simple relationship in terms of $\epsilon$.

\begin{equation}
\tau^{-1} \sim {\mathcal O}(\epsilon^2\omega) \sim {\mathcal O}(\epsilon^3\Omega_0)
\end{equation}

\subsection{Method of multiple scales}

The method of multiple scales (see for instance \cite{bender}) is characterized by the formal introduction of independent variables to describe dependence on different scales.  The equation describing the system of interest is then perturbatively expanded and solved in these independent variables.  



For our purposes, the distribution function, $f({\bf r}, t)$, has dependence on space and time (and velocity).  Here, however, we formally assume

\begin{equation}
f = f({\bf r_s}, {\bf r}, t_s, t)
\end{equation}

\noindent with $t_s \equiv \epsilon^2\;t$ (``slow time'') and ${\bf r_s} \equiv \epsilon\;{\bf r}$ (``slow space'') taken to be independent variables.  (The fast time scale dependence will only enter at the level of single particle dyanics.  In gyrokinetics, this behavior is not included in the collective dynamics captured by the distribution function and associated fields.)  Now we expand the distribution function, electric field and magnetic field in powers of $\epsilon$

\begin{equation}
f = F_0 + \delta f_1 + \delta f_2\;\;  ...
\end{equation}

\noindent and 

\begin{equation}
{\bf B} = {\bf B}_0 + \delta {\bf B},\;\;\;\; {\bf E} = {\bf\delta E}
\end{equation}

\noindent with $\delta f_n/F_0 \sim {\mathcal O}(\epsilon^{n})$, $|\delta {\bf B}|/|{\bf B_0}| \sim {\mathcal O}(\epsilon)$, etc.  Additionally we take $|\delta{\bf E} |/|v_{thi}{\bf B}_0|  \sim {\mathcal O}(\epsilon)$ where $v_{thi}$ is the ion thermal speed which will be taken to be $\sqrt{T_i/m_i}$.  This expansion is the starting point for the method of multiple scales.  Now we must specify the assumptions of scale dependence for the problem of interest, gyro-kinetic transport theory.  We assume the equilibrium quantities $F_0$ and ${\bf B_0}$ have dependence {\em only} on the slow (transport) time variable $t_s$, the slow (macroscopic) spatial vector ${\bf r_s}$ and velocity ${\bf v}$.  The perturbative quantities ${\bf \delta B}$, ${\bf \delta E}$ and $\delta f$ have additional dependence on the turbulent scale variables denoted simply by $t$ and ${\bf r}$.  Furthermore, as with standard gyro-kinetic ordering, the microscopic spatial dependence is assumed only in the direction perpendicular to the equilibrium magnetic field ${\bf B_0}$.  We can summarize these statements as follows:

\begin{eqnarray}
&F_0 = F_0({\bf r_s, v}, t_s)\nonumber
\\&{\bf B_0} = {\bf B_0}({\bf r_s, v}, t_s)\nonumber
\\&\delta f = \delta f({\bf r}_{\perp}, {\bf r_s, v}, t, t_s)\nonumber
\\&{\bf\delta E} = {\bf\delta E}({\bf r}_{\perp}, {\bf r_s, v}, t, t_s)\nonumber
\\&{\bf\delta B} = {\bf\delta B}({\bf r}_{\perp}, {\bf r_s, v}, t, t_s)
\end{eqnarray}

\noindent These assumptions are important for all that follows.  However, we find it possible to avoid the cumbersome subscript notation entirely.  This is done by first explicitly stating the ordering of spatial and temporal variations (described in section \ref{ordering-sec}) and by introducing spatial and temporal averages that serve to separate scale dependence (section \ref{averaging-sec}).  In the following analysis, we thus need only refer to a single space vector ${\bf r}$ and time variable $t$.  As a final note, velocity space dependence is here assumed to occur on the scale of the ion thermal velocity.  There are subtleties involved in the formation of velocity structure and these will be explored in Chapter \ref{phase-space-turbulence-chapter} of this thesis.

\section{Gyrokinetic ordering}

\label{ordering-sec}
Here we describe the ordering of spatial and temporal variation that results directly from our assumptions of scale dependence.  In brief, the slow scale dependence of equilibrium quantities results in space and time derivatives that are an order smaller than the same derivatives performed on perturbative quantities.  These orderings are essential for solving the Fokker-Planck equation and are summarized in the following table.

\begin{center}
\fbox{
\begin{minipage}{4in}
\medskip
\begin{center}{\bf Ordering of Spatial and Temporal Variation}\end{center}
\hrule
\begin{itemize}
\item $\nabla \sim  {\mathcal O} (\frac{1}{L}) \;\;\;$ acting on $F_0$ or ${\bf B_0}$
\item $\frac{\partial}{\partial t} \sim {\mathcal O} (\epsilon^2\omega)\;\;\;$ acting on $F_0$ or ${\bf B_0}$
\item $\nabla_{\perp} \sim {\mathcal O}(\frac{1}{\rho})\;\;\;$ acting on $\delta f_1$, ${\bf \delta B}$ or ${\bf \delta E}$
\item $\nabla_{\parallel} \sim {\mathcal O}(\frac{1}{L})\;\;\;$ acting on $\delta f_1$, ${\bf \delta B}$ or ${\bf \delta E}$
\item $\frac{\partial}{\partial t} \sim  {\mathcal O}(\omega)\;\;\;$ acting on $\delta f_1$, ${\bf \delta B}$ or ${\bf \delta E}$
\end{itemize}
\medskip
\end{minipage}
}
\end{center}

\noindent where $\nabla_{\perp}$ and $\nabla_{\parallel}$ denote the gradient perpendicular and parallel to ${\bf B_0}$.

\subsection{Electron and ion orderings}
\label{electron-ordering-sec}

Because $\rho$ is taken to be the ion Larmor radius, we are considering ion scale turbulence.  We may imagine, for instance, that the electron-larmor-radius-scale instability (ETG) is not present in the system of interest.  (In practice, it is believed that the feedback of electron scale turbulence onto the ion scale is minimal Ref. \cite{waltz, gorler-jenko}.)  The turbulent frequency is defined $\omega = v_{thi}/L$ relative to the ion thermal speed.  It is important to make these assumptions explicit at this point because they will be implicit during the analysis.  In the case of electron dynamics (for ion-scale turbulence), we will be able to exploit an additional small parameter: the electron-ion mass ratio is taken to be the order of the expansion parameter:

\begin{equation}
\frac{m_e}{m_i} \sim {\mathcal O}(\epsilon)
\end{equation}

\noindent Formally, this will require fractional ordering of the equations when square roots of the mass ratio appear.  However, there is nothing fundamentally new about the analysis and the results follow easily from the ion case, as we will see.

\section{Averaging and scale separation}
\label{averaging-sec}

As previously mentioned, the multiple scale approach aims to obtain both turbulent and classical (or neoclassical) theory simultaneously.  To this end, we use smoothing averaging to separate the equations for equilibrium-scale quantities from the rapid turbulent dynamics described by gyro-kinetic theory.  This section introduces time and space smoothing operators which averaging over regions of space and time intermediate to the cyclotron, turbulent and transport scales.  It should be stressed that these are tools of the formalism and the final equations do not depend on a specific

\subsection{Gyro-average and single particle motion}

The gyro-average exploits the disparity between the cyclotron frequency and the turbulent frequency.  The rapid rotation of individual particles sample turbulent electromagnetic fluctuations which are static on this timescale.  This means that turbulent timescale dynamics are influenced by the orbit or gyro-average of the fluctuating fields.  Let's examines individual particle motion more closely.  The velocity vector is defined using cylindrical coordinates aligned with the magnetic field.

\begin{equation}
{\bf v} = v_\parallel{\bf b}_0 + v_\perp(\cos\vartheta \,{\bf e}_1 + \sin\vartheta\,{\bf e}_2).
\label{vdef}
\end{equation}

\noindent where the unit vectors ${\bf b}_0$,  ${\bf e}_1$ and ${\bf e}_2$ form a local right handed coordinate basis {\em i.e.} ${\bf e}_1\times{\bf e}_2 = {\bf b}_0$.  (Because these basis vectors are defined by the equilibrium field geometry, it should be apparent that they vary on the macroscopic, $L$, spacial scale and the slow, $\tau$, time scale.)  The fastest motion is the gyro-motion:

\begin{equation}
\frac{d\vartheta}{dt} = -\Omega_0 + {\mathcal O}(\epsilon\Omega_0).
\label{phase}
\end{equation}

\noindent where $\Omega_0 = qB_0/m$ is the cyclotron frequency for the species of interest.  An individual particle's velocity is thus fluctuating rapidly on the fast time scale.  If we look at the motion of the gyro-center, however, one finds it's fluctuations are small.  One can use the gyro-average to separate smooth turbulent scale motion of the gyro-center to obtain the drift behavior of particle motion in the gyro-kinetic limit.  The gyro-center position is defined by the the {\em Catto Transformation}

\begin{equation}
{\bf R} = {\bf r} - \rhovec 
\label{catto}
\end{equation}

\noindent where $\rhovec$ is the larmor radius vector defined by

\begin{equation}
\rhovec = \frac{{\bf b_0}\times{\bf v}}{\Omega_0}
\label{larmor-eqn}
\end{equation}

\noindent The gyro-average is defined as follows

\begin{equation}
\gyroavg{A({\bf r}, {\bf v}, t)} = \frac{1}{2\pi}\int^{2\pi}_0 A({\bf R} - \frac{{\bf v}\times{\bf b}_0}{\Omega_0}, {\bf v}, t)d\vartheta.
\label{gyroav}
\end{equation}

\noindent It is an average over the gyro-angle $\vartheta$ with the gyro-center position {\bf R} held fixed.  Note that the parallel velocity, magnitude of the perpendicular velocity and all equilibrium quantities are fixed since they do not significantly vary over a particle's trajectory during a gyro-period.  If one applies the gyro-average to the rate of change of the gyro-center position, $d{\bf R}/dt$, one can obtain the drift velocity of a particle's gyro-center (see section \ref{gyro-sec}).  To reiterate, this quantity varies on the medium turbulent time scale, with cyclotron time scale behavior having been smoothed away.

\subsection{Spatial averages}
\label{space-av-sec}

\begin{figure}
\includegraphics[width=5in]{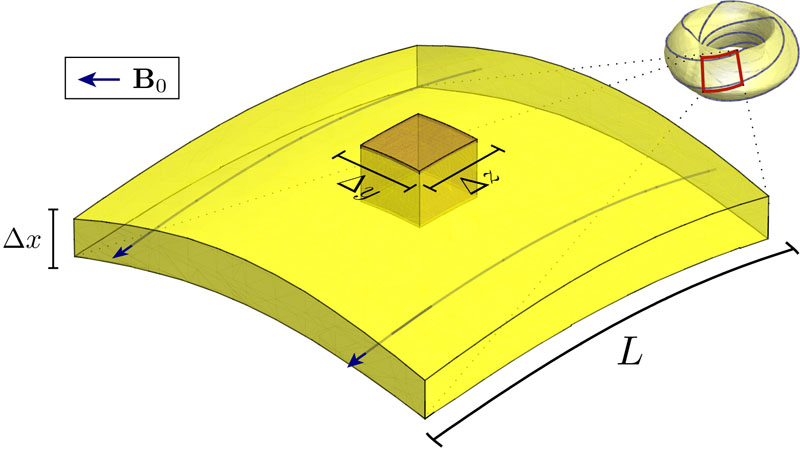}
\caption{Patch Volume and Scale Separation: The patch volume $V_p$ (colored brown) is chosen to be small enough such that the equilibrium geometry is flat but large enough to contain many turbulent scale (correlation) lengths.}
\label{local-patch-fig}
\end{figure}

To separate turbulent spatial scale from the equilibrium scale we employ a {\bf patch volume average}.  The patch volume $V_p$ is defined around a point $(x, y, z)$ defined using locally flat coordinates as indicated in Fig.~\ref{local-patch-fig}.

\begin{equation}
\patch{A} = \frac{\int_{V_p} d^3{\bf r^{\prime}}A}{V_p} = \frac{\int_{x-\Delta x}^{x+\Delta x}dx^{\prime}\int_{y-\Delta y}^{y+\Delta y}dy^{\prime}\int_{z-\Delta z}^{z+\Delta z}dz^{\prime} A(x^{\prime},y^{\prime},z^{\prime})}{\Delta x\Delta y \Delta z}
\label{patch-eqn}
\end{equation}

\noindent where the intervals $\Delta x$, $\Delta y$ and $\Delta z$ define a region with dimensions intermediate to the microscopic and macroscopic spatial scales--i.e. we require 

\begin{equation}
\rho \ll \Delta x, \;\; \Delta y, \;\; \Delta z \ll L.
\end{equation}

\noindent Formally, we take $\Delta x \sim \Delta y \sim \Delta z \sim \sqrt{\epsilon}L$. The details of the patch average definition are less important than its practical utility.  As stated, it smooths away turbulent scale dependence.  For example, if one assumes that $|\nabla A| \sim {\mathcal O}(A/\rho)$ it is easily shown that $|\nabla(\patch{A})| \ll |\nabla A|$.  In analogy to the gyro-average which smoothes out fast time scale behavior, the patch average smoothes microscopic turbulent spatial structure.  We will use it often to separate quantities into equilibrium and turbulent parts (see Sec.~\ref{maxwell-sec}). 

\subsection{Time average}
The last average acts between the slow transport time scale and the medium turbulent time scale, separating out turbulent time scale structure.  The time average is defined

\begin{equation}
{\bar A} = \frac{1}{\Delta t}\int_{t-\Delta t/2}^{t+\Delta t/2}A \;\;dt^{\prime},\;\;\;\; \omega^{-1} \ll \Delta t \ll \tau
\label{t-average}
\end{equation}

\noindent Formally, we take $\Delta t \sim \epsilon^{3/2}\tau$. 

\section{Maxwell's equations and potentials}
\label{maxwell-sec}

In this section we demonstrate use of the ordering assumptions to obtain some simple results from Maxwell's Equations.  Let's start with Faraday's Law

\begin{equation}
\frac{\partial {\bf B}}{\partial t} = -\nabla \times \delta {\bf E}.
\end{equation}

\noindent Upon ordering, the dominant equation is

\begin{equation} 
\nabla \times \delta {\bf E} = 0
\end{equation}

\noindent This means that the inductive part of the electric field must be higher order than the electrostatic part so we may write

\begin{eqnarray}
\delta  {\bf E} = -\nabla (\varphi) + {\mathcal O}(\epsilon^2)
\end{eqnarray}

\noindent This suggests using potentials to express the fields.  Thus we define

\begin{equation}
{\bf A = A_0 + \delta A}
\end{equation}

\noindent where 

\begin{eqnarray}
&{\bf B_0} = \nabla\times{\bf A_0}
\\&{\bf\delta B} = \nabla\times{\bf \delta A}
\end{eqnarray}

\noindent For definiteness we use the coulomb gauge, $\nabla \cdot{\bf A} = 0$.  Note that our ordering requires ${\bf\delta A} \sim {\mathcal O}(\epsilon^2 {\bf A_0})$.  Additionally, it should be apparent that ${\bf A_0}$ is an equilibrium quantity and thus has only slow scale dependence like ${\bf B_0}$.  Likewise, ${\bf \delta A}$ has turbulent scale dependence.  The perturbative electric field becomes

\begin{equation}
\delta {\bf E} = - \nabla\varphi  - \frac{\partial {\bf A}}{\partial t}
\end{equation}

\noindent where $\partial {\bf A}/\partial t$ is composed of equilibrium and perturbative parts of the same order in size.  We now show that ${\bf \delta E}$ and ${\bf \delta B}$ both have small patch averages.  Since ${\bf \delta E}$ is mostly electrostatic we may write

\begin{equation}
\patch{\bf \delta E} \sim -\patch{\nabla\varphi}
\end{equation}

\noindent To demonstrate use of the patch average, we apply it to the electrostatic field:

\begin{equation}
\patch{\nabla \varphi} = \frac{\int dx\;dy\;dz\; \nabla\varphi}{\Delta x\Delta y\Delta z}
\end{equation}

\noindent recalling that $\rho \ll \Delta x, \Delta y, \Delta z \ll L$.  Then we have

\begin{equation}
\patch{{\bf \delta E}} \sim {\mathcal O}(-{\bf\hat{x}}\frac{\varphi}{\Delta x} - {\bf\hat{y}}\frac{\varphi}{\Delta y} - {\bf\hat{z}}\frac{\varphi}{\Delta z}) \ll {\bf \delta E}
\end{equation}

\noindent The proof works likewise for ${\bf \delta B}$.

\section{Ordered Fokker-Planck equation for ion species}

\label{ordered-ions-sec}

In Secs.~\ref{ordered-ions-sec} and \ref{ordered-e-sec} we will be solving the ordered Fokker-Planck equation with the goal of arriving at the gyro-kinetic equation and classical equation which are used to solve for $\delta f_1$.  The ion case is treated in more detail because the electron case can be approached analogously.  Using the orderings listed in Sec.~\ref{ordering-sec}, we write down the 
Fokker-Planck equation with terms ordered in powers of $\epsilon$ relative to ${v_{thi}}F_0/{L}$. Each set of terms with the same power of $\epsilon$ will yield an equation to be solved.  In this section, we will solve the Fokker-Planck equation for ions at each power of $\epsilon$ for the three largest orders.  

\begin{equation}
\frac{\partial f}{\partial t} + {\bf v}\cdot\nabla f +
\frac{q}{m}({\bf E} + {\bf v}\times{\bf B})\cdot \frac{\partial
  f}{\partial {\bf v}} = C(f, f) \label{FPeq}
\end{equation}

\subsection{${\mathcal O}(\epsilon^{-1})$ equation:}
\label{zero-sec}

One term in the Fokker-Planck equation is larger than all
others.  

\begin{equation}
{\bf v\times B}\cdot\frac{\partial F_0}{\partial {\bf v}} =
-\Omega_0\left(\frac{\partial F_0}{\partial \vartheta}\right)_{{\bf r},
  v, v_\perp} =0 
\end{equation}
from which we deduce that $F_0$ is independent of gyro-angle $\vartheta$ so
that:
\begin{equation}
F_0 = F_0(r, v, v_\perp, t) \label{MaxTheta}
\end{equation}

\noindent Now we proceed to ${\mathcal O}(1)$:

\subsection{${\mathcal O}(1)$ equation:}
\label{first-sec}
The next order ion equation is
\begin{equation}
{\bf v}\cdot\nabla F_0 + {\bf v}_\perp\cdot\nabla \delta f_1 +
\frac{q}{m}\left( -\nabla\varphi + {\bf v\times\delta
  B}\right)\cdot\frac{\partial F_0}{\partial {\bf v}}
-\Omega_0\left(\frac{\partial \delta f_1}{\partial \vartheta}\right)_{{\bf
    r}, v, v_\perp} = C(F_0, F_0)
\label{order1}
\end{equation}

\noindent Applying the patch average to this equation, we obtain a separate equation containing the slowly terms.

\begin{equation}
{\bf v}\cdot\nabla F_0  + -\Omega_0\left(\frac{\partial \patch{\delta f_1}}{\partial \vartheta}\right)_{{\bf
    r}, v, v_\perp} = C(F_0, F_0)
\label{order1_slow}
\end{equation}

\noindent From this equation we use Boltzman's H-theorem to show that $F_0$ must be
Maxwellian.  To do this, we multiply Eqn~(\ref{order1}) by $1 + \log
F_0$, integrate over velocity space.  We find that upon integrating,
most of the terms on the left side vanish, leaving us with

\begin{eqnarray}
\int d^3{\bf v} v_{\parallel} {\bf b}_0\cdot\nabla F_0 =  \int d^3{\bf v} C(F_0, F_0)\ln F_0 = 0
 \label{btrick}
\end{eqnarray}

\noindent We make the additional assumption that ${\bf b}_0\cdot\nabla F_0 = 0$.  This assumption is valid for systems with closed flux-surface geometry (as proven in the Appendix) such as toroidal fusion devices because rapid transport along the equilibrium magnetic field lines enforces constant equilibrium temperature and density on each flux surface.  What we are left with from Eqn~(\ref{btrick}), to ${\mathcal O}(1)$, is 

\begin{equation}
\int d^3{\bf v}(\ln F_0)C(F_0, F_0) = 0\label{boltz}
\end{equation}

\noindent Eqn~(\ref{boltz}) and the H theorem tell us there is no entropy production at this order.  Our first significant result in this section is found because we know that the entropy production 
can only be zero if $F_0$ is a local Maxwellian.

\begin{equation}
F_0 = \frac{n}{(\sqrt{2\pi}v_{th})^3}\exp{-(\frac{v^2}{2 v^2_{th}})}\label{localMaxwellian}
\end{equation}

\noindent Given this result, we can obtain information about $\delta f_1$ by
using our solution for $F_0$ in Eqn~(\ref{order1}).  

\begin{equation}
{\bf v}_\perp\cdot\nabla \delta f_1  -\Omega_0\left(\frac{\partial
  \delta f_1}{\partial \vartheta}\right)_{{\bf r}, v, v_\perp} = -{\bf
  v_{\perp}}\cdot\nabla(\frac{q\varphi}{T})\; F_0 - {\bf v_{\perp}\cdot\nabla} F_0\label{order11}
\end{equation}

\noindent We have dropped the factor $v_\parallel{\bf b}_0\cdot\nabla(\frac{q\varphi}{T})\; F_0$ from the right hand side because it is negligible at this order.  The following particular solution satisfies the eqaution at this order:

\begin{equation}
\delta f_{1p} = -(\frac{q\varphi}{T})\; F_0 - {\bf\rhovec\cdot\nabla} F_0.\label{particular}
\end{equation}

\noindent where $\rhovec$ is the larmor radius vector defined by Eqn.~\ref{larmor-eqn}.  The first term is the perturbed Boltzmann response of the particles to the potential.  The potential is static on the gyration time scale and therefore the energy ${\mathcal E} = (1/2)mv^2 + q\varphi$ is conserved to this order.  The second term can be recognized as the beginning of an expansion of $F_0$ about the gyro-center position:

\begin{equation}
F_0({\bf R}) \sim F_0({\bf r}) - {\bf\rhovec}\cdot\nabla F_0({\bf r})
\end{equation}

\noindent We absorb these terms into a new definition of $F_0$ in gyro-center coordinates, {\em i.e.}

\begin{equation}
F_0 = F_m({\bf R})\exp{ -(\frac{q\varphi}{T})}\label{boltz2}
\end{equation}

\noindent where $F_m$ is the maxwellian distribution with real-space coordinates

\begin{equation}
F_m = n(t, {\bf r})\left(\frac{m}{2\pi T(t, {\bf r})}\right)^{3/2}\exp{\left[-\frac{(1/2)mv^2}{T(t, {\bf r})}\right]}
\label{modmax}
\end{equation}  

\noindent To obtain the homogeneous solution for $\delta f_1$ in Eqn.~(\ref{order11}), we rewrite the left hand side of this equation using the following identity:

\begin{equation}
\Omega_0\left(\frac{\partial}{\partial \vartheta}\right)_{{\bf R}} =
\Omega_0\left(\frac{\partial}{\partial \vartheta}\right)_{{\bf r}} - {\bf
  v}_\perp\cdot\nabla\label{deexi}
\end{equation}

\noindent Dropping the right hand side of Eqn.~(\ref{order11}) will give us the equation for the homogeneous part of $\delta f_1$, $\delta f_{1h}$:

\begin{equation}
\left(\frac{\partial \delta f_{1h},}{\partial \vartheta}\right)_{{\bf R}} = 0.
\label{deexi1}
\end{equation}

\noindent Thus the homogeneous part is independent of gyro-angle at fixed ${\bf R}$ {\em i.e.}, 

\begin{equation}
\delta f_{1h} = h({\bf R}, v, v_\perp, t)
\label{deexi2}
\end{equation}

\noindent Sometimes $h({\bf R}, v, v_\perp, t)$ is called the {\bf Guiding center distribution}.  As we show in next order $h$ satisfies the {\bf gyro-kinetic equation}.  Combining all ${\mathcal O}(1)$ results gives the following distribution function:

\begin{equation}
f({\bf r}, {\bf v}, t) = F_0 ({\bf r}, {\bf v}, t) + h({\bf R}, v, v_\perp, t) + \delta f_2({\bf r}, {\bf v}, t) + ...\label{fform}
\end{equation}

\noindent This fixes the form of the solution for the distribution function $f$.  However, we still need to derive equations to describe the evolution of $h({\bf R}, v, v_\perp, t)$, $n(t, \psi)$ and $T(t, \psi)$.  Now we proceed to ${\mathcal O}(\epsilon)$ where we obtain the gyro-kinetic equation and classical equation to solve for $h$:

\subsection{${\mathcal O}(\epsilon)$: Gyrokinetic equation}
\label{gyro-sec}

It is now straightforward to take Eqn~(\ref{fform}) and Eqn~(\ref{modmax}) and proceed to the next order in the Fokker-Planck equation.  We would then gyro-average this equation to obtain the gyrokinetic equation.  This process is somewhat tedious.  We find an improved route by recasting
the Fokker-Planck equation in gyro-center variables:

\begin{eqnarray}
{\bf R} = {\bf r} + \frac{{\bf v}\times{\bf b_0}}{\Omega_0}, & {\mathcal E} = \frac{1}{2}mv^2 + q\varphi, &  \;\; \mu = \frac{mv_\perp ^2}{2B_0}\nonumber
\end{eqnarray}

\noindent The Fokker-Planck equation takes the form

\begin{equation}
\frac{\partial f}{\partial t} + \frac{d{\bf R}}{d
  t}\cdot\frac{\partial f}{\partial {\bf R}} +
\frac{d\mu}{dt}\frac{\partial f}{\partial\mu} + \frac{d{\mathcal E
}}{dt}\frac{\partial f}{\partial {\mathcal E}} +
\frac{d\vartheta}{dt}\frac{\partial f}{\partial \vartheta} = C(f, f)
\end{equation}

\noindent Our final equation will involve the gyro-average defined in Eqn~(\ref{gyroav}) of $d{\bf R}$/$dt$, $d {\mathcal E}$/$ dt$, $d\vartheta$/$dt$ and $d \mu$/$ dt$.   Calculated to necessary order, these quantities are

\begin{eqnarray}
&\gyroavg{\frac{d{\bf R}}{dt}} = v_\parallel{\bf b}_0 + \frac{{\bf b}_0}{\Omega}\times\left(\frac{q\gyroavg{\nabla\chi}}{m} + v^{2}_{\parallel}{\bf b}_0\cdot\nabla{\bf b}_0 + \frac{v^{2}_{\perp}}{2}\nabla\ln B_0 \right)
\label{Rdot2} \\
&\gyroavg{\frac{d{\mathcal E}}{dt}} =  q\left(\frac{\partial\gyroavg{ \chi}}{\partial t} - {\bf v_{\parallel}}\cdot\frac{\partial {\bf A_0}}{\partial t}\right)
\label{energy}\\
&\gyroavg{\frac{d\vartheta}{dt}} = -\Omega_0\\
&\gyroavg{\frac{d\mu}{dt}} = 0 
\end{eqnarray}

\noindent where we define $\chi = \varphi - {\bf v}\cdot{\bf\delta A}$.  To be explicit, we note that in these variables $f = F_0 ({\bf R}, {\mathcal E}, t) + h({\bf R}, {\mathcal E}, \mu, t) + \delta f_2({\bf R}, {\mathcal E}, \mu, \vartheta, t) + ...$  The ${\mathcal O}(\epsilon)$ equation is found to be 

\begin{eqnarray}
&\frac{\partial h}{\partial t} + \frac{d{\bf R}}{d
  t}\cdot\frac{\partial}{\partial {\bf R}}(h + F_0) +
\frac{d\mu}{dt}\frac{\partial h}{\partial\mu} - C(h - \rhovec\cdot \nabla F_m, F_m) \nonumber \\
& = \Omega_0\left(\frac{\partial \delta
  f_2}{\partial \vartheta}\right)_{{\bf R}} + \frac{1}{T} \frac{d{\mathcal E
  }}{d t}F_0 
\end{eqnarray}

\noindent $F_m$ inside the collision operator refers to Eqn.~\ref{localMaxwellian}, and we will write $C(., F_m)$ as $C(.)$ from now on.  Now we can eliminate $\delta f_2$ to obtain an equation in $h$ only -- to do this we gyro-average the equation.  With some algebra we arrive at the following.

\begin{eqnarray} 
&\frac{\partial h}{\partial t} + v_{\parallel}{\bf b}_0\cdot \nabla h + {\bf v}_D + {\bf v}_{\chi} \cdot\nabla h -
\gyroavg{C(h)} \nonumber \\
& = q\frac{F_0}{T_0} \left(\frac{\partial\gyroavg{ \chi}}{\partial t} - {\bf v_{\parallel}}\cdot\frac{\partial {\bf A_0}}{\partial t}\right) -
       ({\bf v}_D + {\bf v}_{\chi}) \cdot\nabla F_0
\label{pre_gyro}
\end{eqnarray}

\noindent where the equilibrium and perturbative parts of the drift velocity are split as follows

\begin{align}
{\bf v}_D& = \frac{{\bf b}_0}{\Omega}\times\left(v^{2}_{\parallel}{\bf b}_0\cdot\nabla{\bf b}_0 + \frac{v^{2}_{\perp}}{2}\nabla\ln B_0\right)\nonumber \\
{\bf v}_{\chi}& =  \frac{{\bf b}_0}{B_0}\times\gyroavg{\nabla\chi}
\label{drift_velocity}
\end{align}

\noindent Note that the term $C(-\rhovec\cdot\nabla F_m,F_m)$ in Eqn.~\ref{pre_gyro} was annihilated by the gyro-average.  Hidden in this equation are actually two independent equations.  This is due to the multi-scale dependence of h.  We recall the patch operator and time average from Sec.~\ref{space-av-sec} and define their action on $h$ as follows.  

\begin{equation}
{\mathcal P}[\bar{h}] = h_{cl}
\end{equation}

\noindent This allows us to split up $h$ into two parts: $h = h_{cl} + h_{gk}$, where ``cl'' stands for classical (which will be the neoclassical response in toroidal geometry) and ``gk'' stand for gyro-kinetic.  Applying ${\mathcal P}$ and the time average to equation (\ref{pre_gyro}) gives the equation for the ``classical'' part of $h$ (an equation of slow varying quantities).  


\begin{equation}
v_{\parallel}{\bf b}_0\cdot \nabla h_{cl} - C(h_{cl}) = -{\bf v}_D \cdot\nabla F_m - q\frac{F_m}{T_0 }{\bf v_{\parallel}}\cdot\frac{\partial {\bf A_0}}{\partial t}
\label{gyro-slow}
\end{equation}


\noindent Now subtracting equation (\ref{gyro-slow}) from equation (\ref{pre_gyro}) above gives the {\bf Gyro-kinetic equation}:

\begin{equation}
\frac{\partial h_{gk}}{\partial t} + (v_{\parallel}{\bf b}_0 + {\bf v}_D + {\bf v}_{\chi})\cdot\nabla h_{gk} - \gyroavg{C(h_{gk})} = q\frac{F_0}{T_0 }\frac{\partial\gyroavg{\chi}}{\partial t} - {\bf v}_{\chi} \cdot\nabla F_0
\label{the-gyro-equation}
\end{equation}



\noindent In some loose sense the {\bf Gyro-kinetic equation} is the kinetic equation for rings of charge centered at ${\bf R}(t)$ of radius $v_\perp/\Omega$.

\subsection{Kinetic and classical equations for electron species}

\label{ordered-e-sec}

\noindent As described in Sec.~\ref{electron-ordering-sec}, we assume  gradients
and time derivatives act on the electron distribution function at the same order as ions.  The significant
difference when deriving the kinetic equation for electrons is the smaller mass.  When re-examining terms in the Fokker-Planck equation, Eqn~(\ref{FPeq}), we order the square root of the mass ratio as $\sqrt{m_e/m_i} \sim \sqrt{\epsilon}$ which formally introduces fractional ordered terms in our perturbative expansion.  However it is much more convenient to take a maximal ordering approach and include root mass ratio terms until the final equation (the kinetic equation for electrons) is obtained at which point we can introduce the effects of the mass ratio as a subsidiary ordering.  Thus we can immediately write the gyrokinetic and classical equations from the previous section for the electron species and include only dominant terms in the root mass ratio:

\begin{equation}
v_{\parallel}{\bf b}_0\cdot\nabla h_{cl} - C(h_{cl}) = -q \frac{F_m}{T_0}{\bf v_\parallel}\cdot\frac{\partial {\bf A_0}}{\partial t}
\label{gyro-slow-e}
\end{equation} 

\noindent and

\begin{equation}
(v_{\parallel}{\bf b}_0 + {\bf v}_e)\cdot\nabla h_{gk} - C(h_{gk}) = -{\bf v}_e \cdot\nabla
F_0 - q \frac{F_0}{T_0}{\bf v_\parallel}\cdot\frac{\partial {\bf \delta A}}{\partial t}
\label{gyro-e}
\end{equation} 

\noindent where ${\bf v}_e$ is the perturbative magnetic field drift velocity for electrons

\begin{equation}
{\bf v}_e = -\frac{b_0\times \nabla(v_\parallel \delta A_\parallel)}{B_0}
\end{equation}



\section{Maxwell's equations for gyrokinetics}
\label{gyro-maxwell-sec}

The gyro-kinetic equation from the previous section determines how to
evolve the turbulent distribution function.  This section describes
the equations that will evolve the turbulent potentials, and
consequently, the turbulent fields.  

\subsection{Quasi-neutrality}

\begin{eqnarray} 
&-\frac{n_iq^2\varphi}{T_i} + 2\pi q\int\int v_\perp dv_\perp
 dv_\parallel \angleavg{h_{i,gk}({\bf R}, v_\perp, v_\parallel,t)} \nonumber \\
 & = \frac{n_ie^2\varphi}{T_e} + 2\pi e\int\int v_\perp dv_\perp
 dv_\parallel\angleavg{ h_{e,gk}({\bf R}, v_\perp, v_\parallel,  t)}
\label{qn}
\end{eqnarray}

\subsection{Parallel Ampere's law}

\begin{equation} 
{\bf b_0}\cdot\nabla^2{\bf A} = \mu_0 { J}_{\parallel} = \int d^3{\bf v}v_{\parallel}[q\;f_i - e\;f_e]
\end{equation}

\noindent which results in two equations upon patch averaging.

\begin{equation}
\nabla^2\delta A_{\parallel} = \mu_02\pi \int\int dv_\perp dv_\parallel v_\parallel v_\perp [q\;h_{i,gk} - e\;h_{e,gk}] 
\label{apara}
\end{equation}

\noindent and

\begin{equation}
{\bf b_0}\cdot\nabla^2{\bf A_0} = \mu_02\pi \int\int dv_\perp dv_\parallel v_\parallel v_\perp [q\;h_{i,cl} - e\;h_{e,cl}] 
\label{apara0}
\end{equation}

\subsection{Perpendicular Ampere's law}

\begin{equation}
(\nabla\times{\bf B})\times{\bf b_0} = \mu_0\int d^3{\bf v}({\bf v}\times{\bf b_0})[q\;f_i - e\;f_e]
\end{equation}

\noindent again, giving two equations.

\begin{equation}
-{\bf \nabla}_\perp \delta B_{\parallel} = \mu_0\int d^3{\bf v}({\bf v}\times{\bf b_0})[q\;h_{i,gk} - e\;h_{e,gk}]
\end{equation}

\noindent and

\begin{equation}
B_0^2{\bf b_0 \cdot\nabla b_0} - \frac{1}{2}{\bf \nabla_{\perp}}B_0^2 = \mu_0\int d^3{\bf v}({\bf v}\times{\bf B_0})[q(-\rhovec_i\cdot\nabla F_{mi}) - e(-\rhovec_e\cdot\nabla F_{me})]
\label{perp-ampere-eq}
\end{equation}

\subsection{Pressure balance}

\noindent The right hand side of this equation, $\mu_0{\bf J_0}\times{\bf B_0}$, can be manipulated to give the familiar equilibrium force balance.  The surviving terms after velocity space integration can be identified as the polarization current (recall that we have assumed there is no equilibrium scale electrostatic potential).

\begin{align}
{\bf J_0}\times{\bf B_0} &= B_0\int d^3{\bf v}({\bf v}\times{\bf b_0})[q(-\rhovec_i\cdot\nabla F_{mi}) - e(-\rhovec_e\cdot\nabla F_{me})]\nonumber
\\ & = \int d^3{\bf v}({\bf v}\times{\bf b_0})({\bf v}\times{\bf b_0})\cdot\nabla[m_i F_{mi} + m_e F_{me}]\nonumber
\\ &= \int d^3{\bf v}\frac{v_{\perp}^2}{2}{\bf \nabla}_{\perp}[m_iF_{mi} + m_eF_{me}]\nonumber
\\ &= \frac{2}{3}\int d^3{\bf v}{\bf \nabla}_{\perp}[\frac{m_iv^2}{2}F_{mi} + \frac{m_ev^2}{2}F_{me}]\nonumber
\\ &= {\bf \nabla}P_0
\end{align}

\noindent where we recall $\rhovec_s = \frac{{\bf b}_0\times{\bf v}}{\Omega_{0s}}$ and used the fixed-${\bf r}$ ring average between lines one and two and use that $P_0 = \frac{2}{3}\int[\frac{m_iv^2}{2}F_{mi} + \frac{m_ev^2}{2}F_{me}]$.  Substituting this into Eqn.~\ref{perp-ampere-eq} we arrive at the following equation

\begin{equation}
B_0^2{\bf b_0 \cdot\nabla b_0} - \frac{1}{2}{\bf \nabla_{\perp}}B_0^2 = \mu_0{\bf \nabla}P_0
\label{p-bal}
\end{equation}

\noindent This equation describes the macroscopic equilibrium of the plasma.  This completes the solution on the medium timescale fluctuations and we now go one more order to obtain the slow evolution of the equilibrium.

\section{${\mathcal O}(\epsilon^2)$: Transport equations in the slab limit}

To proceed to the problem of transport and evolution of the equilibrium, a full description of the equilibrium geometry is necessary.  The full axisymmetric toroidal case is treated in the Appendix.  Here we calculate transport in an illustrative and simple limit of gyrokinetics, the straight field line ``slab'' geometry.  The equilibrium magnetic field lines have a constant direction ($\hat{z}$), and the curvature drift is eliminated from the gyrokinetic equation.  The $x$ coordinate is taken as the flux surface label so the equilibrium has variation only in this direction.  Also note that for this case the classical part $h_{cl}$ will be neglected as will the the evolution of the background magnetic field ($\partial {\bf A_0}/\partial t$).  The system is taken to be periodic in the $y$ and $z$ directions.  In place of the time average and patch average we will use a single smoothing operator which will combine an intermediate-scale average in the $x$-direction with a full average over the system in the $y$ and $z$ directions and the intermediate time-scale average defined earlier by Eqn.~\ref{t-average}.


\begin{equation}
\smooth{A} = \frac{\int_{t - \Delta t/2}^{t + \Delta t/2} dt^{\prime}\int dz dy \int^{x+\Delta x/2}_{x-\Delta x/2} dx^{\prime} }{\Delta t\Delta x L^2}A(x^{\prime},y,z,t^{\prime})
\end{equation}

\noindent where the system extends a distance $L$ in the $y$ and $z$ directions and, thus, $V = \Delta x L^2$ is the volume over which the smoothing operator averages.  To obtain the slow time dependence of $n_0$ and $T_0$ we consider the moment equations of the full kinetic equation and apply the smoothing average.  Many terms will be zero outright and we can avoid tabulating all the ${\mathcal O}({\epsilon}^2)$ terms from the expanded equations.  Also, since the moments are taken in real-space coordinates, it will often be convenient to refer to terms explicitly in these real-space coordinates (as opposed to gyro-center coordinates).  I.e. we will make reference to $\delta f_1$ which we recall is the first order correction to $F_0$ in non gyro-center coordinates: $f = F_m({\bf r}, {\bf v}) + \delta f_1 + {\mathcal O}(\epsilon^2)$ where $\delta f_1 = - \rhovec\cdot\nabla F_{m} - q\varphi F_{m}/T_{0} + h.$

\subsection{\bf Particle transport}

To obtain the time evolution of the density, we will integrate the Fokker-Planck equation over velocity and apply the smoothing operator .

\begin{equation}
\smooth{\int d^{3}{\bf v} \frac{\partial f_{s}}{\partial t} + {\bf v}\cdot\nabla f_s + 
\frac{q}{m}({\bf E} + {\bf v}\times{\bf B})\cdot \frac{\partial
  f_s}{\partial {\bf v}}} = 
\smooth{\int d^{3}{\bf v} C(f,f)}
\end{equation}

\noindent The integral of the collisions over velocity will be zero to conserve
particles.  The $({\bf E} + {\bf v}\times{\bf B})\cdot \frac{\partial f_s}{\partial {\bf v}}$ terms will be zero because they are a perfect divergence in velocity space.  What is left is the continuity equation.

\begin{equation}
\smooth{\int d^{3}{\bf v} 
\frac{\partial f_{s}}{\partial t} + \nabla\cdot ({\bf v}f_s)}= 0
\label{particle_FP}
\end{equation}

\noindent The y-derivative and z-derivative parts of $\nabla\cdot ({\bf v} f_s)$ will spatially average to zero due to periodic boundary conditions in those directions.  Finally, the x component of the $\nabla\cdot ({\bf v} f_s)$ term can be rewritten by considering the following: 

\begin{align}
&\int d^{3}{\bf v}({\bf
  v}\times{\bf b}_0)\cdot {\hat x}\frac{\partial}{\partial {\bf
  v}}\cdot ({\bf v}\times{\bf b}_0 f_s) \nonumber \\
&= \int d^{3}{\bf v} \frac{\partial}{\partial {\bf v}}\cdot\left[({\bf
    v}\times{\bf b}_0)\cdot{\hat x}({\bf v}\times{\bf b}_0 f_s)\right]
  - \int d^{3}{\bf v}({\bf v}\times{\bf
  b}_0 f_s)\cdot\frac{\partial}{\partial {\bf v}}({\bf v}\times{\bf
  b}_0)\cdot {\hat x} \nonumber \\
&= -\int d^{3}{\bf v}\left[({\bf v}\times{\bf b}_0)\times{\bf b}_0\right] \cdot{\hat
  x}f_s \nonumber\\ 
&= \int d^{3}{\bf v} ({\bf v} \cdot {\hat x}) f_s
\label{v_grad}
\end{align}

\noindent Thus, we can write

\begin{align}
&\int d^{3}{\bf v}\frac{\partial}{\partial x}(v_x f_s) = \frac{\partial}{\partial x} \int d^{3}{\bf v}(v_x f_s) \nonumber \\
&= {\bf \partial_x}\cdot\int d^{3}{\bf v}({\bf
  v}\times{\bf b}_0)\frac{\partial}{\partial {\bf
  v}}\cdot ({\bf v}\times{\bf b}_0f_s) \nonumber \\
&=  {\bf \partial_x}\cdot\int d^{3}{\bf v}({\bf
  v}\times{\bf b}_0)\left[({\bf v}\times{\bf b}_0)\cdot
  \frac{\partial f_s}{\partial {\bf v}}\right]\nonumber\\
&=  {\bf \partial_x}\cdot\int d^{3}{\bf v}\frac{({\bf
  v}\times{\bf b}_0)}{\Omega}\left[\frac{q}{m}({\bf
  v}\times{\bf B}_0)\cdot
  \frac{\partial f_s}{\partial {\bf v}}\right]
\end{align}

\noindent where we used $\frac{\partial}{\partial {\bf v}}\cdot({\bf v}\times {\bf b}_0) = 0$ and introducted the notation ${\bf \partial_x} = {\hat x}\partial/\partial x$.  We can rewrite $\left[\frac{q}{m}({\bf v}\times{\bf B}_0)\cdot \frac{\partial f_s}{\partial {\bf v}}\right]$ using the full Fokker-Planck equation.

\begin{align}
& {\bf \partial_x}\cdot\int d^{3}{\bf v}\frac{({\bf
  v}\times{\bf b}_0)}{\Omega}\left[\frac{q}{m}({\bf
  v}\times{\bf B}_0)\cdot
  \frac{\partial f_s}{\partial {\bf v}}\right] \nonumber\\
&= - {\bf \partial_x}\cdot\int d^{3}{\bf v}\frac{({\bf
  v}\times{\bf b}_0)}{\Omega}\left[\frac{\partial
  f_{s}}{\partial t} + {\bf v}\cdot\nabla f_s + 
\frac{q}{m}({\bf E} + {\bf v}\times\delta{\bf B})\cdot \frac{\partial
  f_s}{\partial {\bf v}} - C(f,f)\right] 
\label{fdsa}
\end{align}

\noindent The advantage of writing the ${\bf v}\cdot \nabla f_s$ term in this
form is that the factor of $\frac{({\bf v}\times{\bf b}_0)\cdot {\hat x}}{\Omega}$ in
front of the Fokker-Planck terms will reduce the order of the terms
inside since ($\Omega^{-1} \sim {\mathcal O}(\frac{\epsilon}{\omega})$).  This is as far as we can get before using the solution for $f_s$.  Now, we can rewrite Eqn \ref{particle_FP} as

\begin{eqnarray}
&\smooth{\int d^{3}{\bf v} \underbrace{\frac{\partial f_{s}}{\partial t}}_1}
= \nonumber \\
& {\bf \partial_x}\cdot\int d^{3}{\bf v}\frac{({\bf
  v}\times{\bf b}_0)}{\Omega}\smooth{\underbrace{\frac{\partial
  f_{s}}{\partial t}}_2 + \underbrace{{\bf v}\cdot\nabla f_s}_3 + 
\underbrace{\frac{q}{m}({\bf E} + {\bf v}\times\delta{\bf B})\cdot \frac{\partial
  f_s}{\partial {\bf v}}}_4 - \underbrace{C(f,f)}_5} 
\label{density_moment}
\end{eqnarray}

\noindent If we plug our solution of $f_s$ into term 1 of Eqn
\ref{density_moment}, we obtain up to ${\mathcal O}({\epsilon^2})$

\begin{equation}
\int d^{3}{\bf v} 
\smooth{\frac{\partial f_{s}}{\partial t}} = \frac{\partial n_{0}}{\partial t} +
\smooth{\int d^{3}{\bf v} \frac{\partial \delta f_1}{\partial t} + \frac{\partial \delta f_2}{\partial t}}
\label{density_time}
\end{equation}

\noindent We identify $\int d^{3}{\bf v}\frac{\partial F_{m}}{\partial t} = \frac{\partial n_{0s}}{\partial t}$ as the quantity we are solving for, namely the time derivative of the equilibrium density.  The $\delta f_1$ terms are negligible as follows: firstly, the boltzmann response vanishes after space averaging (we have assumed there is no equilibrium-scale electrostatic potential) and second, gyro-center distribution function $h$ contributes negligibly after smoothing (note that we have neglected to include $h_{cl}$ in this derivation but contribution from this term would be obvioiusly be an order epsilon smaller than $\partial F_0/\partial t$).  Finally, the $\delta f_2$ term in Eqn \ref{density_time} will drop out after performing the time average. Term 2 in Eqn \ref{density_moment} is treated analogously to term 1 but the order of each term is one smaller because of the preceding factor of $\frac{({\bf v}\times{\bf b}_0)\cdot {\hat x}}{\Omega}$.  Thus, it contributes negligibly. The
third term in Eqn \ref{density_moment} has derivatives in $y$ and $z$ which vanish by periodicity after averaging.  What remains is the $x$-gradient,

\begin{equation}
\smooth{\int d^{3}{\bf v} 
\frac{({\bf v}\times{\bf b}_0)\cdot {\hat x}}{\Omega}(v_x
\frac{\partial}{\partial x})(F_0 + h + \delta f_2)}
\label{asdf}
\end{equation}

\noindent None of these terms will enter at ${\mathcal O}(\epsilon^2)$.  To show
this we first note that $h$ and $F_0$, are functions of gyro-center position ${\bf R}$.  These terms can be first spatially averaged over y.  This removes any odd dependence in $v_x$ due to gyro-center dependence.  The resulting terms are odd in $v_x$ and can be integrated away.  The $\delta f_2$ term will drop in ordering after averaging it over space.  The fourth term in Eqn \ref{density_moment} can be written to ${\mathcal O}(\epsilon^2)$ as 

\begin{equation}
\smooth{\int d^{3}{\bf v} 
\frac{({\bf v}\times{\bf b}_0)\cdot {\hat x}}{\Omega}
\frac{q}{m}({\bf E} + {\bf v}\times\delta{\bf B})\cdot
\frac{\partial}{\partial {\bf v}} (F_0 + h)}
\end{equation}

\noindent The parts of $F_0$ are negligible as follows: The Boltzmann response term is zero under velocity differentiation.  The gyor-center correction $-\rhovec\cdot\nabla F_m$ is an order smaller after smoothing over the turbulent fields ${\bf E} + {\bf v}\times\delta{\bf B}$.  The Maxwellian part $F_m$ is argued away by a combination of periodicity in $y$ and $z$, time averaging and oddness in velocity velocity space.  We now write ${\bf E} + {\bf v}\times\delta{\bf B}$ as $-\nabla\chi - \frac{d{\bf A}}{dt}$.  To simplify the h term, we integrate by parts in cartesian velocity coordinates, so that

\begin{equation}
\smooth{\int d^{3}{\bf v} 
\frac{v_y}{\Omega} \frac{q}{m}(-\nabla\chi - \frac{d{\bf A}}{dt})\cdot \frac{\partial}{\partial {\bf v}}h} = 
\smooth{\int d^{3}{\bf v} \frac{1}{B_0}\hat{x}\cdot(\nabla\chi\times{\bf b}_0) h}
\label{density_field}
\end{equation}

\noindent Where the $\frac{d{\bf A}}{dt}$ term is removed by by noting that $\frac{\partial {\bf A}}{\partial t}$ is small in our ordering and employing Eqn \ref{deexi} after integration by parts to remove the rest.  The final term in Eqn \ref{density_moment} simply involves plugging in values for the distribution function.   

\begin{equation}
-\int d^{3}{\bf v} 
\frac{({\bf v}\times{\bf b}_0)\cdot {\hat x}}{\Omega}C(- \rhovec \cdot\nabla F_m, F_m)
\label{density_collision}
\end{equation}

\noindent Recall that the Maxwellian is notated $F_m$ so as to distinguish it from the modified Maxwellian $F_0$ defined in Eqn \ref{modmax}.  Collecting terms, and recalling the $\frac{\partial}{\partial x}$ in front of the integral from Eqns \ref{density_field} and \ref{density_collision} from Eqn \ref{density_moment} , we have the transport equation for particles:

\begin{equation}
\frac{\partial n_{0}}{\partial t} 
= {\bf \partial_x}\cdot \int
   d^{3}{\bf v} \;
\left[\smooth{(\nabla \chi \times\frac{{\bf b}_0}{B_0})\;h} + \frac{({\bf v}\times{\bf b}_0)}{\Omega} C(\rhovec \cdot\nabla F_m, F_m)\right]
\label{density_transport}
\end{equation}

\subsection{\bf Heat transport equation}

\noindent To calculate heat transport , we multiply the Fokker-Planck equation by $\frac{1}{2}m{\bf v}^2$, integrate over velocity, and smooth.  Following the same methodology as used with the particle transport, 

\begin{equation}
\int d^{3}{\bf v}\frac{mv^2}{2} 
\smooth{\frac{\partial f_{s}}{\partial t} + {\bf v}\cdot\nabla f_s + 
\frac{q}{m}({\bf E} + {\bf v}\times{\bf B})\cdot \frac{\partial f_s}{\partial {\bf v}}} = 
\int d^{3}{\bf v} \frac{mv^2}{2}\smooth{C(f,f)}
\label{heating_moment}
\end{equation}

\noindent The first term of Eqn \ref{heating_moment} yields

\begin{equation}
\frac{3}{2}\frac{\partial (T_{0} n_{0)}}{\partial t}  + \smooth{\int d^{3}{\bf v} 
\frac{1}{2}m{\bf v }^{2}\frac{\partial}{\partial t}(\delta f_1 + \delta f_2)} 
\label{heating_time}
\end{equation}

\noindent The first term is equilibrium pressure evolution $\frac{3}{2}T_{0} n_{0} = \int d^{3}{\bf v}\frac{mv^2}{2} F_m$.  The other terms in the time derivative will be negligible after smoothing.  The second term in Eqn \ref{heating_moment} can be rewritten in the same manner as Eqn \ref{v_grad}.  

\begin{align}
&\int d^{3}{\bf v}\frac{mv^2}{2}({\bf v}\cdot\nabla_x f_s) = \frac{\partial}{\partial
  x} \int d^{3}{\bf
  v}\frac{mv^2}{2}(v_x f_s) \nonumber \\
&=  {\bf \partial_x}\cdot\int d^{3}{\bf v}\frac{m}{2}({\bf
  v}\times{\bf b}_0)\frac{\partial}{\partial {\bf
  v}}\cdot ({\bf v}\times{\bf b}_0f_s v^2) \nonumber \\
&=  \frac{m}{2}{\bf \partial_x}\cdot\int d^{3}{\bf v}\frac{({\bf
  v}\times{\bf b}_0)}{\Omega}\left[\frac{q}{m}({\bf
  v}\times{\bf B}_0)\cdot
  (v^2\frac{\partial f_s}{\partial {\bf v}})\right]
\end{align}

\noindent The last line was obtained using $({\bf v}\times{\bf b}_0)\cdot \frac{\partial v^2}{\partial {\bf v}} = 0$. Once again, we are able to substitute in the Fokker-Planck equation, and obtain 

\begin{align}
& \frac{m}{2}{\bf \partial_x}\cdot\int d^{3}{\bf v}\;v^2\frac{({\bf
  v}\times{\bf b}_0)}{\Omega}\left[\frac{q}{m}({\bf
  v}\times{\bf B}_0)\cdot
  \frac{\partial f_s}{\partial {\bf v}}\right] \nonumber\\
&= - \frac{m}{2}{\bf \partial_x}\cdot\int d^{3}{\bf v}\; v^2\frac{({\bf
  v}\times{\bf b}_0)}{\Omega}\left[\frac{\partial
  f_{s}}{\partial t} + {\bf v}\cdot\nabla f_s + 
\frac{q}{m}({\bf E} + {\bf v}\times\delta{\bf B})\cdot \frac{\partial
  f_s}{\partial {\bf v}} - C(f,f)\right] 
\label{jkl}
\end{align}

\noindent The difference between this equation and Eqn \ref{fdsa} is obviously the $v^2$ inside the integral.  Fortunately, all the terms that were negligible in density transport will be negligible in this case as well, because the arguments relied on parity in $v_x$ and $v_y$, spatial smoothing and ordering, and $\vartheta$ integrations.  Thus, what we really need to calculate is 

\begin{equation}
-\smooth{\int d^{3}{\bf v}\;
\frac{mv^2}{2} \frac{v_y}{\Omega}
\frac{q}{m}(-\nabla\chi - \frac{d{\bf
  A}}{dt})\cdot \frac{\partial h}{\partial {\bf v}}}
\end{equation}

\noindent Here, the integration by parts will be a little messier than with the analogous term in the particle transport equation.  The $\frac{d{\bf A}}{dt}$ term is negligible as before.  The result is

\begin{eqnarray}
-{\bf \partial_x}\cdot \int d^{3}{\bf v} 
\smooth{\frac{mv^2(\nabla \chi \times {\bf b})}{2B_0}h + \frac{m({\bf v} \times {\bf b}_0)}{B_0}(({\bf v} \cdot \nabla)\phi)h}
\label{heating_convection}
\end{eqnarray}

\noindent We will later see that the second term of Eqn \ref{heating_convection} will cancel.  The collisional terms of Eqn \ref{jkl} work similarly as with density transport.

\begin{eqnarray}
- {\bf \partial_x}\cdot \int d^{3}{\bf v} v^2 \frac{m({\bf v} \times {\bf b}_0)}{2\Omega}C( -\rhovec \cdot \nabla F_m,F_m)
\label{heating_vdot}
\end{eqnarray}

\noindent Unlike the case of particle transport, the electric field term will not integrate away from Eqn \ref{heating_moment}.  First we manipulate as follows:

\begin{equation}
\int d^{3}{\bf v} \frac{mv^2}{2}\frac{q}{m}({\bf E})\cdot \frac{\partial f}{\partial {\bf v}} 
= - \int d^{3}{\bf v}{q}({\bf E}\cdot {\bf v})f
\label{Ev}
\end{equation}

\noindent The scalar potential part of the electric field is an order larger than the $\frac{\partial A}{\partial t}$ term, but as we will see most of it will cancel.  This will result in the part of the scalar potential that does not average away acting on the same  order as $\frac{\partial A}{\partial t}$.

\begin{eqnarray}
&\int d^{3}{\bf v} 
\left[q {\bf v}\cdot\nabla\phi f \right] = - \int d^{3}{\bf v} 
\left[ q \frac{\partial\phi}{\partial t} f\right]  + 
\int d^{3}{\bf v} \left[ q( \frac{\partial\phi}{\partial t} + {\bf v}\cdot\nabla\phi)
  f\right]\nonumber \\
&= -\int d^{3}{\bf v} 
\left[q \frac{\partial\phi}{\partial t}  f\right]  - 
\int d^{3}{\bf v} \left[ q( \frac{\partial f}{\partial t} + {\bf v}\cdot\nabla f) 
\phi \right] + \int d^{3}{\bf v} 
\left[ q \frac{\partial(\phi f)}{\partial t} \right]\nonumber \\ 
& +\int d^{3}{\bf v} q\nabla(\phi f)\cdot {\bf v}
\nonumber \\
&= - \int d^{3}{\bf v} 
\left[ q \frac{\partial\phi}{\partial t} f \right]  
+ \frac{q^{2}}{m} \int d^{3}{\bf v}\frac{\partial}{\partial {\bf v}}\cdot  
\left[( {\bf E} + {\bf v}\times {\bf B})\phi f \right] - \int d^{3}{\bf v} 
\left[ q( C(f,f)) \phi\right] \nonumber \\
& + \int d^{3}{\bf v} 
\left[ q \frac{\partial(\phi f)}{\partial t} \right]
+ \int d^{3}{\bf v} q({\bf v}\cdot \nabla)(\phi f) \nonumber \\
&= - \int d^{3}{\bf v} 
\left[ q \frac{\partial\phi}{\partial t} f\right] + \int d^{3}{\bf v} 
\left[ q \frac{\partial(\phi  f)}{\partial t} \right] 
+  \int d^{3}{\bf v} q({\bf v}\cdot \nabla)(\phi f)
\end{eqnarray}

\noindent We have integrated by parts in time and space between lines one and two.  Between lines two and three we substitute the full Fokker-Planck equation $\frac{df}{dt}=C(f,f)$ and, finally, we use gauss's law in velocity space and that collisions conserve particles to eliminate the velocity divergence and collisional integral to obtain line four. The second term on the final line will be negligible after time average.  We find that the very last term will cancel the last term of Eqn \ref{heating_convection} when we compute the contribution of $h$ from $f$ in the equation.

\begin{align}
&\int d^{3}{\bf v} 
q({\bf v}\cdot \nabla)(\phi h) \nonumber \\
&= \frac{\partial}{\partial x} \int d^{3}{\bf v} 
q v_x(\phi h) 
\nonumber \\
&= - \frac{\partial}{\partial x} \int d^{3}{\bf v} 
q v_y\phi \left({\frac{\partial h}{\partial \vartheta}}\right)_r \nonumber \\
&= - \frac{\partial}{\partial x}\int d^{3}{\bf v} 
q v_y\phi (\frac{{\bf v}\cdot \nabla}{\Omega}h + \left({\frac{\partial
    h}{\partial \vartheta}}\right)_R)
\nonumber\\
&= \frac{\partial}{\partial x} \int d^{3}{\bf v} 
q v_y h \frac{{\bf v}\cdot \nabla}{\Omega}\phi\nonumber\\
&= {\bf \partial_x}\cdot \int d^{3}{\bf v} 
\frac{m({\bf v} \times {\bf b}_0)}{B_0}({\bf v} \cdot \nabla\phi)h
\label{heat_EV}
\end{align}

\noindent Between lines 1 and 2 we integrated by parts in velocity coordinate
$\vartheta$.  Between lines 3 and 4 we have integrated by parts in space
to move the gradient onto the $\phi$.  We now want to combine our
results from the electrostatic part of the electric field in Eqn
\ref{heat_EV} with the inductive part.

\begin{equation}
\frac{m}{2} \int d^{3}{\bf v} \;
v^2\frac{q}{m}{\bf E}\cdot \frac{\partial f}{\partial {\bf v}} = 
\int d^{3}{\bf v} \{ q \frac{\partial}{\partial t}\left[\phi - {\bf v}\cdot {\bf 
A}\right] \delta f_1 \} 
\end{equation}

\noindent where the Maxwellian part of $F_0$ averages away due to velocity parity.  We now plug in our solution for $\delta f_1$.  The adiabatic/Boltzman term gives

\begin{equation}
\smooth{\int d^{3}{\bf v} q \frac{\partial}{\partial t}[\phi - v_{\parallel}A_{\parallel} - {\bf v_{\perp}}\cdot {\bf 
A_{\perp}}] (-q\frac{\phi}{T_{0}}F_{m})} = -\frac{\partial}{\partial t}\smooth{\frac{n_0}{T_0}(\frac{\phi^2}{2})}
\end{equation}

\noindent since:  the $A_{\parallel}$ term is odd in $v_{\parallel}$ and the $A_{\perp}$ 
gyro-averages to zero (or equivalently is odd in $v_{\perp}$).  What is left will time average away, leading to the result that at this order the ${\bf E}\cdot{\bf v}\frac {q\phi}{T}F_m$ term produces no
heating.  Returning to the original moment equation in this section, Eqn.~\ref{heating_moment}, the ${\bf v} \times{\bf B}$ terms will not contribute, as magnetic fields do no work (it can be written as a perfect divergence and eliminated by integration over velocity space).  The final term we need to calculate is the collisional energy exchange terms between species are now included -- note like particle collisions do not produce a loss of energy and thus do not appear (this is easy to prove).  The collisional energy exchange is standard since to this order it is between Maxwellian species.  Specifically:

\begin{equation}
\int d^3{\bf v}\frac{1}{2}mv^{2}C_{ie}(f_i, f_e) = n_{0} \nu_{{\mathcal E}}^{ie}(T_e - T_i). 
\label{heating_collisions}
\end{equation}

\noindent where $\nu_{{\cal E}}^{sr}$ is found in \cite{helander} -- it is $\sqrt{m_e/m_i}$ smaller than the ion-ion collision rate which is itself $\sqrt{m_e/m_i}$ smaller than the electron-ion collision rate.  We now collect all the results and write down what we have so far for the heat transport equation.

\begin{align}
&\frac{3}{2}\frac{\partial n_{0i}T_{0i}}{\partial t} = \nonumber \\ 
& {\bf \partial_x}\cdot \int d^{3}{\bf v}\frac{mv^2}{2} \left(\smooth{\frac{(\nabla \chi \times {\bf b}_0)}{B_0} h} + \frac{({\bf v} \times {\bf b}_0)}{\Omega} \left[C( \rhovec \cdot \nabla F_m,F_m)\right]\right)\nonumber \\ 
& + \int d^{3}{\bf v} q \smooth{\frac{\partial \chi}{\partial t} h} + n_{0} \nu_{{\mathcal E}}^{ie}(T_{0e} - T_{0i})
\end{align}

\noindent We can rewrite the gyrokinetic heating term $q h \partial \chi/\partial t$ (this is a necessary step for establishing entropy balance later).  We write the smoothing operator explicitly as a spatial average (ignoring the time average for now) and use Parseval's identity to write the integral as a sum of Fourier components so that the gyro-angle average may be shifted to $\chi$. The result is

\begin{eqnarray}
&\int\frac{d^{3}{\bf r}}{V}\int d^{3}{\bf v} q \frac{\partial \chi}{\partial t} h \nonumber \\
&= \int\frac{d^{3}{\bf r}}{V}\int d^{3}{\bf v} q \frac{\partial \chi}{\partial t} \angleavg{h} \nonumber \\ 
&= {\displaystyle\sum_{\bf k}}\int d^{3}{\bf v} q \frac{\partial \chi_{\bf k}}{\partial t} J_0(k \rho)h_{\bf -k} \nonumber \\ 
&= \int\frac{d^{3}{\bf R}}{V}\int d^{3}{\bf v} q \frac{\partial \gyroavg{\chi}}{\partial t} h 
\end{eqnarray}

\noindent Now using the gyrokinetic equation, Eqn.~\ref{the-gyro-equation}, this expression becomes

\begin{eqnarray}
& - \int\frac{d^{3}{\bf R}}{V}\int d^{3}{\bf v} 
\frac{\gyroavg{\nabla \chi}\times {\bf b}_0}{B_0}\cdot \nabla F_0\frac{hT}{F_0}
- \int\frac{d^{3}{\bf R}}{V}\int d^{3}{\bf v} \frac{hT}{F_0}\gyroavg{C(h,F_m)}\nonumber \\
& + \int\frac{d^{3}{\bf R}}{V}\int d^{3}{\bf v} \frac{\partial}{\partial t}(\frac{h^2T}{2F_0})
\label{heating_electric}
\end {eqnarray}

\noindent where we have multiplied the gyrokinetic equation by $\frac{Th}{F_0}$, and applied the indicated averages.  The final term will average away upon time-averaging.  We may reverse the change of variables trick and revert back to the smoothing operator form to plug into the heating equation. 


\noindent Thus, combining all the surviving terms from Eqns \ref{heating_time}, \ref{heating_convection}, 
\ref{heating_vdot}, \ref{heating_electric}, and
\ref{heating_collisions} will give us the heating equation.

\begin{align}
&\frac{3}{2}\frac{\partial n_{0i}T_{0i}}{\partial t} = \nonumber \\ 
& {\bf \partial_x}\cdot \int d^{3}{\bf v} \left(\smooth{\frac{mv^2(\nabla \chi \times {\bf b}_0)}{2B_0} h} + v^2 \frac{m({\bf v} \times {\bf b}_0)}{2\Omega} \left[C(\rhovec \cdot \nabla F_m,F_m)\right]\right)\nonumber \\ 
& - \int d^{3}{\bf v} \smooth{\frac{\nabla \chi \times {\bf b}_0}{B}\cdot \nabla F_m\frac{ T}{F_m}h} - \int d^{3}{\bf v} \smooth{\frac{hT}{F_m}C(h, F_m)}\nonumber \\
& + n_{0} \nu_{{\mathcal E}}^{ie}(T_e - T_i).
\label{avheating} 
\end{align}

\subsection{Entropy balance}

Establishing entropy balance is a good way to ground the results of the gyrokinetic transport calculation to an exact result from statistical physics.  A tokamak (or any other plasma) is ultimately a collisional (dissipative) system.  The generation of entropy by the collision operator provides the mechanism for the ultimate dissipation of turbulent fluctuations and in particular prevents the formation of unphysically small scale features of the distribution function.  Also, calculation of entropy balance has the added benefit of a consistency check to confirm that the terms in our transport equations have been calculated correctly.  The time rate of change of the mean entropy per unit volume (contained in the volume over which we are smoothing) is given by

\begin{equation}
\frac{\partial S}{\partial t}= - \int d^{3}{\bf v}\smooth{\frac{\partial}{\partial t}f \ln f}
\end{equation} 

\noindent We multiply the Fokker-Planck equation by $1+\ln f$ and integrate to obtain the relation

\begin{equation}
  \int d^{3}{\bf v}\frac{\partial f \ln f}{\partial t} +   \int d^{3}{\bf v} \; \smooth{{\bf v} \cdot \nabla (f \ln f)} =  \int d^{3}{\bf v} \smooth{C(f, f) \ln f}
\end{equation}

\noindent We will now use heat and particle transport equations to demonstrate that entropy balance does indeed take the form

\begin{equation}
\frac{\partial S}{\partial t} = {\bf \partial_x}\cdot \int d^{3}{\bf v} \smooth{{\bf v}f \ln f}  -  \int d^{3}{\bf v} \smooth{C(f, f) \ln f}
\end{equation}

\noindent It is not hard to show that $\frac{\partial S}{\partial t}$ is mostly due to the background Maxwellian, i.e.

\begin{equation}
\frac{\partial S}{\partial t} \sim -\frac{\partial}{\partial t} \int d^{3}{\bf v} F_m \ln F_m
\end{equation}

\noindent Performing the integrations give

\begin{equation}
\frac{\partial S}{\partial t}= -\left[\ln n_0 + \frac{3}{2} \ln (\frac{m}{2 \pi T_0}) + 1\right]\frac{\partial n_0}{\partial t} + \frac{3}{2T} \frac{\partial n_0T_0}{\partial t}
\end{equation}

\noindent We can now substitute in from Eqns \ref{density_transport} and
\ref{avheating}. 

\begin{eqnarray}
&\frac{\partial S}{\partial t} = - \int d^{3}{\bf v}\left(\ln n_0 + \frac{3}{2}\ln(\frac{m}{2\pi T_0}) + 1\right)
  {\bf \partial_x} \cdot \left[\smooth{\frac{\nabla \chi \times {\bf b_0}}{B_0}h} - \rhovec\; C(\rhovec \cdot F_m, F_m)\right] \nonumber \\
&+ \frac{1}{T}\left[{\bf \partial_x}\cdot
 \int d^{3}{\bf v}\frac{mv^2}{2}\left(\smooth{\frac{\nabla \chi \times b_0}{B_0}h} + \rhovec\;  C(\rhovec \cdot \nabla F_m,
  F_m)\right)\right]\nonumber \\ 
&- \frac{1}{T}[ \int d^{3}{\bf v} \smooth{\frac{\nabla \chi \times b_0}{B_0}h} \cdot \nabla (\ln F_m)T +  
\int d^{3}{\bf v}\smooth{\frac{hT}{F_m}C(h, F_m)} \nonumber \\
& + n_0\nu_{{\mathcal E}}^{ie}\frac{T_{0e} - T_{0i}}{T_{0i}}]
\end{eqnarray} 

\noindent What is left to do is some integration by parts and arrangement of terms followed by evaluation of $\nabla \ln F_m$ on the third line.  The result is


\begin{eqnarray}
& \frac{\partial S}{\partial t} = {\bf \partial_x}\cdot \int d^{3}{\bf v}(\ln F_m + 1)( \smooth{\frac{\nabla \chi \times b_0}{B_0}h} - \rhovec \; C(\rhovec \cdot \nabla F_m, F_m)) \nonumber \\
& -  \int d^{3}{\bf v}(\rhovec\cdot\nabla\ln F_m)C(\rhovec \cdot \nabla F_m, F_m) - \int d^{3}{\bf v}\smooth{\frac{h}{F_m}C(h, F_m)} \nonumber \\
& + n_0\nu_{{\mathcal E}}^{ie}\frac{T_{0e} - T_{0i}}{T_{0i}}
\end{eqnarray}

\noindent which completes entropy balance.  Notice that the finally two lines can be easily seen to come from $\int d^3{\bf v}\ln f C(f, f)$.  The two terms on the second line represent the nonvanishing parts of $\int d^3{\bf v}\frac{-\delta f_1}{F_m}C(\delta f_1, F_m)$ and the final term is simply $\int d^3{\bf v}\ln F_m C(F_{m,i}, F_{m,e})$

\section{Poynting's theorem}
\label{poynting-sec}
A standard derivation of Poynting's Theorem begins taking the inner product of Ampere's law with the electric field

\begin{equation}
{\bf J}\cdot{\bf E} = -\epsilon_0 \frac{\partial {\bf E}}{\partial t}\cdot{\bf E} + \frac{1}{\mu_0}{\bf E}\cdot(\nabla\times{\bf B})
\end{equation}

\noindent One continues by applying the identity $\nabla\cdot({\bf A\times B}) = {\bf B}\cdot\nabla\times{\bf A} - {\bf A}\cdot\nabla\times{\bf B}$ followed by substituting Faraday's Law $\nabla\times{\bf E} = -\frac{\partial {\bf B}}{\partial t}$ to obtain the standard form of Poynting's Theorem:

\begin{equation}
{\bf J}\cdot{\bf E} = -\frac{1}{2}\left(\frac{1}{\mu_0}\frac{\partial B^2}{\partial t} + \epsilon_0\frac{\partial E^2}{\partial t}\right) -\frac{1}{\mu_0}\nabla\cdot({\bf E}\times{\bf B})
\end{equation}

\noindent This gives a complete picture of the electromagnetic energy flow, making no ordering assumptions, but throws away details by lumping together all fields (electrostatic, inductive, equilibrium, perturbative).  We can obtain Poynting's Theorem in three separate parts and use our ordering assumptions to get a detailed picture of electromagnetic energy flows.  We do this by taking the inner product of ampere's law {\em separately} with three different parts of the electric field: the electrostatic part (${\bf E_s} = -\nabla\varphi$), the equilibrium inductive part (${\bf E_{i0}} = -\frac{\partial {\bf A_0}}{\partial t}$) and the perturbative inductive part (${\bf E_{i1}} = -\frac{\partial {\bf \delta A}}{\partial t}$).  We proceed exactly as with the standard Poynting's Theorem derivation above.  Only lowest order terms are retained in each equation.  We order the ${\bf J\cdot E}$ terms using preceding kinetic analysis (see Eqn.~\ref{mess}) and find they are each ${\mathcal O}(\epsilon^2\omega T)$.  We then use a patch volume average and time average to separate the multi-scale dependence.  First the electrostatic part we find

\begin{equation}
{\bf J}\cdot{\bf E}_s = -\frac{1}{\mu_0}\nabla\cdot({\bf E}_s\times{\bf B})
\end{equation}

\noindent Where we have neglected the displacement current term which will enter at a higher order in Poynting's Theorem for both non-relativistic and $\epsilon$ ordering arguments.  We have also used here that $\nabla\times{\bf E_s} = 0$ because ${\bf E_s}$ is the gradient of a scalar.  We note that ${\bf J}\cdot{\bf E}_s$ is an order smaller than the remaining term on the right hand side of the equation.  This gives us the following result, at dominant order:

\begin{equation}
{\bf E}_s\cdot(\nabla\times{\bf B}) = 0
\end{equation}

\noindent We can interpret this to mean that the electrostatic Poynting flux is everywhere zero.  At next order we learn something about ${\bf J}\cdot{\bf E}_s$:

\begin{equation}
{\bf J}\cdot{\bf E}_s = -\frac{1}{\mu_0}\nabla\cdot({\bf E}_s\times{\bf \delta B} + {\bf E}_{s2}\times{\bf B}_0)
\end{equation}

\noindent where ${\bf E}_{s2}$ is the second order electrostatic potential -- for which we do not solve.  However, all terms on the right hand side are linear in the fluctuating fields, so applying the patch average, they are eliminated (increased in order) and we obtain the result

\begin{equation}
\patch{{\bf J}\cdot{\bf E}_s} = 0\label{electrostatic-current-work}
\end{equation}

\noindent which means that the electrostatic field does no work on the total current.  Now, the equilibrium inductive part:

\begin{equation}
{\bf J}\cdot{\bf E}_{i0} = -\frac{1}{\mu_0}\nabla\cdot({\bf E}_{i0}\times{\bf B}) -\frac{1}{2\mu_0}\frac{\partial B_0^2}{\partial t} 
\end{equation}
 
\noindent Again, neglecting the displacement current and higher order terms.  We apply the patch average to remove the ${\bf \delta B}$ perturbative term on the right hand side.

\begin{equation}
{\bf J}\cdot{\bf E}_{i0} = -\frac{1}{\mu_0}\nabla\cdot({\bf E}_{i0}\times{\bf B_0}) -\frac{1}{2\mu_0}\frac{\partial B_0^2}{\partial t} 
\end{equation}

\noindent This equation describes the balance of energy flow due to Ohmic heating.  Lastly, the perturbative inductive part of the electric field.

\begin{equation}
{\bf J}\cdot{\bf E}_{i1} = -\frac{1}{\mu_0}\nabla\cdot({\bf E}_{i1}\times{\bf B}) -\frac{1}{\mu_0}{\bf B_0}\cdot\frac{\partial {\bf \delta B}}{\partial t}
\end{equation}

\noindent The patch average and time average of the right hand side removes all terms due to ordering arguments.

\begin{equation}
\patch{\overline{{\bf J}\cdot{\bf E}_{i1}}} = 0
\end{equation}

\noindent Thus from this result and Eqn.~\ref{electrostatic-current-work} we find that the fluctuating electric field (both inductive and electrostatic) does no mean work on the total current.
\chapter{Primary and Secondary Mode Theory}\label{secondary-chap}

{\em The following statement applies to the figures included in this chapter, which have been taken from the paper \cite{plunk}:  Reprinted with permission from Gabriel Plunk, Physics of Plasmas, Vol. 14, Issue 11, Page 112308, 2007.  Copyright 2007, American Institute of Physics.}

\section{Introduction}

A natural first step in understanding anomalous transport is the calculation of the primary instabilities which drive micro-turbulence.  We will begin this chapter by reviewing a fluid description of the relevant instabilities driving core turbulence and also solve the gyrokinetic linear dispersion relation.  This chapter continues with secondary instability, a powerful and physically intuitive analytic theory that helps to explain mode saturation and transition to turbulence.  The material on secondary instabilities is copied to a large extent from the author's work \cite{plunk} with permission from the publishers.

The ion temperature gradient (ITG) drives modes at spatial scales comparable to the ion Larmor radius and is typically the strongest source of transport.  Electron thermal transport due to turbulence driven by the electron temperature gradient (ETG) and/or trapped electron modes (TEM) is regarded as a strong second candidate.  This chapter investigates ITG and ETG turbulence.  

In this chapter and the following chapter we take the simplification that the electrostatic perturbations are dominant over the magnetic field perturbuations.  In this chapter we also assume that the collisional terms are subdominant.  In plasma turbulence, collisions ultimately must play a critical role even when they are very weak and do not enter at the scales of the primary instability.  As with fluid turbulence, the steady state of a turbulent plasma is characterized by injection and ultimate dissipation, without which entropy production and velocity space distribution would develop unphysical characteristics.  This issue will be explored in more detail in the following chapter.

The system is closed by the quasi-neutrality condition and, for most of the chapter, the Boltzmann or ``adiabatic'' response will be assumed as a simplification.  In particular this means that for ETG driven turbulence, ions are assumed to have a Boltzmann response and for ITG driven turbulence, the electrons have the appropriate Boltzmann response.  The previous chapter has built the foundation for a local treatment, with spatial derivatives of equilibrium quantities taken as constants.  As demonstrated later, the local approximation is remarkably accurate in calculating linear growth rates of micro-instabilities.  In nonlinear regime, we expect the local case to give insight into tokamak turbulence and provide the basis for a more complete treatment in future work.

\section{Normalized equations}
\label{sec_eqns}

It is convenient to now use a normalized version of the gyrokinetic equation, Eqn.~\ref{the-gyro-equation}.  A locally flat orthogonal coordinate system is used where, as is the convention, $x$ is identified with the radial direction, $z$ is the coordinate along the magnetic field and $y$ is the binormal coordinate (normal to both $z$ and $x$).  The gradient lengths are taken to be constants and the magnetic field is taken to have constant curvature (i.e. ${\bf b}\cdot\nabla{\bf b}$ and $\nabla\ln B$ are constant vectors where they appear in the drift velocity below).  The normalized gyrokinetic equation is written

\begin{align}
& \frac{\partial h}{\partial t} + v_{\parallel} \frac{\partial h}{\partial z} + \hat{z}\times\gyroavg{ \nabla\varphi }\cdot\nabla h + {\bf {\bf v}_D} \cdot\nabla h \nonumber \\
& = \left(\frac{\partial\gyroavg{\varphi}}{\partial t} -  \hat{z}\times\gyroavg{ \nabla\varphi } \cdot  {\bf k}_{\star}\right) F_0
\label{gyro-norm}
\end{align}

\noindent with

\begin{eqnarray}
& {\bf k}_{\star} = \hat{x}(\eta^{-1} + v^2/2 - 3/2) \nonumber \\
& {\bf v}_D = \hat{z}\times(v^{2}_{\parallel}L_T{\bf b}\cdot\nabla{\bf b} + \frac{v^{2}_{\perp}}{2}L_T\nabla\ln B)
\label{gyro_defs}
\end{eqnarray}

\noindent with $\eta = L_n/L_T$.  The bracket average $\gyroavg{.}$ in Eqn (\ref{gyro-norm}) denotes the gyro-average, an average over gyro angle at fixed gyro center ${\bf R}$.  In this normalization the gyrocenter variable is defined ${\bf R} = {\bf r} - \rhovec = z{\hat z} + Y{\hat y} + X{\hat x} =  z{\hat z} + (y - v_{\perp}\sin{\vartheta}){\hat y} + (x + v_{\perp}\cos{\vartheta}){\hat x}$.   The normalized quasi-neutrality condition is

\begin{equation}
2\pi\int\int v_\perp dv_\perp dv_\parallel  \angleavg{ h } = (1 + \tau)\varphi - \left(\delta_{(ITG)}\right)\tau \bar{\varphi}
\label{quasi_neutrality}
\end{equation}

\noindent with 

\begin{equation}
\delta_{(ITG)} \equiv
\begin{cases}
1 & \text{if ITG}\\
0 & \text{if ETG}
\end{cases}
\nonumber
\end{equation}

\noindent The flux surface average $\bar{\varphi} = \int\int \varphi\; dy dz/(L_yL_z)$ is an average over the periodic box with sides $L_y$ and $L_z$.  The notation $\angleavg{.}$ is used for an average over gyro-angle $\vartheta$ at fixed spatial coordinate ${\bf r}$.  Note that the flux surface averaged potential enters crucially for the Boltzmann (adiabatic) electron response in ITG turbulence but is absent for the Boltzmann ion response for ETG turbulence.  This is the only difference between the equations governing electrostatic ITG and ETG turbulence.  This results from the different physical mechanisms that create the boltzmann response.  For ETG, the ion Larmor motion provides rapid motion perpendicular to the field lines that allow the ions to ``sense'' electrostatic perturbations.  In the case of ITG, the Larmor radius of electrons is small compared to the fluctuations and it is the rapid motion along the field lines that gives the mobility to sense potential fluctuations.  Thus the fluctuations are felt relative to the flux surface that contains the field line and the effective potential is $\varphi - \bar{\varphi}$.

The normalization used is as follows:

\vspace{1pt}
\begin{center}\begin{tabular}{cc}
$t = t_{\mbox{\scriptsize{phys}}}v_{\mbox{\scriptsize{th}}}/L_T$ & $v = v_{\mbox{\scriptsize{phys}}}/v_{\mbox{\scriptsize{th}}}$ \\ 
$x = x_{\mbox{\scriptsize{phys}}}/\rho$ & $h = h_{\mbox{\scriptsize{phys}}}\frac{v_{\mbox{\scriptsize{th}}}^3 L_T}{n \rho}$ \\
$y = y_{\mbox{\scriptsize{phys}}}/\rho$ & $\varphi = \varphi_{\mbox{\scriptsize{phys}}}\frac{q L_T}{T \rho}$ \\
$z = z_{\mbox{\scriptsize{phys}}}/L_T$ & $F_0 = F_{0\mbox{\scriptsize{phys}}}v_{\mbox{\scriptsize{th}}}^3/n = \frac{e^{-v^2/2}}{(2\pi)^{3/2}}$ \\
\label{normalization}
\end{tabular}\end{center}
\vspace{1pt}

\noindent where $\rho = v_{\mbox{\scriptsize{th}}}/\Omega$, $v_{\mbox{\scriptsize{th}}} = (T/m)^{1/2}$, $\Omega = qB/m$ and $L_T^{-1} = \partial \ln T/\partial x$ and dimensional quantities have subscript `phys' (but only where it is necessary to distinguish them with the dimensionless versions).  The temperature ratio is defined $\tau_e = \frac{1}{\tau_i}= Z\frac{T_e}{T_i}$ and $Z = q_i/e$.

Note that quantities are normalized in accordance with the turbulent species, dropping subscripts.  For example, it should be understood that ITG length scales are normalized to $\rho_i$ and $L_{T_i}$  in the perpendicular and parallel directions while ETG has normalization relative to $\rho_e$ and $L_{T_e}$.  Also it is worth making explicit that here $\Omega_e = q_eB/m_e = -eB/m_e$ while $\Omega_i = q_iB/m_i = ZeB/m_i$.  (This convention gives identical normalized equations for the linear theories of ETG and ITG).

\subsection{The primary mode}

The instabilities considered here are the ITG/ETG slab and toroidal modes.  As described in Ref.~\cite{cowley}, the slab instability can be physically understood in terms of phase differences in the perturbed quantities that lead to an unstable feedback process.  The slab mode has variation on the $y$ and $z$ directions.  It is a three dimensional instability since it relies on a temperature gradient in the radial direction, mode structure and propogation in the poloidal direction and flows along the magnetic field.  The toroidal or curvature-driven mode is much simpler as it is unstable in the two dimensional limit and can be understood in terms of equilibrium drift (${\bf v}_D$) reinforcing density and pressure perturbations.

Returning to local gyrokinetic theory, we write the general primary mode as a plane wave with variation in the $y$ and $z$ directions.  The form of the solution is assumed to be

\begin{equation}
h_p = \varphi_{p0} \; f_p(v_{\parallel},v_{\perp}) \; \exp{\dot{\imath}(k_p Y + k_{\parallel}z - \omega_p t)}\label{primary_kinetic_distr}
\end{equation}

\begin{equation}
\varphi_p = \varphi_{p0} \; \exp{\dot{\imath}(k_p y + k_{\parallel}z - \omega_p t)}
\label{primary_phi}
\end{equation}

\noindent where $\varphi_{p0}$ is a real amplitude.  Substituting this into the gyrokinetic Eqn (\ref{gyro-norm}) yields an expression for the kinetic response $f_p$

\begin{equation}
f_p = J_0(k_pv_{\perp}) F_0 \left[\frac{k_p(\eta^{-1} + v^2/2 - 3/2) - \omega_p}{k_p \epsilon_T(v^2_{\parallel} + v^2_{\perp}/2) + v_{\parallel}k_{\parallel} - \omega_p}\right]
\label{f_p}
\end{equation}

\noindent where the approximation $\hat{x}\cdot{\bf b}\cdot\nabla{\bf b} \approx \hat{x}\cdot\nabla\ln{B}$ allows one to write

\begin{equation}
{\bf {\bf v}_D}\cdot\hat{y} = \epsilon_T(v^2_{\parallel} + v^2_{\perp}/2)\nonumber
\end{equation}

\noindent The quantity $\epsilon_T$ is interpreted as $L_T/R$ when considering modes localized to the outboard midplane.  The dispersion relation is obtained by combining Eqn (\ref{f_p}) and Eqn(\ref{quasi_neutrality}) noting the flux surface average is zero for the primary mode solution.

\begin{equation}
1 = \frac{2\pi}{1+\tau}\int_0^{\infty}\int_{-\infty}^{\infty}J_0(k_pv_{\perp})\;f_p\;v_{\perp}dv_{\perp}dv_{\parallel}
\label{dispersion}
\end{equation}

Obviously, Eqn.~\ref{dispersion} must be solved for the complex frequency $\omega_p$.  One approach is to employ a numerical search algorithm over the complex plane as a two dimensional space.  However, since the equation does not depend in general on two variables, but on the single complex number $\omega_p$, an improved route is found in Muller's method (see \cite{frank}).  Muller's method is used to find complex roots of transcendental equations using quadratic interpolation.  The convergence depends on the transcendental equation being analytic in the neighborhood of the root.  The quickest method employed by the author has been to initialize the search algorithm to a solution from fluid theory and use Muller's method to locate and polish the solution to the kinetic dispersion relation.  Growth rate plots can then be obtained quite efficiently by making small adjustments to the parameters of the primary mode and using the previous solution as an initial guess for the new root.

\begin{figure}
\begin{center}
\ifthenelse{\boolean{JpgFigs}}
{\includegraphics[width= 12 cm]{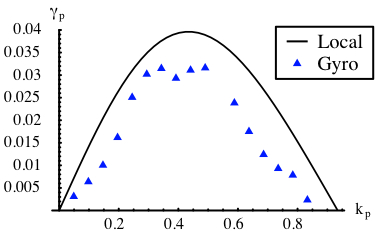}}
{\includegraphics[width= 12 cm]{Figures/Secondary-Fig-1.pdf}}
\end{center}

\caption{{\bf Primary growth rate comparison:} Taking $\eta = 3.14$, $\epsilon_T = .14$ and $\tau=1$, the local dispersion relation is solved and compared with results from linear run\supercite{wang} from GYRO.\supercite{candy}  (color figure)}
\label{primary_compare}
\end{figure}

An example from a Mathematica\texttrademark\; calculation is given below.  It is remarkable how closely the solution to the local dispersion relation agrees with exact (full geometry) eigenmode calculations.  Fig \ref{primary_compare} compares numerical solutions of this dispersion relation with results from linear runs\supercite{wang} on GYRO\supercite{candy} with the equilibrium corresponding to the Cyclone\supercite{dimits, nevins} base case.

It is worthwhile to note that this linear solution to the gyrokinetic equation is actually an exact nonlinear solution -- linear modes are not self-interacting in the local approximation.  To recover the fluid limit, the kinetic response given in Eqn.~\ref{f_p} can be expanded in the slab and toroidal limits.  In the slab limit we take $\epsilon_T = 0$ (low-beta limit) and $v_{\parallel}k_{\parallel} \ll \omega_p \ll k_p(\eta^{-1} + v^2/2 - 3/2)$ (strong temperature gradient limit).  This leaves gaussian integrals which can be performed analytically.  For the toroidal branch we set $k_{\parallel} = 0$ and $k_p \epsilon_T(v^2_{\parallel} + v^2_{\perp}/2) \ll \omega_p \ll k_p(\eta^{-1} + v^2/2 - 3/2)$ so that the temperature gradient again dominates but is made unstable by bad curvature.


\section{Introduction to secondary instability theory}

The instabilities that drive turbulence and transport in tokamaks themselves become unstable at finite amplitude to secondary instabilities.  These ``secondaries'' are a key part of the nonlinear physics.  The following sections in this chapter present a fully gyrokinetic secondary instability theory for electron temperature gradient (ETG) and ion temperature gradient (ITG) driven turbulence, drawing from the author's work \cite{plunk}.  We continue in the local electrostatic limit to derive an integral equation for the secondary mode.  The assumptions made in deriving the equation are designed to find ``fast'' secondary modes which satisfy $\gamma_s \gg \gamma_p$ and can therefore lead to mode saturation.  Finite Larmor radius (FLR) and other kinetic effects are treated exactly capturing $k\rho > 1$ as well as $k\rho \ll 1$ quasi-singular behavior.


Aside from their potentially important role in anomalous transport in fusion plasmas, secondary instabilities are of theoretical interest as a mechanism for the transition to turbulence and as a simplification of the fully nonlinear state of plasma turbulence.

\subsection{Motivation}

ITG turbulence is characterized by a self regulating process whereby zonal flows ($E\times B$ flows due to potential perturbations varying only in the radial direction) are strongly excited leading to the shearing of turbulent eddies which both inhibits transport and reduces the linear growth drive.  Internal transport barriers (ITB) are characterized by the presence of spontaneously created sheared flows.  Although the role of turbulence in generating these flows is debated, they do have the effect of suppressing ITG turbulence and reducing ion thermal transport to neoclassical levels.  However, within these barriers, the electron thermal transport persists at anomalous levels with a temperature gradient close to the critical gradient of ETG linear theory.\supercite{stallard}

The presence of persistent high electron thermal transport in both experiments and simulations has been a challenge to explain.  Initial mixing length estimates for ETG driven turbulence gave $\chi_e \sim \sqrt{m_e/m_i}\;\chi_i$ which caused many to believe that electron thermal transport due to ETG should be negligibly small compared to the transport from ITG.  These estimates were motivated by the notion that ETG and ITG turbulence are nearly isomorphic and might be expected to exhibit analogous spectrums -- i.e. that typical eddy size and turnover times, etc., could be obtained by simple transformation from ITG values.  With the discovery of streamer structures in ETG simulations,\supercite{dorland1} it became clear that fully developed ETG turbulence is qualitatively different and has the potential to deliver non-negligible thermal transport.

Mixing length theory is used to estimate transport in plasma turbulence.  It does so by assuming a representative eddy (e.g. an ETG streamer) which stirs plasma in the presence of an equilibrium temperature and density gradient.  There are questions which naturally arise if we are to interpret ETG or ITG turbulence in this framework.  What typical temporal and spatial scales should one expect?  Do primary mode structures exist which are stable or quasi-stable?  If so, do they have an important role in transport?  At what amplitude do primary modes `saturate' and by what mechanism?  In other words, what are the properties the ``typical'' eddies which cause transport?  From this information, one can obtain estimates and the scaling properties for transport\supercite{cowley}.  We will show how secondary instability theory can be used to obtain these estimates and also gain insight into underlying physical processes.

To get a sense of what is meant by ``secondary instability'', one may consider the following idealized lifetime of a primary mode.  It is broken into three stages as follows.  In the first stage the mode is growing linearly.  The second stage begins when the amplitude of this primary instability gets large (along with its associated gradients) and a secondary mode is destabilized with a growth rate substantially larger than the primary growth rate.  The secondary growth rate is proportional to the amplitude of the primary and its time dependence is therefore commonly described as the exponential of an exponential.  Despite the fact that the growth rate of this secondary is large, the perturbation itself is still small and therefore the primary mode growth is basically unchanged.  In the third stage, the secondary mode has rapidly ``caught up'' to the primary and the dynamics become fully nonlinear leading to a ``turning over'' of both the primary and secondary modes.  

The secondary instability described in stage two is the subject of the remainder of this chapter.  The true non-linear evolution of an ITG/ETG mode is not as simple as the above description.  Coupling with other modes is always present to some degree.  It is therefore only in the most controlled simulations where one actually encounters a single linear mode growing in a quiescent background.  This scenario is, nonetheless, of fundamental interest.  In studying secondary instability theory, we hope to find dominant effects which determine the features (scales, amplitudes, etc.) of fully developed turbulence.

Here we consider what will be referred to as ``fast'' secondaries which are normal modes that satisfy the condition $|\omega_s| \gg |\omega_p|$ where $\omega_s$ and $\omega_p$ are the secondary and primary mode frequencies.  In particular, the modes of interest also satisfy $\gamma_s \gg \gamma_p$.  Such modes are desirable to study because they are readily observable in simulations and are regarded as an important mechanism to bring about mode saturation because they have the ability to ``catch up'' to the primary.  Thus the primary mode functions as a ``frozen'' equilibrium which drives the secondary.  The key assumption leading to the destabilization of a ``fast'' secondary is that the gradients of the primary mode (pressure, density, etc.) become so large that they dominate over the background gradients.  Note that this does not violate the assumptions of gyrokinetics (which require for example $\delta n \ll n_0$) since the spatial scales of perturbed quantities are formally an order smaller than the background scales allowing perturbed and equilibrium gradients of comparable magnitude.  In the presence of such a strong primary, the effective total gradient becomes oriented in the ${\bf k}_{\perp}$ direction\supercite{cowley} (on the outboard midplane, this would be in the poloidal direction for the most unstable linear modes) instead of the radial direction.  Concurrently, the electrostatic perturbation is causing $E \times B$ and parallel flows which are sheared and can give rise to a Kelvin-Helmholtz type instability.  Thus there are several ingredients present which can, given the appropriate primary mode phasing, create a strong secondary instability.  The assumption of large perturbative gradients also results in a simple version of the gyrokinetic equation where the background linear drive terms do not enter.  (Also, geodesic acoustic modes will be absent as a result of ignoring background gradients.)

The interest in secondary instability theory for tokamak turbulence has been strong in recent years.  Authors have explored both ``fast''\supercite{drake, dorland1, cowley} and ``slow''\supercite{diamond:1991, diamond00, diamond1, chen, lashmore} ($|\omega_s| \ll |\omega_p|$) instabilities of primary mode structures.  The break up of primary mode structures via secondary instabilities has been shown by many authors to be tied closely to zonal flow generation (see for instance Refs \cite{diamond00, chen, rogers}), leading to a predator prey-type understanding of ITG turbulence.\supercite{diamond0}  Secondary instability theory has also been used successfully to estimate saturation amplitudes and explain the qualitative differences between ETG and ITG driven turbulence.\supercite{dorland3}  Thus secondary instability theory has enjoyed success in advancing our fundamental understanding of tokamak turbulence.

Refs \cite{dorland1, cowley, diamond1, chen} investigate modes destabilized by both linear and nonlinear turbulent structures.  We broadly refer to these modes as secondary instabilities.  They employ a variety of model equations and ordering assumptions but have a common aim of investigating instabilities which arise from the presence of primary mode structures.  The secondary instability considered in the present work is closely related to the one considered in Ref \cite{dorland1} and to the secondary instability described in Ref \cite{cowley}.  It can be thought of as a fully gyrokinetic extension of those works.

\subsection{Content}

The remaining content of the chapter is as follows.  In Sec (\ref{sec_eqns}) the equations for the secondary instability are derived.  The primary mode is calculated using the local kinetic dispersion relation (which is justified for large range of perpendicular wavenumbers).  The amplitude of the primary is assumed to be large enough (as described above) that the secondary instability is driven only by the primary.  Therefore, the background equilibrium only enters in the secondary calculation through the dependence of the primary mode solution on the parameters $\tau_e = Z T_e/T_i$, $\eta = L_n/L_T$ and $\epsilon_T = L_T/R$.

In Sec (\ref{sec_tor}) results are discussed for the case of the toroidal branch primary mode ($k_{\parallel} = 0$).  At small primary wavenumber, the maximum ETG and ITG secondary growth rates are proportional to $k_p^4$ and $k_p$ respectively in agreement with Ref \cite{dorland1}.  The secondary instability dependence on the parameters $\eta = L_n/L_T$ and $\epsilon_T = L_T/R$ is studied.  The parametric dependence of the ETG secondary instability is qualitatively different and significantly more sensitive than the ITG case.

Sec (\ref{sec_mix}) continues the discussion of the toroidal case.  Mixing length theory is used to explore the consequences of secondary instability theory on transport in tokamaks.  In both ITG and ETG, the well known Dimits shift\supercite{dimits} above the critical gradient parameter $(R/L_{T})_{\mbox{\scriptsize{crit}}}$ can be seen to follow from secondary theory.  The difference between the ITG and ETG cases suggest that stiffness may play a weaker role in the determination of the electron temperature profile due to an unexpected weakening of the ETG secondary at strong temperature gradient (more specifically, small $\epsilon_T$).  We draw attention to recent experiments in electron cyclotron heating which did not find a critical gradient in the electron temperature.\supercite{deboo}  The weakening of ETG secondaries at high temperature gradient also may be a key element in the understanding of electron ITB formation.

Sec (\ref{sec_slab}) presents results for the secondary instability for slab primary modes ($\epsilon_T = 0$ and $k_{\parallel} \neq 0$) initially investigated in Ref \cite{cowley}.  The ITG and ETG theories are identical, in agreement with Ref \cite{dorland3}.  The mode is strong at small primary wavenumber $k_p$ in contrast with the toroidal case.  Thus the role of this instability deserves a more detailed investigation especially in light of the fact that the parallel wavenumber of toroidal ETG/ITG primary modes becomes important in the small $k_p$ limit.

Finally, Sec (\ref{sec_ki}) takes a brief look at secondary instability theory applied to the the intermediate ITG/ETG coupling regime that has been explored in recent numerical simulations.\supercite{waltz, jenko}  The regime $\rho_i^{-1} < k_{\perp} < \rho_e^{-1}$ is especially challenging because the adiabatic (aka Boltzmann) response is no longer valid and the kinetic treatment of ions is dominated by FLR effects.  This makes gyrokinetic secondary theory well suited to explore the transition and interaction between ITG and ETG driven turbulence.

\subsection{Secondary instability}

The secondary instability equation is derived by considering a small perturbation to the primary mode

\begin{equation}
h = h_p + h_s \;\;\; and \;\;\; \varphi = \varphi_p + \varphi_s
\end{equation}

\noindent with $\varphi_s/\varphi_p \ll 1$ and $h_s/h_p \ll 1$.

The solution for the primary mode is given by Eqn (\ref{primary_kinetic_distr} - \ref{dispersion}).  At next order one obtains a linear equation for $h_s$ and $\varphi_s$.  As discussed, we are looking for ``fast'' secondaries and assume the primary mode dominates over the background.  Specifically we take $\nabla h_p \gg \nabla F_0$ and terms involving the equilibrium do not appear in the equation that follows.  Therefore the secondary mode draws energy directly from the primary mode and can serve as a saturating mechanism.  The time dependence ($\omega_p t$) of the primary mode is taken as a constant phase which is set to zero without loss of generality.  We thus restrict the calculation to secondary modes with frequency $\omega_s$ satisfying $|\omega_s| \gg |\omega_p|$.  This includes the most important modes with $\gamma_s > \gamma_p$ that are capable of ``catching up'' with the primary mode.  In summary, the primary mode serves as a ``frozen'' fine-scale equilibrium that drives the secondary instability.  The gyrokinetic equation for the secondary is

\begin{align}
 \frac{\partial h_s}{\partial t} + v_{\parallel} \frac{\partial h_s}{\partial z} + \hat{z}\times\gyroavg{ \nabla\varphi_s }\cdot\nabla \mbox{Re}[h_p]
 +  \hat{z}\times\gyroavg{ \nabla \mbox{Re}[\varphi_p] }\cdot\nabla h_s = F_0\frac{\partial\gyroavg{\varphi_s}}{\partial t}
\label{gyro_2}
\end{align}

\noindent The third and fourth terms on the left hand side result from the Poisson bracket nonlinearity in the gyrokinetic equation.  The first of these represents the action of the primary as a perturbed gradient (i.e. with $\nabla h_p$ in place of $\nabla F_0$).  The second term represents the action of the sheared $E\times B$ primary flow on the secondary mode.  The secondary is next written as a fourier mode in $z$ (because $|\omega_s| \gg |\omega_p|$, see Sec \ref{sec_slab}) and $x$ multiplied by what will loosely be referred to as an eigenfunction, depending on $y$

\begin{equation}
\varphi_s = \tilde{\varphi} (y) \;\exp{\dot{\imath} (k_x x + k_{\parallel s} z - \omega_s t)}
\end{equation}

\noindent with the gyrocenter distribution function given by

\begin{equation}
h_s = \tilde{h} (Y, v_{\perp}, v_{\parallel}) \;\exp{\dot{\imath} (k_x X + k_{\parallel s} z - \omega_s t)}
\end{equation}

\noindent Combining this with quasi-neutrality Eqn (\ref{quasi_neutrality}) gives an integral equation for $\tilde{\varphi}(y)$:

\begin{align}
\tilde{\varphi}(y) = &\int d^4{\bf v} \; {\mathcal H}(v_{\perp},\vartheta,\vartheta^{\prime})\;\tilde{\varphi}(y - v_{\perp}(\sin\vartheta - \sin\vartheta^{\prime})) + \delta_{(ITG)}\frac{\tau}{\tau+1}\overline{\tilde{\varphi}(y)}
\label{secondary_eqn}
\end{align}

\noindent with

\begin{eqnarray}
& {\mathcal H}(v_{\perp},\vartheta,\vartheta^{\prime}) \equiv \frac{\mbox{Re}[-\dot{\imath}f_p \; e^{\dot{\imath}k_p(y - v_{\perp}\sin\vartheta)}]  \; - \; \omega_z F_0}{\sin[k_p(y - v_{\perp}\sin\vartheta)] J_0(k_p v_{\perp}) + v_{\parallel}k_{\parallel z} - \omega_z} \; e^{\dot{\imath} k_x v_{\perp}(\cos\vartheta - \cos\vartheta^{\prime})} \nonumber \\
& d^4{\bf v} \equiv \frac{dv_{\parallel}v_{\perp}dv_{\perp} d\vartheta d\vartheta^{\prime}}{2\pi(1+\tau)} \nonumber \\
&  \omega_z \equiv \frac{\omega_s}{k_pk_x\varphi_{p0}} \nonumber \\
& k_{\parallel z} \equiv \frac{k_{\parallel s}}{k_pk_x\varphi_{p0}}
\label{secondary_eqn_defs}
\end{eqnarray}

\noindent The secondary stability Eqn (\ref{secondary_eqn})  is solved for $\tilde{\varphi}(y)$ and $\omega_z$ for fixed parameters $k_p$, $\tau$, $\eta$, $\epsilon_T$, $k_x$ and $k_{\parallel z}$.  First assume a truncated fourier series expansion (as used for instance in Refs \cite{thess,frenkel}).

\begin{equation}
\tilde{\varphi} \sim \tilde{\varphi}_M = e^{\dot{\imath} k_fy} \displaystyle \sum_{n = -M}^M c_n\; e^{\dot{\imath} nk_py}
\end{equation}

\noindent where $k_f$ is the Floquet exponent.  Eqn (\ref{secondary_eqn}) then leads to the matrix equation  

\begin{equation}
{\bf c} = L\; {\bf c} + D\; {\bf c}
\end{equation}

\noindent with ${\bf c}$ the column vector with elements $\{c_{-M}...c_{M}\}$.  The operator $D$ carries through the flux surface average term for ITG.  The elements of D are

\begin{equation}
D_{mn} = \delta_m\delta_n\left(\delta_{(ITG)}\delta_{k_f}\delta_{k_{\parallel s}}\right)\frac{\tau}{\tau+1}
\end{equation}

\noindent The factors $\delta_{k_f}$ and $\delta_{k_{\parallel z}}$ simply account for the fact that non-zero $k_f$ or $k_{\parallel z}$ yield a vanishing flux surface average of $\varphi_s$.  The elements of $L$ are

\begin{align}
&L_{mn} = \frac{2\pi}{1 + \tau}\int dv_{\parallel}v_{\perp}dv_{\perp}\;Z_{mn}J_0(k_mv_{\perp})J_0(k_nv_{\perp}) \nonumber\\
&Z_{mn} = \frac{1}{2\pi}\int_0^{2\pi} dy \frac{(\mbox{Re}[-\dot{\imath}f_pe^{\dot{\imath}y}] - \omega_zF_0)\;e^{\dot{\imath}(n-m)y}}{\sin(y)J_0(k_pv_{\perp}) + v_{\parallel}k_{\parallel z} - \omega_z} \nonumber\\
&=  \left [ \frac{\dot{\imath}\mbox{Re}[-\dot{\imath}f_pe^{\dot{\imath}y}] - \omega_zF_0}{\cos(y)J_0(k_pv_{\perp})}e^{\dot{\imath}(n-m)y} \right ]_{y=y_{pole}}
\label{matrix_elms}
\end{align}

\noindent with

\begin{eqnarray}
&k_m = \sqrt{k_x^2 + (k_f + mk_p)^2}\nonumber \\
&k_n = \sqrt{k_x^2 + (k_f + nk_p)^2}\nonumber \\
&y_{pole} = \sin^{-1}[\frac{\omega_z - v_{\parallel}k_{\parallel z}}{J_0(k_pv_{\perp})}] \nonumber
\end{eqnarray}

The matrix approach with a truncated fourier series has the advantage that the $\vartheta$ and $\vartheta^{\prime}$ integrals are done analytically to obtain Bessel functions and the integration over $y$ is handled by contour integration in the final line of Eqn \ref{matrix_elms}.  The system is solved by finding roots $\omega_z$ which correspond to functions $\tilde{\varphi}$ possessing convergent fourier expansions ($\tilde{\varphi}_M \rightarrow \tilde{\varphi}$ as $M \rightarrow \infty$.)

\section{Secondary instability for toroidal branch primary mode}
\label{sec_tor}

Consider the case $k_{\parallel} = 0$.  This corresponds to a primary mode driven by magnetic curvature and the temperature gradient.  This mode belongs to the toroidal branch and its secondary instability will be referred to as the Rogers-Dorland (R-D) secondary instability.  This is the instability considered in Ref \cite{dorland1}.

The authors of Ref \cite{dorland1} identify the physical mechanism driving this secondary as the shear in the radial ExB flows generated by the primary.  Thus the R-D secondary is closely related to the Kelvin-Helmholtz instability from fluid theory.  This statement is justified by comparing the secondary instability equation derived from the gyro-fluid equations to the Rayleigh Stability Equation -- the equations differ by an additive constant due to the ``ion response.''\supercite{dorland2}  (However, the difference is significant as the conventional Kelvin-Helmholtz instability is markedly weaker than the R-D ITG secondary but stronger than the R-D ETG secondary.)

We find for the case of ETG that $\omega_s$ is weakly dependent on the Floquet exponent $k_f$ and that $k_f$ acts to stabilize the secondary instability.  For ITG non-zero $k_f$ is strongly damping as it causes the flux average term to vanish.  We therefore set $k_f$ to zero.  The same appears to be true for the parallel wavenumber of the secondary $k_{\parallel z}$ and this parameter is also set to zero.

With $k_{\parallel} = k_{\parallel z} = 0$, the symmetry of Eqn (\ref{secondary_eqn}) ensures that complex frequencies $\omega_z$ appear in sets of four.  If $\omega_z$ is a solution then $\omega_z^{\star}$, $-\omega_z$ and $-\omega_z^{\star}$ are also solutions.  We can therefore confine the search for solutions to the upper right quadrant of the complex plane.

\begin{figure}
\subfigure[Growth rate of primary mode]{
\ifthenelse{\boolean{JpgFigs}}
{\includegraphics[width= .5\columnwidth]{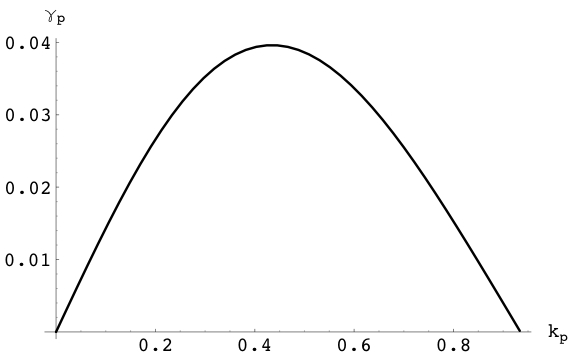}}
{\includegraphics[width= .5\columnwidth]{Figures/Pgrowth.pdf}}
}
\subfigure[Growth rate of ITG secondary mode]{
\ifthenelse{\boolean{JpgFigs}}
{\includegraphics[width= .5\columnwidth]{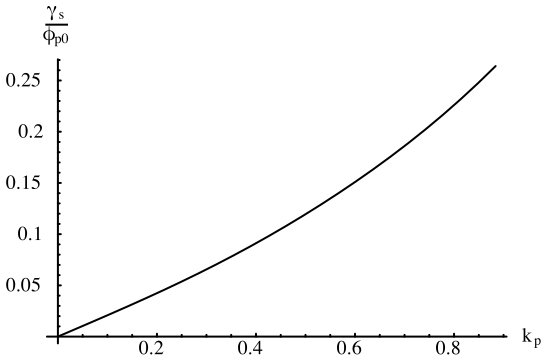}}
{\includegraphics[width= .5\columnwidth]{Figures/ITGgrowth.pdf}}
}
\subfigure[Growth rate of ETG secondary mode]{
\ifthenelse{\boolean{JpgFigs}}
{\includegraphics[width= .5\columnwidth]{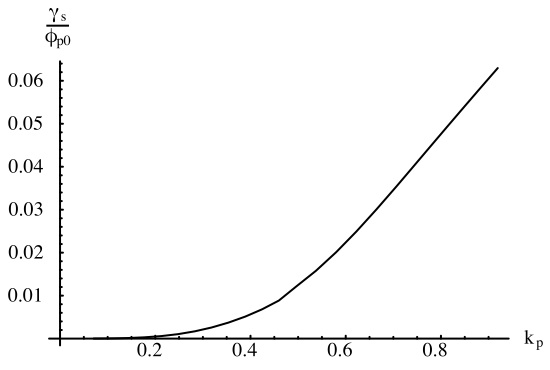}}
{\includegraphics[width= .5\columnwidth]{Figures/ETGgrowth.pdf}}
}
\subfigure[Radial wavenumber at maximum growth rate]{
\ifthenelse{\boolean{JpgFigs}}
{\includegraphics[width= .5\columnwidth]{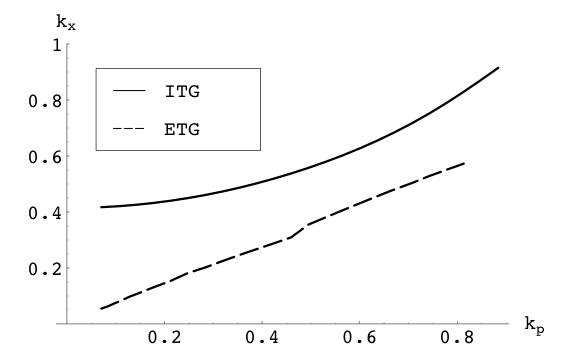}}
{\includegraphics[width= .5\columnwidth]{Figures/RogersKx.pdf}}
\label{rogers_kx_plot}
}

\caption{{\bf Rogers-Dorland secondary instability:} Growth rates for the toroidal (R-D) case are plotted as a function of primary wavenumber $k_p$.  The secondary growth rates correspond to the fastest growing $k_x$ mode for each value of $k_p$ and are given relative to the amplitude of the primary mode.}
\label{growth_plots}
\end{figure}

Growth rate curves are computed by maximizing the secondary growth rate over $k_x$.  Fig \ref{growth_plots} shows the primary growth rate curve as well as secondary growth rates for the standard (Cyclone) parameters $\eta = 3.14$, $\epsilon_T = .14$ and $\tau = 1$.

\subsection{ETG}

For ETG secondaries, we find that {\it both} the large and small $k_p$ limit introduce kinetic effects that cannot be treated perturbatively and make the integral equation fundamentally non-local.  In these cases, one introduces inaccuracies in modeling the secondary instability with a differential equation.

For short wavelengths, $k_p \apprge 1$, the exact instability equation can be efficiently solved with a small number of fourier modes yielding a healthy secondary growth rate across the domain of instability for the primary mode.

\begin{figure}
\begin{center}
\ifthenelse{\boolean{JpgFigs}}
{\includegraphics[width= 13cm]{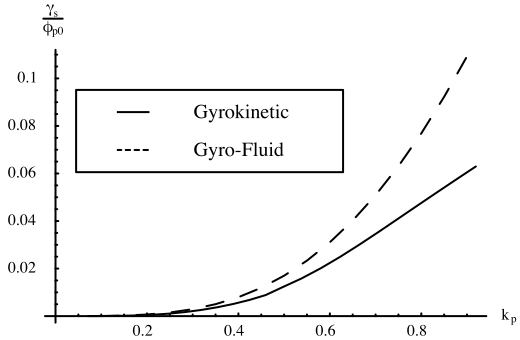}}
{\includegraphics[width= 13cm]{Figures/Secondary-Fig-3.pdf}}
\end{center}

\caption{{\bf Secondary growth rate comparison:}  For the R-D secondary, the growth rate from gyorkinetic secondary theory, Eqn (\ref{secondary_eqn}), is compared with values computed from the gyro-fluid model of Ref \cite{dorland2}.  For the gyrokinetic case parameters are: $\tau = 1$, $\eta = 3.14$ and $\epsilon_T = 0.14$.}
\label{rogers_dorland_comparison_plot}
\end{figure}

\begin{figure}
\subfigure[$k_x$ spectrum ($k_p = .14$)]{
\ifthenelse{\boolean{JpgFigs}}
{\includegraphics[width= .45 \columnwidth]{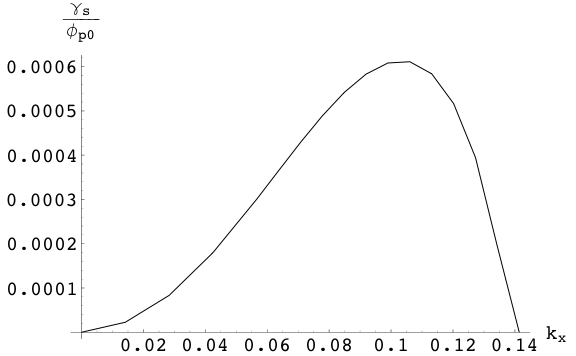}}
{\includegraphics[width= .45 \columnwidth]{Figures/1kxspectrum.pdf}}
}
\subfigure[$k_x$ spectrum ($k_p = .35$)]{
\ifthenelse{\boolean{JpgFigs}}
{\includegraphics[width= .45 \columnwidth]{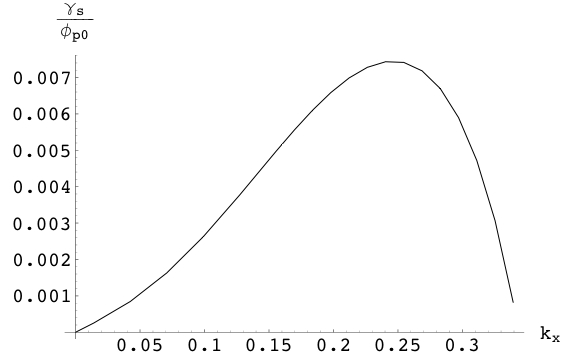}}
{\includegraphics[width= .45 \columnwidth]{Figures/3kxspectrum.pdf}}
}
\subfigure[$k_x$ spectrum ($k_p = .7$)]{
\ifthenelse{\boolean{JpgFigs}}
{\includegraphics[width= .45 \columnwidth]{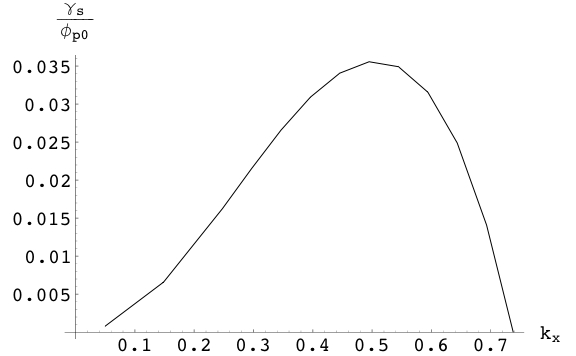}}
{\includegraphics[width= .45 \columnwidth]{Figures/7kxspectrum.pdf}}
}

\caption{{\bf ETG $k_x$ spectrum:} Growth rates for the R-D secondary are plotted as a function of radial wavenumber $k_x$ for three values of $k_p$.  Parameters are $\tau = 1$, $\eta = 3.14$ and $\epsilon_T = 0.14$}
\label{kx_plots}
\end{figure}

For modes with $k_p^2 \ll 1$ the eigenvalue $\omega_z$ is approximately proportional to $k_p^2$ and becomes very small.  Since these modes are unstable for $k_x \sim k_p$, it follows from Eqn (\ref{secondary_eqn_defs}) that the growth rate maximized over $k_x$ (see Fig \ref{kx_plots}) is approximately of the form $\gamma_s \propto k_p^4$ and can be exceedingly small, as also reported in Ref \cite{dorland1}.  Fig \ref{rogers_dorland_comparison_plot} compares the gyrokinetic secondary growth rate calculated from Eqn (\ref{secondary_eqn}) with the growth rate computed from the gyro-fluid model, Eqn (2) of Ref \cite{dorland2}.  The gyro-fluid model has parameters $k_p$, $k_x$ and $\tau$.  Here we take $\tau = 1$ and maximize the growth rate over $k_x$.  The gyrokinetic secondary is solved for Cyclone parameters and maximized over $k_x$.  The agreement is reasonable with the gyro-fluid growth rate differing by about 30\% from the gyrokinetic value.  As we will see, however, the gyrokinetic secondary growth rate is sensitive to the parameters $\eta$ and $\epsilon_T$ and the agreement with the gyro-fluid model will only hold when these parameters are carefully chosen.

In the small $k_p$ limit, the secondary stability equation becomes quasi-singular with a resonant denominator.  The number of fourier modes needed to accurately resolve the eigenfunction $\tilde{\varphi}$ increases rapidly as $k_p$ tends to zero and the matrices in the numerical calculation become large.  Fortunately, it is possible to expand the integrals in the matrix elements in the $k_p^2 \ll 1$ limit which decouples the $v_{\perp}$ integral and makes the computation less intensive.  Solutions obtained from this approximate method agree with those obtained from the exact computation. 
 
\begin{figure}
\ifthenelse{\boolean{JpgFigs}}
{\includegraphics[width=.9 \columnwidth]{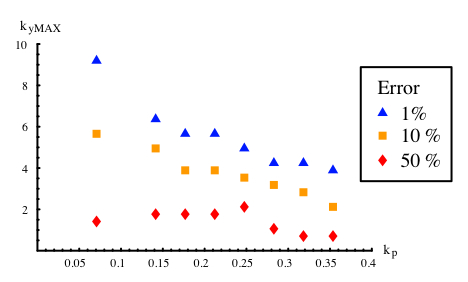}}
{\includegraphics[width=.9 \columnwidth]{Figures/Secondary-Fig-5.pdf}}

\caption{Resolution in $k_y$ needed for convergence of $\omega_s$ to within 50\%, 10\% and 1\% of it's fully converged value for R-D ETG secondary instability.  (color figure)}
\label{theta_res}
\end{figure}

Fig \ref{theta_res} shows the resolution in $k_y$ needed to give properly convergent growth rates.  The quasi-singularity occurs as the eigenvalue $\omega_z$ of Eqn \ref{secondary_eqn} tends to zero at small $k_p$.  In particular, the denominator in the kernel of the integral equation can become small in boundary layers around the nulls of the electric field (where flow shear is maximum).  Quasi-singularity was observed in Ref \cite{dorland1} but was found to weakly affect the secondary growth rate calculation.  As Fig \ref{theta_res} demonstrates, the gyrokinetic secondary instability exhibits more sensitivity of the secondary growth rate to properly resolving the fine scale structure.  Such sensitivity underscores the need for sufficient resolution in simulations where secondaries play an important role.

\begin{figure}

\subfigure[$k_p = .14$]{
\ifthenelse{\boolean{JpgFigs}}
{\includegraphics[width= .45 \columnwidth]{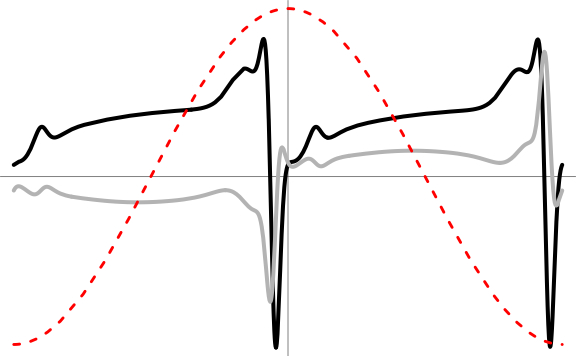}}
{\includegraphics[width= .45 \columnwidth]{Figures/1eigenfunction.pdf}}
}
\subfigure[$k_p = .35$]{
\ifthenelse{\boolean{JpgFigs}}
{\includegraphics[width= .45 \columnwidth]{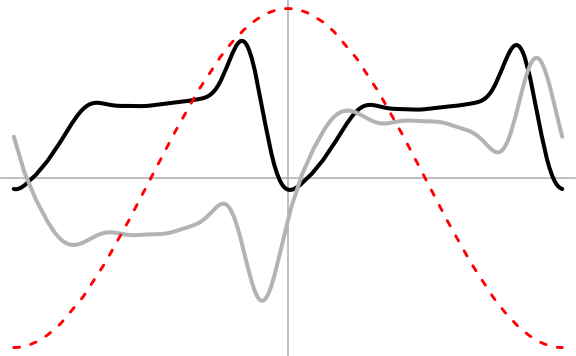}}
{\includegraphics[width= .45 \columnwidth]{Figures/3eigenfunction.pdf}}
}
\subfigure[$k_p = .7$]{
\ifthenelse{\boolean{JpgFigs}}
{\includegraphics[width= .45 \columnwidth]{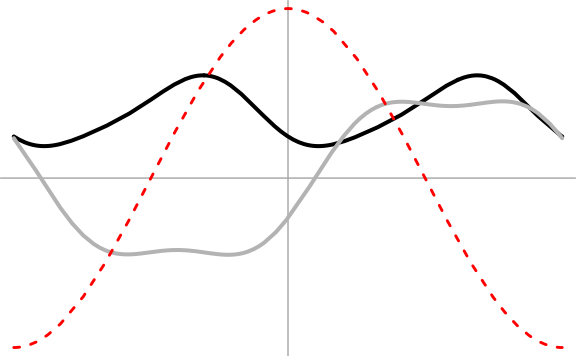}}
{\includegraphics[width= .45 \columnwidth]{Figures/7eigenfunction.pdf}}
}

\caption{{\bf Rogers-Dorland secondary eigenfunctions:} $\mbox{Re}[\tilde{\varphi(y)}]$ (black-solid) and $\mbox{Im}[\tilde{\varphi(y)}]$ (grey-solid) are plotted for the fastest growing modes.  The primary mode $\varphi_p = \varphi_{p0}\; \cos(k_p y)$ appears red-dashed.  Parameters are $\tau = 1$, $\eta = 3.14$ and $\epsilon_T = 0.14$.  (color figure)}
\label{eigenfunctions}
\end{figure}

The fine scale structure can be observed in plots of the eigenfunctions in Fig \ref{eigenfunctions}.  Here we compare eigenfunctions for different values of $k_p$.  One noteworthy feature of these eigenfunctions is that the sharp features appear close to the region where the magnitude of the shear in the $E\times B$ flow is maximum (where $\partial^2\varphi_p/\partial y^2$ is largest in magnitude), pointing to shear flow as a central driving force of the instability.
 
\subsection{ITG}

The R-D secondary instability for ITG is easily computed for all values of $k_p$ across the range of primary instability.  The eigenvalue $\omega_z$ tends to a constant in the limit $k_p \rightarrow 0$ and the secondary instability equation is non-singular.  As a result, the secondary growth rate has dependence $\gamma_s \propto k_p$ for small $k_p$, as can be seen in Fig \ref{growth_plots}.  Additionally, the value of $k_x$ at which $\gamma_s$ is a maximum ($k_{x,max}$) also tends to a constant.  Of course this means that for $k_p \ll 1$, the secondaries have $k_{x,max} \gg k_p$ so that long wavelength modes tend to be highly unstable to short radial wavelength perturbations.  Fig \ref{growth_plots}(d) plots $k_{x, max}$ as a function of $k_p$ for the Cyclone parameters.  In contrast to ETG, ITG secondaries have the feature $k_{x,max} > k_p$.  This is an indication that ITG turbulence should develop short radial wavelength structure in contrast to the ``streamers'' found in ETG turbulence.  However, one cannot be sure that the fastest growing secondaries may are the most effective mechanism of mode saturation and it is possible that ``tertiary'' instabilities\supercite{dorland2} play a role in determining which secondary modes dominate the saturation process.

\subsection{Parametric dependence and contour plots}

\begin{figure}


\subfigure[ETG]{
\ifthenelse{\boolean{JpgFigs}}
{\includegraphics[width= .45 \columnwidth]{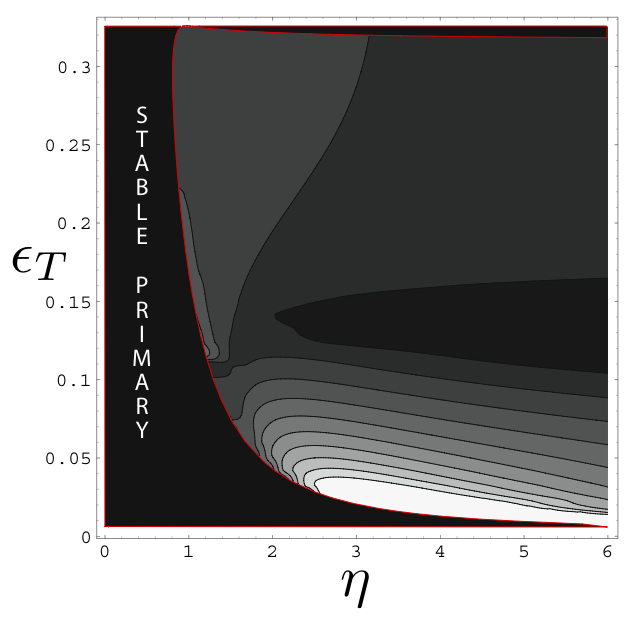}}
{\includegraphics[width= .45 \columnwidth]{Figures/Contour2ndary3.pdf}}
}
\subfigure[ITG]{
\ifthenelse{\boolean{JpgFigs}}
{\includegraphics[width= .45 \columnwidth]{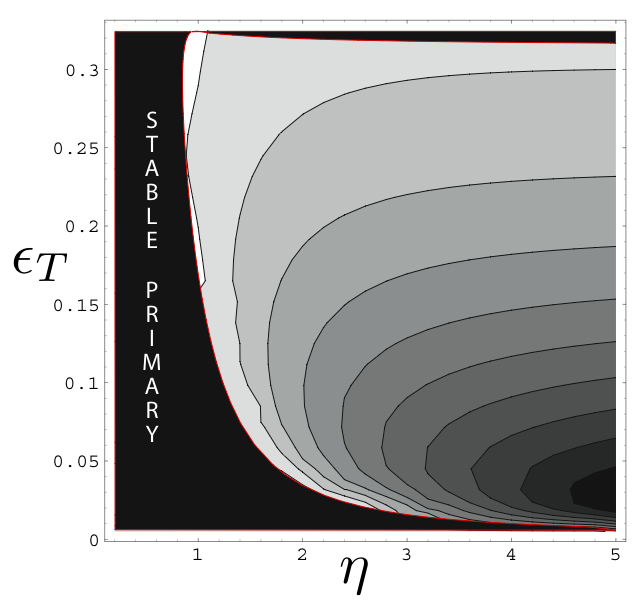}}
{\includegraphics[width= .45 \columnwidth]{Figures/Contour2ndaryITG3.pdf}}
}

\caption{{\bf Rogers-Dorland secondary parameter scan:} The growth rate $\gamma_s/\varphi_{p0}$ is plotted for $\tau = 1$ and $k_p = 0.21$.  The red line gives the approximate marginal stability contour and the black region outside this contour corresponds to $\gamma_p \leq 0$.  Also at the top of the plot the primary growth rate goes to zero at the critical marginal stability parameter which is roughly $\epsilon_{T,crit} \sim 0.3$ (as calculated from the local kinetic dispersion relation).   Secondary growth rates are calculated for the region to the right of the red contour where $\gamma_p > 0$.  The color scales linearly for ITG from black $< 0.035$ to white $> 0.05$.  ETG has a more sensitive parametric dependence with black $< 0.0003$ and white $> 0.003$.}
\label{par_scan_gamma}
\end{figure}

The qualitative features of the R-D secondary instability described in the previous section hold across the range of parameters $\eta$ and $\epsilon_T$ investigated in this work.  Namely for small $k_p$, $\gamma_s \propto k_p^4$ for ETG and $\gamma_s \propto k_p$ for ITG and the secondary is unstable for all $k_p$ such that $\gamma_p > 0$.  Parameter scans across the $\eta$-$\epsilon_T$ plane give quantitative growth rate comparisons.  Fig \ref{par_scan_gamma} plots the growth rate of ETG and ITG secondaries across the $\eta$-$\epsilon_T$ ($\tau = 1$) instability region ($\gamma_p > 0$) for a fixed value of the primary wavenumber, $k_p = 0.21$.  This value of $k_p$ was chosen to correspond to the typical poloidal wavenumber of streamer structures in ETG simulations.  The lines in these plots correspond to curves of constant $\gamma_s/\varphi_{p0}$.  The qualitative differences between ETG and ITG are apparent: the ITG secondary growth rate is generally bigger close to marginal stability of the primary mode, whereas the ETG secondary appears strongest at small $\epsilon_T$.  Also we note that the strength of the ETG secondary varies by an order of magnitude whereas the ITG growth has a less sensitive parametric dependence.

\begin{figure}

\subfigure[ETG]{
\ifthenelse{\boolean{JpgFigs}}
{\includegraphics[width= .45 \columnwidth]{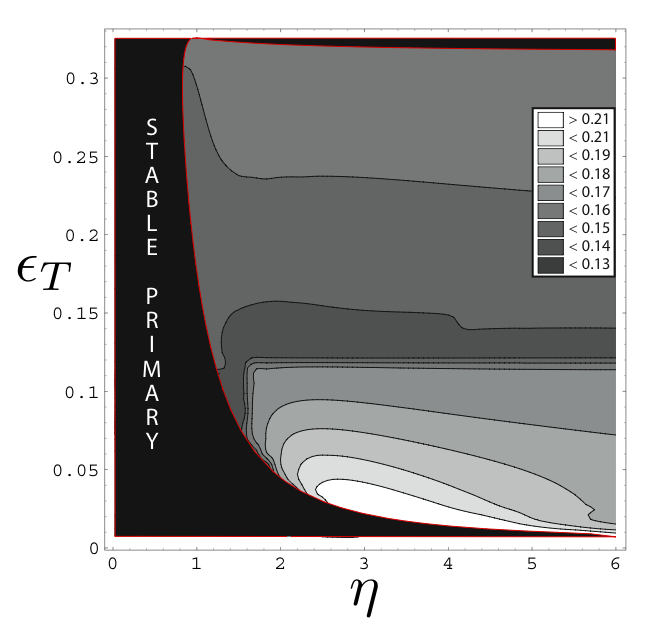}}
{\includegraphics[width= .45 \columnwidth]{Figures/KxContour3_key.pdf}}
}
\subfigure[ITG]{
\ifthenelse{\boolean{JpgFigs}}
{\includegraphics[width= .45 \columnwidth]{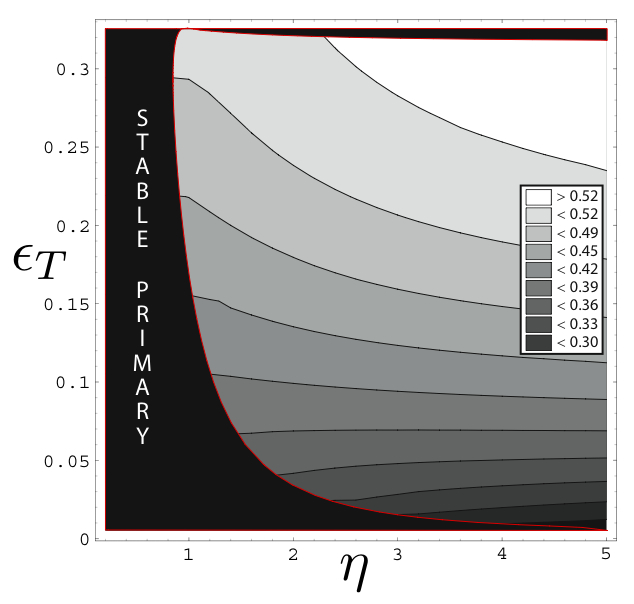}}
{\includegraphics[width= .45 \columnwidth]{Figures/KxContour3ITG_key.pdf}}
}

\caption{{\bf Rogers-Dorland $k_{x, max}$ parameter scan:} Contour plot parametric dependence of $k_{x, max}$, the value of $k_x$ corresponding to the fastest growing secondary.  Here we have $k_p = 0.21$.  The red line gives the approximate marginal stability contour and the black region outside this contour corresponds to $\gamma_p \leq 0$.  Also at the top of the plot the primary growth rate goes to zero at the critical marginal stability parameter which is $\epsilon_{T,crit} \sim 0.3$ (as calculated from the local kinetic dispersion relation).}
\label{par_scan_kx}
\end{figure}

From the plot of the maximizing radial wavenumber $k_{x,max}$ in Fig \ref{par_scan_kx}, it can be seen that the enhancement of growth rate at small $\epsilon_T$ for ETG is accompanied by a qualitative change in the radial wavenumber ($k_x$) spectrum.  For large $\epsilon_T$, the instability condition $k_x < k_p$ is approximately true.  But in the enhanced growth rate region there are unstable modes with $k_x > k_p$.  Furthermore Fig \ref{par_scan_kx}, shows there is a small region (white) where the fastest secondary modes satisfy $k_{x,\mbox{\scriptsize{max}}} > k_p$.  This may manifest as a qualitative change in the nature of turbulence for systems in this regime.

\begin{figure}


\subfigure[ETG]{
\ifthenelse{\boolean{JpgFigs}}
{\includegraphics[width= .45 \columnwidth]{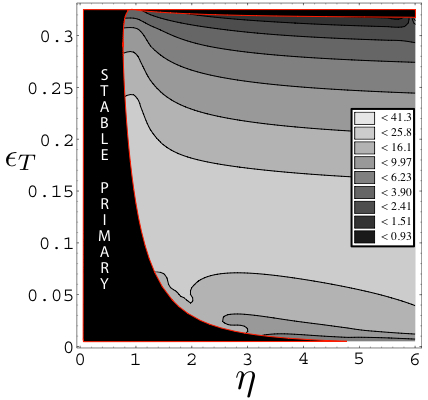}}
{\includegraphics[width= .45 \columnwidth]{Figures/AspectRatioScan3ETG_key.pdf}}
\label{xi_scan_etg}
}
\subfigure[ITG]{
\ifthenelse{\boolean{JpgFigs}}
{\includegraphics[width= .45 \columnwidth]{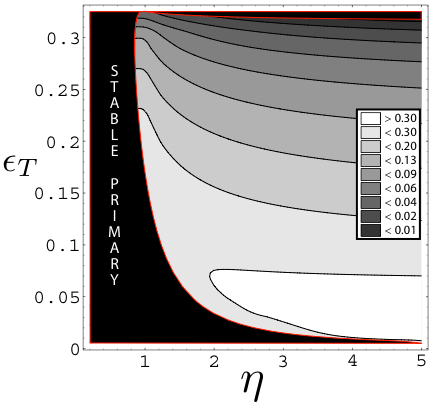}}
{\includegraphics[width= .45 \columnwidth]{Figures/AspectRatioScan3ITG_key.pdf}}
\label{xi_scan_itg}
}

\caption{{\bf Effective aspect ratio parameter scan:} Constant contours of the effective aspect ratio at saturation $\xi_{sat}/\lambda_p$ are plotted.  The primary wavenumber and temperature ratio are held constant: $k_p \sim 0.21$ and $\tau = 1$.  As expected, ETG exhibits the signature large aspect ratio ``streamers,'' while the ITG aspect ratio is kept below unity aspect ratio.  Both cases indicate a strong suppression of turbulent transport near the linear marginal stability cutoff $(L_T/R)_{crit}$ (top of plots) giving a theoretical understanding of the Dimits shift.}
\label{xi_scan}
\end{figure}

\section{Transport and mixing length theory}
\label{sec_mix}
Although ITG shows a much more modest variation in the secondary growth rate shown in Fig \ref{par_scan_gamma}(b), in both the cases of ETG and ITG, it is interesting to consider how the variation can effect transport.  To this end we consider the mixing length estimate of diffusivity which gives $\chi \sim \ell_r^2/\tau$ where $\ell_r$ is a radial step length and $\tau$ is a nonlinear step time.  

As in Ref \cite{cowley} we have in mind a typical electrostatic eddy of radial extent $\ell_r$ and lifetime $\tau$.  Such a structure convects plasma by $E\times B$ flow.  The distance that the plasma can ``step'' is limited by the radial extent $\ell_r$ but also limited by the maximum fluid displacement determined by the strength of the $E\times B$ flow and the linear timescales associated with the eddy.  Here we define the fluid displacement $\xi$ as the time integral of the primary mode $E \times B$ velocity (which in our normalization is simply $v_E = k_p\varphi_p$): $\xi = \int k_p \varphi_p dt$.  To obtain an estimate for the maximum fluid displacement we first estimate the maximum amplitude of the primary (``saturation amplitude'') as the amplitude at which $\gamma_s = \gamma_p$.\supercite{dorland1}  From Eqn \ref{secondary_eqn_defs}, the secondary growth rate may be written $\gamma_s = \bar{\gamma_s} \varphi_{p0}$.  Then the saturation amplitude is $\varphi_{sat} =  \gamma_p/\bar{\gamma_s}$.  Finally the fluid displacement at mode saturation is

\begin{equation}
\xi_{sat} = \xi(\varphi_{p} = \varphi_{sat}) \sim \frac{k_p \varphi_{sat}}{|\omega_p|}
\end{equation}

This quantity can be interpreted within the mixing length framework as a radial step length.  The growth rates of Figs \ref{par_scan_gamma}(a) and \ref{par_scan_gamma}(b) are reinterpreted in terms of $\xi_{sat}$ in Fig \ref{xi_scan}.  Constant contours of the quantity $k_p\xi_{sat}/2\pi = \xi_{sat}/\lambda_p$ (``effective aspect ratio'') are plotted in the $\eta-\epsilon_T$ plane.

Note that the color scale is logarithmic and the variation in the effective aspect ratio is quite large.  In both cases, a mechanism for the nonlinear shift away from marginal stability (Dimits shift\supercite{dimits}) is evident.  At the top of the plot, the growth rate of the primary goes to zero at a critical value of $\epsilon_T$.  It follows that the secondary growth rate can exceed the primary growth rate at small primary amplitude close to this marginal stability boundary.  The result is a very small effective step length for the turbulent mixing.  Therefore, transport would not be expected to ``turn on'' until the temperature gradient is shifted some amount above marginal stability.

A surprising feature of the ETG case of Fig \ref{xi_scan}(a) is the reduction of the effective aspect ratio at $\epsilon_T < 0.1$ where the temperature gradient gets very large.  This is due to the strengthening of the secondary instability.  Coming back to Fig \ref{xi_scan}(a), one may imagine a scenario where heating is causing a steepening of the temperature gradient.  Keeping the density gradient and major radius constant, the local parameters would tend toward small $\epsilon_T$ and increasing $\eta$.  Initially close to marginal stability, turbulent transport is effectively turned off and heating pushes the temperature gradient up easily.  However, the effective radial mixing length $\xi_{sat}$ plateaus at around $\epsilon_T \sim 0.12$.  After this point, $\xi_{sat}$ now actually decreases as the temperature gradient climbs.  Such behavior supports experimental results that the electron temperature profile is not strongly stiff.\supercite{deboo}  Also, note that values of $\epsilon_T$ less than $0.1$ are observed in electron internal transport barriers (see for instance Ref \cite{stallard} or Ref \cite{maget}).  The strengthening of the ETG secondary at low $\epsilon_T$ may therefore be a key element in the formation of electron ITBs.

\begin{figure}[t!]

\subfigure[Primary mode growth rate]{
\ifthenelse{\boolean{JpgFigs}}
{\includegraphics[width= .45 \columnwidth]{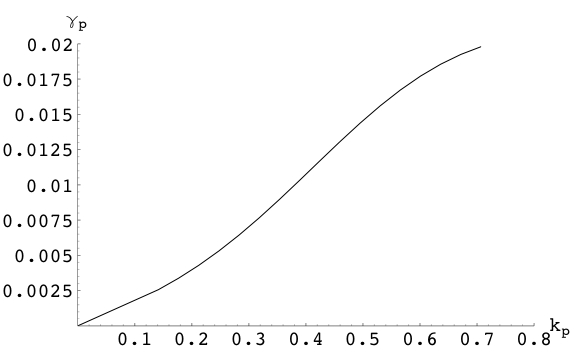}}
{\includegraphics[width= .45 \columnwidth]{Figures/SlabPrimaryPlot.pdf}}
}
\subfigure[Primary mode parallel wavenumber at maximum growth rate]{
\ifthenelse{\boolean{JpgFigs}}
{\includegraphics[width= .45 \columnwidth]{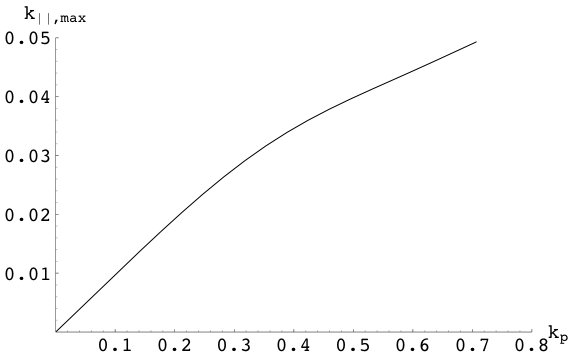}}
{\includegraphics[width= .45 \columnwidth]{Figures/SlabPrimaryK3.pdf}}
}
\subfigure[Secondary mode growth rate]{
\ifthenelse{\boolean{JpgFigs}}
{\includegraphics[width= .45 \columnwidth]{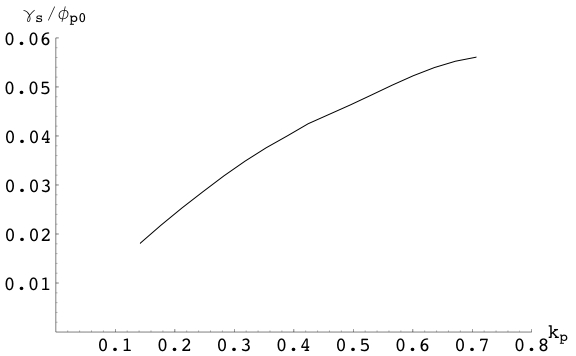}}
{\includegraphics[width= .45 \columnwidth]{Figures/SlabSecondary.pdf}}
}
\subfigure[Secondary mode parallel and radial wavenumbers at maximum growth rate]{
\ifthenelse{\boolean{JpgFigs}}
{\includegraphics[width= .45 \columnwidth]{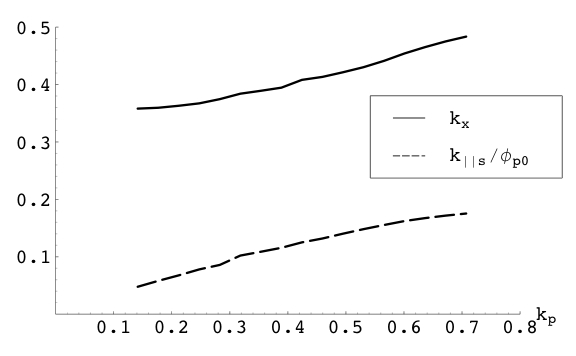}}
{\includegraphics[width= .45 \columnwidth]{Figures/SlabSecondaryK.pdf}}
}

\caption{{\bf Cowley (slab branch) secondary:} $\eta = 3.14$.  The primary growth rate is maximized over $k_{\parallel}$.  The secondary growth rate is maximized over both $k_{\parallel s}$ and $k_x$.}
\label{cowley_secondary}
\end{figure}

\section{Secondary instability for the slab branch primary mode}
\label{sec_slab}

For most of the spectrum of unstable wavenumbers $k_p$, the toroidal branch is the dominant primary instability branch for ETG/ITG tokamak turbulence.  However, the slab branch may play a role in long wavelength turbulence\supercite{kim}, especially at wavelengths where the toroidal branch is stabilized.  The primary mode with $\epsilon_T = 0$ but $k_{\parallel} \neq 0$ belongs to the slab branch of ETG/ITG.  It is characterized by sheared parallel flows which are unstable to a secondary instability first studied by Cowley in the fluid limit\supercite{cowley}.  For this reason it is referred to as the Cowley secondary instability.  

Because the primary mode now has z-dependence, the eigenfunction ${\tilde \varphi}$ should formally now have both $z$ and $y$ dependence.  This complication can be avoided by noting that the $z$-dependence of $\varphi_s$ is unimportant unless $v_{\mbox{\scriptsize{th}}} \partial \ln \varphi_s /\partial z \sim {\cal O}(\omega_s)$.  The assumption that $\omega_s \gg \omega_p$ then guarantees that $\partial \ln \varphi_s /\partial z \gg k_{\parallel}$ is satisfied.  The equation to be solved is again Eqn (\ref{secondary_eqn}).  The flux surface average found in the ITG equation is now zero and the ITG and ETG secondaries are the same.  This is in agreement with simulations by Jenko and Dorland\supercite{dorland3} that showed slab branch driven turbulence having the same normalized heat flux for ETG and ITG.

Fig \ref{cowley_secondary} shows results for $\eta = 3.14$.  The growth rate $\gamma_s$ is maximized over both $k_x$ and $k_{\parallel s}$ in the plot.  Like the ITG R-D secondary, the Cowley secondary has dependence $\gamma_s \propto k_p$ for small $k_p$ and therefore is much stronger than the ETG R-D secondary.  The fastest growing secondaries also have finite $k_{x, max}$ in the limit $k_p \rightarrow 0$ and have $k_{\parallel s, max} \propto k_p$.  By comparing the plots of Fig \ref{cowley_secondary} it can also be seen that the assumption $\omega_s \gg \omega_p$ does indeed guarantee that $k_{\parallel s} \gg k_{\parallel}$ and the Fourier mode dependence in $z$ is justified.  (An amplitude $\varphi_{p0}$ which gives $\omega_s \gg \omega_p$ also gives $k_{\parallel s} \gg k_{\parallel}$.)

\section{ETG secondary with kinetic ion response}
\label{sec_ki}

Recent simulations indicate that the adiabatic ion response (assumed so far in this analysis) may not capture the physics needed for resolving the non-linear behavior of long wavelength ETG turbulence.\supercite{waltz, jenko} ETG turbulence simulations conducted on independently developed codes employing the adiabatic ion response exhibit a divergence of the electron heat flux at high normalized shear.  This increased heat flux is associated with the secular growth of long wavelength fluctuations.\supercite{nevins}  Inclusion of a non-zero kinetic ion distribution function has been shown to give convergent electron heat transport.  It is possible to examine this issue from the perspective of gyrokinetic secondary theory because the FLR effects which dominate the kinetic ions can be treated exactly.

The mode investigated here is the R-D Secondary.  The dispersion relation is modified to include two kinetic species and the gyrokinetic equation for the ions is normalized relative to the ETG scales with the exception of velocities in the ion kinetic response which are normalized to the ion thermal velocity.  The secondary instability equation follows as before:

\begin{align}
\tilde{\varphi}(y) = \int d^4{\bf v} \; {\mathcal H}_e \;\tilde{\varphi}(y - v_{\perp}(\sin\vartheta - \sin\vartheta^{\prime})) -{\mathcal H}_i \;\tilde{\varphi}(y - v_{\perp}\mu (\sin\vartheta - \sin\vartheta^{\prime})) \label{secondary_eqn_ki}
\end{align}

\noindent where

\begin{eqnarray}
&{\mathcal H}_e \equiv \frac{\mbox{Re}[-\dot{\imath}f_{pe} \; e^{\dot{\imath}k_p(y - v_{\perp}\sin\vartheta)}]  \; - \; \omega_z F_{0e}}{\sin[k_p(y - v_{\perp}\sin\vartheta)] J_0(k_p v_{\perp}) + v_{\parallel}k_{\parallel z} - \omega_z} \; e^{\dot{\imath} k_x v_{\perp}(\cos\vartheta - \cos\vartheta^{\prime})} \nonumber \\
&{\mathcal H}_i \equiv \frac{\mbox{Re}[-\dot{\imath} f_{pi} \; e^{\dot{\imath}k_p(y - v_{\perp}\mu \sin\vartheta)}]  \; + \; \tau\omega_z F_{0i}}{\sin[k_p(y - v_{\perp}\mu \sin\vartheta)] J_0(k_p v_{\perp}\mu ) + v_{\parallel}k_{\parallel z} - \omega_z} \; e^{\dot{\imath} k_x v_{\perp}\mu (\cos\vartheta - \cos\vartheta^{\prime})} \nonumber \\
&\mu \equiv |\rho_i/\rho_e| = Z\sqrt{m_iT_i/(T_em_e)}
\end{eqnarray}

Gyrokinetic simulations capturing both electron and ion scale turbulence are made possible by setting the larmor radius ratio $\mu$ to be smaller than the actual physical value (e.g. $20-30$ instead of $40-60$).  This technique yields a vast improvement in computational efficiency by essentially squeezing electron and ion scales together and reducing the range of temporal and spatial scales that need to be resolved.  

\begin{figure*}
\begin{center}
\ifthenelse{\boolean{JpgFigs}}
{\includegraphics[width=\columnwidth]{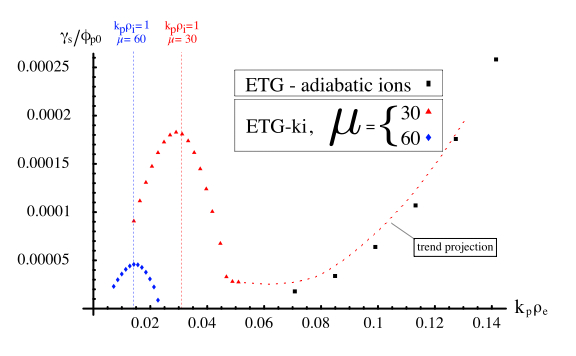}}
{\includegraphics[width=\columnwidth]{Figures/Secondary-Fig-11.pdf}}
\end{center}

\caption{{\bf Rogers-Dorland secondary instability with kinetic ions:} Parameters are $\eta_i = \eta_e = 3.14$, $\epsilon_{Ti} = \epsilon_{Te} = .14$ and $\tau = 1$.  (color figure)}
\label{kinetic_ions}
\end{figure*}

Here we calculate (again with Cyclone parameters) the effect of including the full kinetic response on the secondary instability growth rate.  Fig \ref{kinetic_ions} plots three cases: the ETG R-D secondary with adiabatic ion (ETG-ai) response, the kinetic ion (ETG-ki) case with the unphysically small $\mu \sim 30$ and the ETG-ki case with the physical (corresponding to deuterium) value $\mu \sim 60$.  Although the data for the ETG-ki and ETG-ai do not overlap, both ETG-ki cases should agree with the ETG-ai plot well before $k_p\rho_e = 1$.  (The approximate continuation of the $\mu = 30$ ETG-ki case is drawn in dashed red and labeled ``trend projection.'')  Starting from the right hand side of the plot, we see the familiar weak ETG-ai secondary that goes like $k_p^4$.  Then there is a recovery of the secondary growth rate due to the kinetic response of the ions.  However it is clear that the value of $\mu$ has a dominant effect on the strength of the kinetic ion secondary (relative to the adiabatic ion secondary).

The simulations of Ref \cite{waltz} are invariant under scaling of $\mu$ in the sense that the electron thermal diffusivity measured in {\em ion} Gyro-Bohm units does not change significantly for several cases ($\mu = 20$, $\mu =30$, $\mu = 40$).  However, electron thermal transport at ETG scales does increase sharply with $\mu$ when measured in {\em electron} Gyro-Bohm units.  We interpret this to indicate that ETG driven turbulence becomes more intense as its coupling with ITG turbulence weakens and as the kinetic-ion secondary weakens (relative to ETG scales)--but this is accompanied by an overall downward scaling ETG transport as the ETG spatial scales shrink relative to those of the ITG turbulence (observe that $\chi_{GBe} \sim \chi_{GBi}/ \mu$).  So the net result is an invariance of electron thermal transport in absolute terms (ion Gyro-Bohms).

\begin{figure}
\begin{center}
\ifthenelse{\boolean{JpgFigs}}
{\includegraphics[width= \columnwidth]{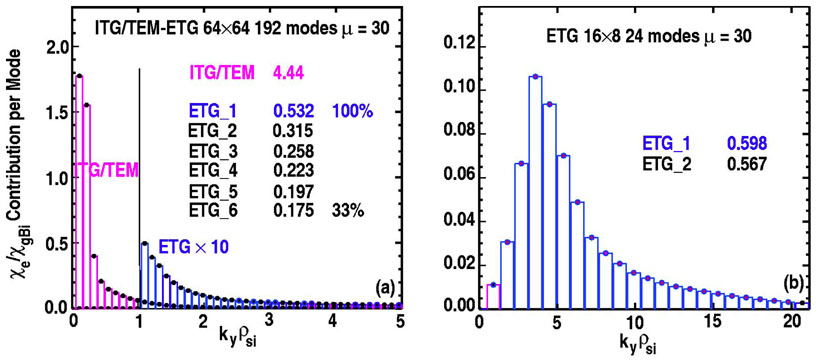}}
{\includegraphics[width= \columnwidth]{Figures/Secondary-Fig-12.pdf}}
\end{center}
\caption{Spectrum of electron energy transport versus wavenumber.  (a) Large-box ITG/TEM-ETG coupled simulation; (b) small-box ETG-ki uncoupled (no ITG cascade).  Reproduced with permission from Ref \cite{waltz}.  (color figure)}
\label{waltz_fig}
\end{figure}

We can gain insight into the role of secondary instabilities in long wavelength ETG turbulence by comparing the kinetic secondary instability results with Fig \ref{waltz_fig}(b) taken with permission from Ref \cite{waltz}.  There is a rough correspondence in the location of the trough of the secondary growth rate curve and the peak of the ``small box'' ETG turbulence spectrum.  In this case it seems plausible that the secondaries play a role in determining this spectrum.  (Actually, the quantity being plotted in Fig \ref{waltz_fig}(b) is thermal diffusivity per mode and should be compared with something like a mixing length estimate of diffusion using $\xi_{sat}(k_p)$ as the radial step length--in this case the error of the location of the peak is still ``rough,'' at about 50\%.)  However Fig \ref{waltz_fig}(a) shows that the full ETG-ITG coupled simulation exhibit a monotonically decreasing transport spectrum that is dominated by ITG wavelengths and does not show special behavior at the location of weakest secondary growth rate.  This seems to suggest another mechanism is important for bringing about saturation of long wavelength ETG modes.

A prevalent view is that the growth of long wavelength ETG modes is moderated by ambient shear flow resulting from the ITG turbulence cascade--and therefore we might conclude that the kinetic-ion secondary instabilities do not play a critical role in the full ETG-ITG coupled scenario.

However, one should not count out kinetic ion secondaries entirely.  Indeed, it is useful to remember that one of the original motivating factors to giving attention to ETG turbulence was the persistence of electron thermal transport in cases where ITG turbulence was supressed\supercite{stallard} (i.e. internal transport barriers).  In such a case we may expect these kinetic ETG secondary instabilities to be quite important and lead to a transport spectrum like the one described in Fig \ref{waltz_fig}(b).

\section{Conclusion}

Secondary instability theory is a key part of the nonlinear physics of tokamak turbulence.  This work has shown how the local electrostatic gyrokinetic theory of secondary instabilities has bearing on important topics in the theory of tokamak turbulence.

First we have seen that the gyrokinetic theory compares well with the gyro-fluid model, yielding $k_p^4$ and $k_p$ growth rate spectra for ETG and ITG R-D secondaries.  At wavelengths significantly shorter than the electron larmor radius, the gyrokinetic secondary growth rate is healthy.  (This eliminates the possibility of stable ultra-fine-scale streamers.)  Quasi-singular behavior of the secondary stability equation Eqn (\ref{secondary_eqn}) for long-wavelength primaries (small $k_p$) calls for high poloidal resolution for an accurate secondary growth rate.  This is relevant for simulations where resolving saturating mechanisms is important.

We have seen that estimating saturation amplitudes from secondary growth rates can yield estimates for radial mixing lengths and turnover times which can be used to estimate transport.  The parametric ($\epsilon_T$ and $\eta$) dependence of the radial mixing length (or ``effective aspect ratio'') yields a novel way to estimate the effect of varying local equilibrium scale lengths ($L_T$, $L_n$, $R$) on transport.  These parameter scans suggest that the suppression of turbulent transport close to marginal stability (Dimits shift) can be understood in terms of the relationship between secondary and primary instability growth rates.  It is worthwhile to emphasize that the parametric sensitivity of the gyrokinetic secondary is a somewhat unexpected result.  It indicates that the gyrokinetic secondary is not driven exclusively by $E\times B$ shear flow and is not as closely related to the Kelvin-Helmholtz instability as the gyro-fluid secondary of Ref \cite{dorland2}.

One of the central results of the work \cite{plunk} is the surprising strengthening of the R-D ETG secondary at small $\epsilon_T$.  First this suggests that the phenomenon of profile stiffness may play a weaker role for an electron temperature profile controlled by ETG turbulence than the an ion temperature profile controlled by ITG turbulence.  A strong ETG secondary at small $\epsilon_T$ also immediately suggests the possibility that secondary instabilities play a role in the formation of electron ITBs.  However, there are questions which remain.  Small or reverse magnetic shear is known to be a important ingredient in the formation of electron ITBs.  An issue not addressed in this work is how magnetic shear effects secondaries.  It is known that the gyro-fluid secondary is modestly destabilized by shear.\supercite{dorland1}  However, since the effect of shear on gyrokinetic secondaries is not included, further work would be useful to determine their role in electron ITBs.

Finally, we calculate the ETG secondary employing a kinetic ion response.  This calculation cannot be done in the fluid limit because the ion gyro-average dominates the ion kinetic response.  Comparison of results with coupled simulations suggest that kinetic secondaries are an important mechanism for saturation of long wavelength ETG modes in cases such as ITBs where the ITG cascade is quenched.
\chapter{The phase space cascade in two dimensional gyrokinetics}
\label{phase-space-turbulence-chapter}

\noindent {\em The paper on which this chapter is based has been substantially updated (with some errors corrected).  Please find the full article online: arXiv:0904.0243}

\section{Introduction}

The gyrokinetic system of equations \cite[]{taylor-hastie, rutherford, catto1977, antonsen, catto, frieman, brizard-hahm} are used to describe plasma dynamics on time-scales much larger than the ion Larmor period.  This system is essentially a kinetic theory of charged rings--rings formed by the fast Larmor motion of particles around the magnetic field lines.  These equations have been developed by the magnetic fusion community to describe plasma turbulence which causes the transport of heat and particles in fusion devices.  As demonstrated in the series of recent works \cite{schekochihin, howes, howes3}, gyrokinetics is also appropriate for a wide range of astrophysical plasmas.  A review of the theoretical framework of turbulence in magnetised plasmas is given by \cite{krommes}.

In this chapter, we study the simplest gyrokinetic system--the two dimensional electrostatic gyrokinetic system driven by a statistically homogeneous and isotropic fluctuating force.  The most important scale in this problem is the Larmor radius scale.  For electrostatic fluctuations on scales much larger than the Larmor radius, all particles move together with the same $E \times B$ drift and a fluid description is correct.  At turbulent scales at the Larmor radius scale and smaller, particles of distinct velocities have different effective $E \times B$ drifts (as they sample different regions of electric field during their rapid Larmor motion) and a kinetic description is required.  At such scales, gyrokinetic turbulence exhibits strongly kinetic behaviour in phase-space due to the presence of so-called nonlinear phase-mixing \cite[]{dorland-hammett-93}.  Thus fluctuations in the distribution function nonlinearly cascade to create fine scale structure in velocity-space in addition to the the two real-space dimensions--and we may think of velocity-space as an additional dimension to be treated on equal footing to position-space, although in this case having a somewhat subsidiary role to position-space as we shall see.

In this chapter, we borrow traditional methods from fluid turbulence.  This is made possible in part due to the fact that the gyrokinetic system bears some resemblance to equations of incompressible fluid turbulence.  The nonlinearity has a nearly divergence-free convective velocity (the $E\times B$ drift velocity due to fluctuating electrostatic fields) while dissipation occurs via a collisional operator which acts via second-order derivatives \cite[see the appendix of][]{schekochihin} in phase-space and, therefore, acts on fine scales in analogy to viscosity in fluid turbulence.  In the proper limit (see section \ref{chm-sec}), the electrostatic (that is, magnetic fluctuations are neglected) gyrokinetic system can be reduced to the Charney--Hasegawa--Mima (CHM) equation \cite[]{charney, hasegawa}, or with additional assumptions reduces to the the inviscid vorticity equation \cite[the relationship between plasma dynamics and two dimensional fluid turbulence was demonstrated first by][]{taylor} which describes two dimensional Euler turbulence \cite[]{kraichnan1967, batchelor}.  Thus from a fluid turbulence perspective, two-dimensional electrostatic gyrokinetics can be seen as a simple kinetic extension of fluid equations which have been extensively studied.

The chapter is structured as follows.  We begin in section \ref{eqns-sec} by discussing the gyrokinetic model, its applicability and the specific assumptions and approximations we will be using.  We then discuss the dynamical collisionless invariants of the gyrokinetic system in section \ref{invariants-sec}.  It is shown that there are two basic collisionless invariants, one which will correspond to the inversely cascading Euler/CHM energy invariant and the other, which can be chosen with some freedom, is related to the perturbed entropy or free energy of the system.  We argue that this invariant must cascade to fine scales in phase-space, where it is ultimately acted upon by the collisional operator, and incorporated into the background Maxwellian to increase entropy.

In section \ref{chm-sec} we explore the relationship between gyrokinetic and the CHM turbulence.  We derive the CHM/Euler and the viscous CHM/NS equation from gyrokinetics to elucidate their relationship.  We also discuss the two invariants of the CHM/Euler equation and how they relate to the gyrokinetic collisionless invariants.  We find that the inversely cascading gyrokinetic invariant transforms continuously into the inversely cascading CHM/Euler invariant (`energy').  The forward cascading quantity from CHM/Euler turbulence (`enstrophy') is found to feed into the gyrokinetic forward cascade but composes only one part of the forward cascading gyrokinetic invariant.

In section \ref{phenom-sec}, we shift attention to the non-linear phase-mixing regime $k \gg 1$, where the role of phase-space has a strongly kinetic character--this regime will be the focus of the remainder of the chapter.  This section gives a heuristic perspective, beginning with a description of the nonlinear phase-mixing process and a phenomenological analysis based upon the work by \cite{schekochihin}.

Following phenomenology, we begin a more formal analysis of the gyrokinetic system.  Section \ref{statistical-sec} introduces the statistical tools and notation that will be needed, followed by a list of the symmetries of the gyrokinetic system.  Symmetries play an important role, and in subsequent arguments we will appeal to the principle of restored symmetry in the fully developed state.  In section \ref{third-order-result-W-sec} we present a derivation of exact third-order statistical results following from the gyrokinetic equation, in the style of \cite{kolmogorov41c} and \cite{yaglom}.

In sections \ref{free-energy-cascade-sec} and \ref{inverse-cascade-sec} we derive scaling laws for various spectra.  To describe these spectra, we must first provide appropriate definitions and identities, tailored to the needs of gyrokinetic turbulence.  Scales in real-space are treated in the conventional way using a Fourier transform while we find an appropriate treatment of velocity-space scales in terms of a zeroth-order Hankel transform, which has a natural compatibility with the mathematical structure of gyrokinetics.  Given these definitions, we can derive approximate spectral scaling laws, for the forward and inverse cascades, in the limit of turbulent scales much smaller than the Larmor radius ($k\rho \gg 1$) and velocity scales much smaller than the thermal velocity ($pv_{th} \gg 1$).

\section{Equations}\label{eqns-sec}

Following is a brief discussion of the equations, their normalisation and applicability, and the Boltzmann response model.  An important point that emerges from this discussion is that the model studied in this chapter will have a wide range of applicability, subject to specific interpretation of the fluctuating fields and normalisation.  Readers wishing to skip these details are advised to take equations \ref{gyro-g} and \ref{qn-g}, with definitions \ref{gyro-avg-def} and \ref{angle-avg-def} as the model equations.  




In kinetic plasma theory, each species 's' (electrons or ions) of a plasma is described by the distribution function $f_s$ which is a function of space, time and velocity.  As is customary with so-called delta-$f$ gyrokinetics, the full distribution function is split into an equilibrium part and fluctuating parts such that the total distribution function is expressed

\begin{equation}
f_s = F_{0s} - \varphi F_{0s} + h_s 
\end{equation}

\noindent where $\varphi$ is the electrostatic potential (the normalisation is explained below).  The equilibrium distribution $F_{0s}$ is a Maxwellian distribution in velocity-space, $F_{0s} = (2\pi)^{-3/2}e^{-v_{\perp}^2/2 - v_{\parallel}^2/2}$.  The second term the so-called Boltzmann part of the distribution.  And the final term is the gyro-centre distribution function, $h_s$, which depends on the perpendicular velocity (magnitude of the velocity perpendicular to the mean magnetic field), the parallel velocity (velocity parallel to the equilibrium magnetic field) and the gyro-centre position ${\bf R}$ which is the spatial position {\bf r} minus the Larmor radius vector $\rhovec$ (defined in normalised units below).  It can be shown by rigourous asymptotic expansion that the gyro-centre distribution function $h_s$ satisfies the nonlinear gyrokinetic equation.  In the normalisation we choose, the gyrokinetic equation is the same for each species.  Thus we drop the species label.  The electrostatic gyrokinetic equation, without variation in the background magnetic field or temperature, is

\begin{equation}
\frac{\partial h}{\partial t} + v_{\parallel} \frac{\partial h}{\partial z} + {\bf v}_E\cdot\bnabla h - \CollisionOp{h} = \frac{\partial\gyroavg{\varphi}}{\partial t} F_0 + \Source
\label{gyro_full}
\end{equation}

\noindent where the gradient is in the gyro-centre coordinate, \ie $\bnabla = \bnabla_{\bf R}$, and $F_0 = (2\pi)^{-3/2}e^{-v_{\perp}^2/2 - v_{\parallel}^2/2}$.  We have also introduced a general kinetic forcing term $\Source$ which will be left unspecified at this stage.  The normalisation is summarised in the following table, with physical quantities having subscript `p':

\vspace{.5cm}
\begin{center}
\begin{tabular}{cccc}
$t = t_{\mbox{\scriptsize{p}}}v_{\mbox{\scriptsize{th}}}/L_n$ & $x = x_{\mbox{\scriptsize{p}}}/\rho$ & $y = y_{\mbox{\scriptsize{p}}}/\rho$ & $z = z_{\mbox{\scriptsize{p}}}/L_n$  \\ 
$v = v_{\mbox{\scriptsize{p}}}/v_{\mbox{\scriptsize{th}}}$ & $\varphi = \varphi_{\mbox{\scriptsize{p}}}\frac{q L_n}{T_0 \rho}$ & $h = h_{\mbox{\scriptsize{p}}}\frac{v_{\mbox{\scriptsize{th}}}^3 L_n}{n_0 \rho}$  & $F_0 = F_{0\mbox{\scriptsize{p}}}v_{\mbox{\scriptsize{th}}}^3/n_0$ 
\end{tabular}
\end{center}
\vspace{.5cm}

\noindent where we have the following definitions: The equilibrium density and temperature of the species of interest are $n_0$ and $T_0$; the mass and charge are $m$ and $q$; the thermal velocity is $v_{th} = \sqrt{T_0/m_s}$; the Larmor radius is $\rho = v_{\mbox{\scriptsize{th}}}/\Omega_{c}$ where the Larmor frequency is $\Omega_{c} = qB/m$.  The equilibrium density scale length is taken to be $L_n = |d\ln n_0/dx|^{-1}$.  The gyro-average $\gyroavg{.}$ is an average over the gyro-angle $\vartheta$ with gyro-centre position {\bf R} held fixed.  The real space and gyro-centre coordinates are related by the transformation ${\bf r} = {\bf R} + \rhovec$ where the Larmor radius vector is $\rhovec(\vartheta) = {\bf {\hat z}}\times{\bf v} = v_{\perp}({\bf {\hat y}}\cos{\vartheta} - {\bf {\hat x}}\sin{\vartheta})$.  In this normalisation, the gyro-average acting on an arbitrary function of position $A({\bf r})$ is defined 

\begin{equation}
\gyroavg{A({\bf r})} = (2\pi)^{-1}\int d\vartheta A({\bf R} + \rhovec(\vartheta))
\label{gyro-avg-def}
\end{equation}

Lastly, the gyro-averaged $E\times B$ velocity in equation \ref{gyro_full} is

\begin{equation}
{\bf v}_E = \hat{z}\times\bnabla\gyroavg{\varphi}
\end{equation}

The full plasma dynamics are described by the evolution of the gyrokinetic equations for all species, with the additional constraint that charge neutrality must be maintained (the quasi-neutrality constraint).  However, due to numerical and analytic difficulty, one does not typically solve the gyrokinetic equations for multiple species simultaneously.  It is common to solve a single gyrokinetic equation, and exploit the separation in scales between the ion and electron Larmor radii to obtain a Boltzmann (or `adiabatic') approximation for the other species.  Thus, conceptually, one can consider two separate regimes, one corresponding to non-trivial ion-dynamics at the ion Larmor radius scale and ion thermal velocity scale (`ion-scale' turbulence) and the other corresponding to dynamics of the electrons at the electron Larmor radius scale and electron thermal velocity (`electron-scale' turbulence).  (These scales are separated in magnitude by roughly the square root of the ion-electron mass ratio.)

For ion-scale turbulence, the electron gyrokinetic equation is dominated by the parallel streaming term which gives the simple equation $v_{\parallel} \partial h_e/\partial z = 0$.  Thus the electron gyro-centre distribution function must be a constant in the magnetic surface (flux surface) defined by the equilibrium.  We will make the assumption that this constant is zero and thus the perturbed electron distribution function is given only by the electron Boltzmann response\footnote{Note that a modified Boltzmann response \cite[see equation 2.69 of][]{dorland-thesis} would be more correct than this pure Boltzmann response.  The modified response is physically more accurate as it accounts for the fact that the parallel mobility of the electron species only allows them to establish a Boltzmann distribution relative to the flux surface to which they are confined.  However, the additional term in the modified response (containing a flux-surface-average of the potential) formally breaks the isotropy assumption but does not otherwise substantially affect the analysis that follows.  Thus we will omit it for simplicity.} (normalised to ion-scales).

\begin{equation}
\delta f_e \approx -\frac{eT_i}{qT_e}\varphi F_0
\end{equation}

As an alternative to the Boltzmann response, one may consider the `no response' model, $\delta f_e \approx 0$.  This may be more appropriate in the case of exact two dimensional turbulence because a small but non-vanishing parallel wavenumber is required to physically establish the Boltzmann electron response.  For electron-scale turbulence, the ion gyrokinetic equation, expanded in the root-mass-ratio becomes simply $\partial_t h_i = 0$.  Thus $h_i$ is a constant which is set to zero.  And the appropriate ion response model is the Boltzmann ion response (normalised to electron-scales):

\begin{equation}
\delta f_i \approx -\frac{qT_e}{eT_i}\varphi F_0
\end{equation}

We will now give the gyrokinetic system, normalised to the scales of the turbulent species.  The subsequent analysis of this chapter will proceed without reference to a specific species or response model.  As a final point before introducing the normalised system, we note that although it may be useful to think in terms of ion-scale turbulence for comparison with the Hasegawa--Mima turbulence, the results may be even more applicable to electron-scale turbulence as the ion Boltzmann response becomes more exact at high wavenumbers whereas the Boltzmann-electron approximation will be broken for high wavenumbers --  that is, the electrons become kinetic as the scale approaches the electron Larmor radius ($k\rho_e \sim 1$).

We take equation \ref{gyro_full} and neglect the parallel ion inertia term $v_{\parallel} \frac{\partial h}{\partial z}$ (see \ref{future-work-sec} for a discussion of this).  The $v_{\parallel}$-dependence is no longer important (except for its appearance in the collisional operator).  Thus it will be concealed when possible in the remainder of this chapter and it should be assumed that wherever velocity integration is present, integration over $v_{\parallel}$ is implied.\footnote{When written explicitly, the full collisional operator is the gyro-averaged linearised Landau operator \cite[see the appendix of][]{schekochihin}, an integro-differential operator with non-trivial dependence on $v_{\parallel}$-space.}  Henceforth, we will notate the perpendicular velocity $v_{\perp}$ as simply $v_{}$.  We express the gyrokinetic equation compactly in terms of the gyro-centre dependent quantity $g = h - \gyroavg{\varphi}F_0$.  Thus the two dimensional gyrokinetic equation we will be using in this chapter is

\begin{equation}
\frac{\partial g}{\partial t} +  {\bf v}_E \cdot\bnabla g = \CollisionOp{h} + \Source
\label{gyro-g}
\end{equation}


\noindent The gyrokinetic system is closed with the quasi-neutrality constraint, which states that the ion charge and electron charge, which are expressed as integrals over the distribution function, must sum to zero :

\begin{equation}
2\pi\int v dv \angleavg{g} = \alpha\varphi - \Gamma_0\varphi
\label{qn-g}
\end{equation}

\noindent where we define the angle average $\angleavg{.}$ as an average over gyro-angle with real-space coordinate ${\bf r}$ held fixed:

\begin{equation}
\angleavg{A({\bf R})} = (2\pi)^{-1}\int d\vartheta A({\bf r} - \rhovec(\vartheta))
\label{angle-avg-def}
\end{equation}

\noindent The operator $\Gamma_0$ is defined $\Gamma_0\varphi = \int v dv \;\e^{-v^2/2}\angleavg{\gyroavg{\varphi}}$.  This operator is multiplicative in Fourier-space--it is simply $\hat{\Gamma}_0(k) = I_0(k^2)e^{-k^2}$, where $I_0$ is the zeroth-order modified Bessel function.  The constant $\alpha$ in equation \ref{qn-g} varies depending on the choice of scales and response model

\begin{equation}
\label{alpha-def}
\alpha =
\begin{cases}
(1 + \frac{eT_i}{qT_e}) & \text{\em ion-scale, Boltzmann-electron response}\\
(1 + \frac{qT_e}{eT_i}) & \text{\em electron-scale, Boltzmann-ion response}\\
1 & \text{\em no-response model}
\end{cases}
\end{equation}

\noindent (Note that the constant temperature ratios are determined by the fact that we normalise the Boltzmann response to the scales of the turbulent species.)  
\noindent Quasi-neutrality in Fourier space is expressed

\begin{equation}
\hat{\varphi} (k) = \beta(k)\int v_{}dv_{} J_0(k v_{}) \hat{g}({\bf k}, v_{})\label{qn-g-k-1}
\end{equation}


\noindent with

\begin{equation}
\beta(k) = \frac{2\pi}{\alpha - \hat{\Gamma}_0(k)} 
\end{equation}

\noindent We will take the system defined by equations \ref{gyro-g} and \ref{qn-g} (or equivalently equation \ref{qn-g-k-1}) to be our gyrokinetic model.  We propose that this system may be thought of as a simple paradigm for kinetic turbulence.  

\section{Collisionless invariants}\label{invariants-sec}

In two dimensional gyrokinetics, $g^2$ is collisionlessly conserved for each value of $v_{}$ individually.  Multiplying the gyrokinetic equation \ref{gyro-g} by $g_{}$ and averaging over gyro-centre position {\bf R} we have

\begin{equation}
\int \frac{d^2{\bf R}}{V}\frac{1}{2}\frac{\partial g_{}^2}{\partial t} =  \int \frac{d^2{\bf R}}{V}g_{}\;\CollisionOp{h} + \varepsilon
\label{gk-invariant-1}
\end{equation}

\noindent  where the injection rate of kinetic free energy is defined

\begin{equation}
\varepsilon = \int \frac{d^2{\bf R}}{V} g \Source
\end{equation}

Multiplying this equation by $F_0^{-1}$ and integrating over velocity-space yields the global balance for what we will call `generalised free energy':

\begin{equation}
\frac{d W_g}{d t} = \int \frac{d^2{\bf r}}{V} 2\pi\int vdv \angleavg{\frac{g_{}\;\CollisionOp{h}}{F_0}} + \bar{\varepsilon}
\end{equation}

\noindent where the generalised free energy is defined

\begin{subequations}
\begin{align}\label{gen-free-energy-0}
W_g &= 2\pi\int v dv\int \frac{d^2{\bf R}}{V} \frac{g^2}{2F_0} \\
&= 2\pi\int v dv \int \frac{d^2{\bf r}}{V}\frac{\angleavg{g^2}}{2F_0}
\end{align}
\end{subequations}

\noindent and the injection rate of free energy is defined

\begin{equation}
\bar{\varepsilon} = \int \frac{d^2{\bf r}}{V} 2\pi\int vdv \angleavg{\frac{g \Source}{F_0}}
\end{equation}

\noindent Of course, $g_{}^2$ multiplied by any function of $v_{}$ is conserved in the absence of collisions.  (In fact any power of $g$ multiplied by any power of $v_{}$ is conserved.)  Beginning in section \ref{statistical-sec}, when we develop the statistical theory of gyrokinetics, we will re-define $g$ to be multiplied by an arbitrary function of $v$ so that the following analysis will be more general.  Here, we take the preceding definition which relates in a straightforward way to the three dimensional invariant of gyrokinetics as we discuss below.

There is a second invariant in two-dimensional gyrokinetics, which is a real-space invariant.  It is found by multiplying the gyrokinetic equation \ref{gyro-g} by $\gyroavg{\varphi}$ and integrating over real-space coordinate {\bf r} and velocity $v$.  The resulting equation is

\begin{equation}
\frac{d E}{dt} = \frac{1}{2}\int \frac{d^2{\bf r}}{V}\varphi\; 2\pi\int vdv \angleavg{\CollisionOp{h}} + \varepsilonE
\end{equation}

\noindent where the invariant $E$ is defined

\begin{equation}
E_{} = \frac{1}{2}\int \frac{d^2{\bf r}}{V}\left[\alpha\varphi^2 - \varphi\Gamma_0\varphi\right]
\end{equation}

\noindent and the injection rate of $E$ is defined

\begin{equation}
\varepsilonE = \frac{1}{2}\int \frac{d^2{\bf r}}{V} \varphi \; 2\pi \int v dv \angleavg{\Source}
\end{equation}

\noindent In this chapter we will take $W_g$ and $E_{}$ to be the fundamental invariants.  Obviously there is freedom to this choice.  Another choice is the three dimensional invariant in gyrokinetics.  That is, the gyrokinetic equation \ref{gyro_full} retaining the term $v_{\parallel}\partial h/\partial z$, has just one invariant, just as three dimensional fluid turbulence has one invariant (energy).  In the electrostatic limit, the single gyrokinetic invariant (in physical units) is the free energy, $W_{} = T_0\delta S$, where the perturbed entropy is $\delta S_{} = -\sum_s\int d^3{\bf v} (\delta f_s)^2/2F_{0s}$ \cite[see][]{krommes-hu, sugama96}.  (In electromagnetic gyrokinetics, \cite{howes} have derived the corresponding invariant which they refer to as `generalised energy.')  We can easily demonstrate the conservation of (normalised) free energy $W_{}$, for the two dimensional electrostatic case by multiplying the gyrokinetic equation \ref{gyro-g} by $h/F_0$ and integrating over velocity and position ${\bf r}$.  The quantity is expressed in terms of $h$ for comparison with \cite{howes}:

\begin{equation}
W_{} = \frac{1}{2}\int \frac{d^2{\bf r}}{V}\left[2\pi\int vdv\frac{\angleavg{h^2}}{F_0} - \alpha\varphi^2\right]
\end{equation}

\noindent Alternatively, free energy can be expressed as a linear combination of $W_g$ and $E_{}$:

\begin{equation}
W_{} = W_g + E_{}
\end{equation}

\noindent where it may be useful to see that $W_g$ written in terms of $h$ is expressed

\begin{equation*}
W_g = \frac{1}{2}\int \frac{d^2{\bf r}}{V}\left[2\pi \int v dv \frac{\angleavg{h^2}}{F_0} + \varphi \Gamma_0 \varphi - 2\alpha\varphi^2\right]
\end{equation*}

\noindent As we will see, $E_{}$ becomes subdominant to $W_g$ at high-$k$ and we will have $W_{} \approx W_g$ in this limit.

\section{Charney--Hasegawa--Mima/Euler turbulence}\label{chm-sec}

In this section we will derive a single equation which corresponds to the Euler equation or the inviscid CHM equation depending on the choice of a parameter, $\lambda$.  The system is still kinetic in that there is a passive cascade of a kinetic invariant.  We will derive the invariants of this system and provide a complete description of the relationship of these invariants to the gyrokinetic invariants derived in the previous section.

At the end of the section, we will briefly show how raising the ordering of the collision operator can produce the viscous CHM and NS equations.  Thus we verify that the kinetic description is needed in the regime of low-collisionality, while in the high-collisionality limit, the fluid equations are a complete description.

We begin by repeating the gyrokinetic system:

\begin{subequations}
\begin{align}\label{gyro-hm}
& \frac{\partial g}{\partial t} +  {\bf v}_E \cdot\bnabla g = \CollisionOp{h} + \Source \\
& 2\pi\int v dv \angleavg{g} = \alpha\varphi - \Gamma_0\varphi
\end{align}
\end{subequations}

\noindent To arrive at the CHM/Euler equation from the system \ref{gyro-hm}, one takes the small ion Larmor radius limit, $k \ll 1$, as well as the small temperature ratio limit such that $(1 - \alpha) \sim {\cal O}(k^2)$.\footnote{The limit of small temperature ratio is somewhat artificial but leads to an elegantly simple fluid equation, the CHM equation.  We note that a more detailed system of fluid equations may be worked out for a temperature ratio of order unity but, for our purposes, will not be substantially different that the familiar CHM equation.}  In this limit, the gyro-average on the $E\times B$ velocity acts as unity.  Also, the gyro-average on the collision operator can be neglected since its corrections are higher order.  We must keep $\Gamma_0$ to second order in $k$ (the zeroth order part gives no contribution upon substitution into the nonlinearity), so we take $\Gamma_0 \approx 1 + \nabla^2$ in the quasi-neutrality constraint.  The gyrokinetic system in this limit is given by

\begin{subequations}\label{gyro-sys-chm}
\begin{gather}
\frac{\partial g}{\partial t} +   {\bf v}_{E0}\cdot\bnabla g = \CollisionOpZ{g} + \Source\label{gyro-chm} \\ 
2\pi\int v_{} dv_{} \angleavg{g} =  \lambda^2\varphi - \nabla^2\varphi \label{qn-chm}
\end{gather}
\end{subequations}

\noindent where ${\bf v}_{E0} = {\bf {\hat z}}\times\bnabla\varphi$ is the $E\times B$ velocity (without gyro-average) and $\lambda^2 = \alpha - 1$.  The scale given by the inverse of this constant, $\lambda^{-1}$, corresponds to the Rossby deformation radius in the atmosphere or the sound Larmor radius, $\rho_s$, in a plasma.  For two dimensional Euler turbulence the parameter is zero, $\lambda = 0$, corresponding to the no-response model given in equation \ref{alpha-def}.  Now, by integrating equation \ref{gyro-chm} over velocity-space (noting that the integral of the collision operator is zero by particle conservation) and substituting quasi-neutrality, equation \ref{qn-chm}, we obtain a single equation for the electrostatic potential.  This equation can be recognised as the inviscid and CHM equation:

\begin{equation}
\partial_t(\lambda^2 - \nabla^2)\varphi + {\bf v}_{E0}\cdot\bnabla (-\nabla^2\varphi) = \SourceCHM
\label{chm-eqn}
\end{equation}

\noindent where $\SourceCHM = 2\pi\int vdv \angleavg{\Source}$.  Again, the two dimensional Euler equation follows from this equation by taking the no-response model, $\lambda = 0$.  Note that there is traditionally a term $\partial_y\varphi$ due to the background density gradient.  Although we have not included this term explicitly in our gyrokinetic system, it may be included in the source term $\SourceCHM$ if needed.

It is well known that there are two invariants of the CHM/Euler equation.  These are referred to as energy and enstrophy, although their physical interpretation depends upon the specific scale of interest.  Energy and enstrophy, are found by multiplying equation \ref{chm-eqn} by $\varphi$ and $\nabla^2\varphi$ respectively and integrating over the system volume.  The invariants are

\begin{subequations}
\begin{align}
& E_{\mbox{\scriptsize{CHM}}} = \frac{1}{2}\int \frac{d^2{\bf r}}{V}(\lambda^2\varphi^2 + |\bnabla\varphi|^2) \\
& Z_{\mbox{\scriptsize{CHM}}} = \frac{1}{2}\int \frac{d^2{\bf r}}{V}(\lambda^2|\bnabla\varphi|^2 + (\nabla^2\varphi)^2)
\end{align}
\end{subequations}

\noindent It should be clear that the two dimensional Euler invariants are recovered by neglecting the terms proportional to $\lambda^2$.  We now show that there is an additional invariant associated with the (equivalent) gyrokinetic description given by equation \ref{gyro-sys-chm}.  This will complete our discussion of the relationship between the CHM (and two dimensional Euler) cascade and the gyrokinetic cascade.  To derive the new invariant we must solve for $g$ in equation \ref{gyro-chm}.  Without loss of generality, we introduce the following {\em ansatz} for $g$:

\begin{equation}
g = F_0 (\lambda^2 - \nabla^2)\varphi + \tilde{g} \label{g-ansatz}
\end{equation}

\noindent It is clear that quasi-neutrality, equation \ref{qn-chm}, implies

\begin{equation*}
\int v dv \angleavg{\tilde{g}} = 0
\end{equation*}

\noindent Furthermore, by multiplying the CHM equation \ref{chm-eqn} by $F_0$ and subtracting it from equation \ref{gyro-chm}, we obtain a separate equation for $\tilde{g}$:

\begin{equation}
\frac{\partial \tilde{g}}{\partial t} + {\bf v}_{E0} \cdot\bnabla \tilde{g} = \CollisionOp{\tilde{g}} + (\Source - F_0\SourceCHM) \label{gyro-twiddle}
\end{equation}

\noindent This implies another collisionless invariant (in addition to the $W_g$ invariant which must obviously still be conserved):

\begin{equation}
W_{\tilde{g}} = \frac{1}{2}\int\frac{d^2{\bf r}}{V}\int v dv \frac{\angleavg{\tilde{g}^2}}{F_0}
\end{equation}

\noindent Expanding the generalised free energy invariant $W_g$ in terms of the solution for $g$ we can express it in terms of the three gyrokinetic--CHM invariants:

\begin{equation}
W_g = W_{\tilde{g}} + \lambda^2 E_{\mbox{\scriptsize{CHM}}} + Z_{\mbox{\scriptsize{CHM}}}\label{invar-reln}
\end{equation}

\noindent each of which are conserved individually in the CHM/Euler limit, as we have shown.

To complete the picture of how the gyrokinetic cascade continues in the CHM/Euler limit, we note that the the gyrokinetic invariant $E_{}$ reduces to the energy invariant with the CHM approximation $\Gamma_0 \approx 1 + \nabla^2$.  

\begin{equation}
E \approx E_{\mbox{\scriptsize{CHM}}}
\end{equation}

\noindent As a last note, notice that in the CHM/Euler limit, the $\tilde{g}$ distribution function, governed by equation \ref{gyro-twiddle}, takes on the role of a hidden passive scalar, being advected by the $E\times B$ flow.



\subsection{Collisional limit, Navier--Stokes and viscous Charney--Hasegawa-Mima equations}

Viscosity does not appear in the preceding fluid limit of the gyrokinetic system.  This is to be expected, as the collisional dissipation is ordered small in gyrokinetics, \ie the collisionality is sufficiently small that a kinetic description of the plasma is required.  To recover the viscous CHM and the two-dimensional Navier--Stokes equation, we must raise the ordering of the collisional operator.  Specifically we will assume that $\CollisionOpZ{g} \sim {\cal O}(k^{-2}\partial g/\partial t)$.  We apply the same ansatz for the solution of $g$ given in equation \ref{g-ansatz}.  This time, the dominant-order equation for $g$ is simply

\begin{equation}
\CollisionOpZ{\tilde{g}} = 0\label{g-tilde-zero}
\end{equation}


\noindent Note that the collision operator acting on the other part of $g$ is zero since it has been defined to be proportional to a Maxwellian.  As the collision operator has been linearised, the meaning of this equation is that collisions between $\tilde{g}$ and the background Maxwellian are negligible.  Consequently, $\tilde{g}$ must be a perturbed Maxwellian, $F_0(n_0 + \Delta n, T_0 + \Delta T) - F_0(n_0, T_0)$.  Because quasi-neutrality implies $\int v dv \tilde{g} = 0$, there is no density perturbation so we have $\tilde{g} = (v^2/2 - 3/2)F_0 \tilde{T}$, where $\tilde{T} = \Delta T/T_0$ and $\tilde{T}$ satisfies a passive scalar equation.

At next order we obtain the CHM/Euler equation by integrating the remnants of the gyrokinetic equation over velocity.  The integral of the collision term is somewhat involved.  Note that we must retain all gyro-average and angle-averages with the collisional term, as the dominant terms will all be zero and the final result is due only to the so-called finite-Larmor-radius ${\cal O}(k^2)$ corrections.  The result is

\begin{subequations}
\begin{align}
& 2\pi \int v dv \angleavg{\CollisionOp{\gyroavg{\varphi}F_0}} = -\nu \nabla^4 \varphi\\
& \nu \equiv -\int v dv d\vartheta \frac{v^2}{4}(1 + \cos^2\vartheta)\CollisionOpZ{v^2(1/2 + \cos^2\vartheta)F_0}
\end{align}
\end{subequations}

\noindent Thus we arrive at the viscid CHM equation

\begin{equation}
\partial_t(\lambda^2 - \nabla^2)\varphi + {\bf v}_{E0}\cdot\bnabla (-\nabla^2\varphi) = -\nu \nabla^4 \varphi + \SourceCHM
\label{chm-eqn-visc}
\end{equation}

\noindent Again, we note that the two-dimensional NS equation is recovered in the no-response limit, by setting $\lambda=0$.

As we have reviewed in this section, the small-$k$ limit of gyrokinetic turbulence imparts a fluid character to the plasma dynamics.  In particular the phenomenon of nonlinear-phase mixing is crucially absent in this limit.  In this chapter we are especially interested in the case where the kinetic behaviour is nonlinear in character, which is corresponds to $k \sim {\cal O}(1)$ and larger.  In the large-$k$ limit, we will show how statistical self-similarity and scaling relations can be derived for the gyrokinetic turbulence cascade.  In the following section we sketch phenomenological arguments corresponding to the low-$k$ CHM limit, and then proceed to the phenomenology of the large-$k$ limit and build on the previous sections to give a general sketch of the full range of the stationary spectra, from CHM to gyrokinetics.

\section{Phenomenology}
\label{phenom-sec}

We introduce the notation by sketching the phenomenology for stationary, driven CHM turbulence.  We define $\varphi_{\ell}$ to be the electrostatic potential at scale spatial scale $\ell$.  We take this to be a fluctuating quantity--\ie it is a random variable with negligible mean.  We notate the averaged quantity as $\bar{\varphi}_{\ell}$ which may be interpreted as the root-mean-square potential at scale $\ell$ or the square root of the structure function.

We assume injection at a scale $k_i$, giving rise to two inertial ranges, the inverse cascade inertial range corresponding to $k \ll k_i$ and the forward cascade inertial range $k \gg k_i$.  From equation \ref{chm-eqn} we have

\begin{equation}
\tau^{-1}_{NL}(\lambda^2\bar{\varphi}_{\ell} - \ell^{-2}\bar{\varphi}_{\ell}) \sim \ell^{-4}\bar{\varphi}^2_{\ell} \label{CHM-NL-phenom}
\end{equation}

\noindent We multiply this equation by $\ell^{-2}\bar{\varphi}_{\ell}$ and assume constancy of enstrophy flux in the forward cascade range--\ie we take the left hand side equal to a constant, the enstrophy injection rate, $\varepsilon_Z \sim \tau^{-1}_{NL}(\ell^{-2}\lambda^2 - \ell^{-4})\bar{\varphi}^2_{\ell}$.  Thus $\ell^{-6}\bar{\varphi}^3_{\ell} \sim \varepsilon_Z$ so that $\bar{\varphi}_{\ell} \sim \varepsilon_Z^{1/3}\ell^2$ and the $\varphi^2$-spectrum is $E_{\varphi} \sim k^{-5}$.  Likewise, multiplying \ref{CHM-NL-phenom} by $\bar{\varphi}_{\ell}$ and assuming constancy of energy flux in the inverse cascade range gives $\bar{\varphi}_{\ell} \sim \varepsilon^{1/3}_E\ell^{4/3}$ or $E_{\varphi} \sim k^{-11/3}$.  

The cascade of $W_{\tilde{g}}$ is passive \cite[see for example][]{lesieur} and is directed forward.  In the forward cascade range we have $\tilde{\varepsilon} \sim \ell^{-2}\bar{\tilde{g}}^2_{\ell}\bar{\varphi}_{\ell}$.  Therefore we have $\bar{\tilde{g}}_{\ell} \sim \ell\tilde{\varepsilon}^{1/2}\bar{\varphi}_{\ell}^{-1/2} \sim \tilde{\varepsilon}^{1/2}\varepsilon_Z^{-1/6}$ and $W_{\tilde{g}}(k) \sim k^{-1}$.

\subsection{The nonlinear phase-mixing range, $k \gg 1$}

\cite{schekochihin} have produced a phenomenological description of the large-$k$ forward cascade in two-dimensional gyrokinetics \cite[see also ][]{schek-ppcf}.  The scaling results obtained by \cite{schekochihin} are in agreement with the results of this chapter.  We reproduce their arguments here, with an expanded notation, and also give the inverse cascade phenomenology.  The phenomenology will motivate the following material in this chapter and highlight the salient physical processes.

\begin{figure}
\begin{center}
\ifthenelse{\boolean{JpgFigs}}
{\includegraphics{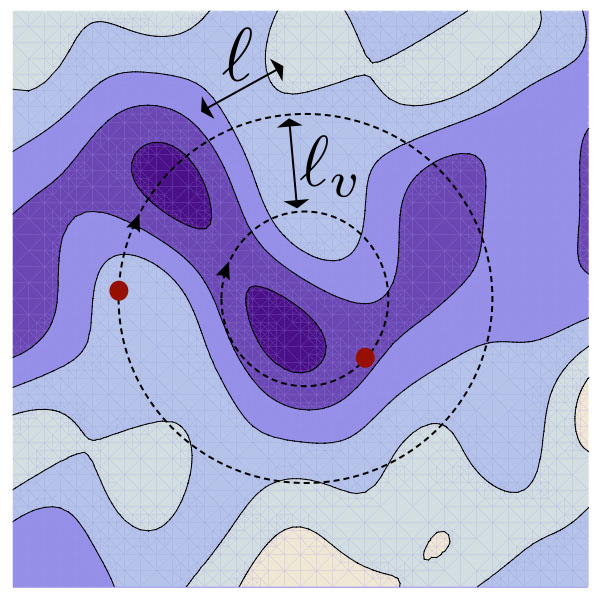}}
{\includegraphics{Figures/vel-correlation.pdf}}
\end{center}
\caption{Correlation in velocity-space}
\label{v-correl-fig}
\end{figure}

In gyrokinetics, ions and electrons are acted upon by fluctuating fields which are averaged along their gyro-orbits.  Two particles sharing the same gyro-centre but having different velocity perpendicular to the equilibrium magnetic field are subject to a different averaged electrostatic potential and thus experience a distinct $E\times B$ drift.  This is illustrated in figure \ref{v-correl-fig}.  One can see from this cartoon that if the $E\times B$ flow has a correlation length $\ell$, then the particle motion must be correlated in velocity-space such that $\ell_v \sim \ell$.

Now let $g_{\ell}$ be the the distribution function $g$ at scale $\ell$.  Again, this quantity is assumed to be random and the scaling we will find will be for the averaged quantity $\bar{g}_{\ell}$.  The central assumption for our treatment of velocity-space dependence is a two-scale dependence.  To be explicit here, we may assume that $g_{\ell}$ may be separated into its fine and smooth-scale dependence as follows

\begin{equation}
g_{\ell} = \bar{g}_{\ell} F(v) \tilde{f}_{\ell}(v)
\end{equation}

where $F(v)$ is a smoothly varying function of velocity which decays strongly at large $v$ so that we may say $\int v^n F^m(v) dv \sim {\cal O}(1)$ for any $m$ and $n$.  We assume $\tilde{f}_{\ell}(v)$ is a random variable which varies on the scale $\ell_v \sim \ell$, as argued above, and has negligible mean and a variance which is of order unity.  To see that $\bar{g}_{\ell}$ functions as the amplitude of $g_{\ell}$ we may calculate the velocity integral of $g_{\ell}^2$, exploiting the scale separation in velocity-space:

\begin{align}
\int v dv g_{\ell}^2 &= \bar{g}_{\ell}^2 \int v dv F^2(v) \tilde{f}_{\ell}^2 \approx \bar{g}_{\ell}^2 \displaystyle{\sum_0^{\infty}} v F^2(v) \int_{v}^{v+1}\tilde{f}_{\ell}^2(v^{\prime}) dv^{\prime}\nonumber\\
&\sim \bar{g}_{\ell}^2 \int v dv F^2(v) \sim \bar{g}_{\ell}^2
\end{align}

\noindent where we have used that $\int_{v}^{v+1}\tilde{f}_{\ell}^2(v^{\prime}) dv^{\prime} \sim 1$.  Now, from quasi-neutrality, \ref{qn-g}, we can determine the scaling of $\bar{\varphi}_{\ell}$ in terms of $\bar{g}_{\ell}$.  Equation \ref{qn-g} gives $\bar{\varphi}_{\ell} \sim \int v dv J_0(v/\ell)g_{\ell}$.  The right hand side is calculated as follows

\begin{align}
\int v dv J_0(v/\ell)g_{\ell} &\sim \ell^{1/2}\int v^{1/2} dv \cos(v/\ell - \pi/4)g_{\ell}\nonumber\\
&\sim \ell^{1/2}\bar{g}_{\ell} \displaystyle{\sum_0^{\infty}} v^{1/2} F(v) \int_{v}^{v+1}\tilde{f}_{\ell}(v^{\prime}) dv^{\prime}\nonumber\\
&\sim \ell\bar{g}_{\ell} \int v^{1/2} dv F(v) \nonumber\\
&\sim \ell \bar{g}_{\ell}
\end{align}

\noindent where we have used the large-argument approximation of the Bessel function $J_0$ on the first line.  Also we argue that the $\tilde{f}$ integral accumulates like a random walk to write $\int_{v}^{v+1}\tilde{f}_{\ell}(v^{\prime}) dv^{\prime} \sim \ell_v^{1/2} \sim \ell^{1/2}$.  So the result is

\begin{equation}
\bar{\varphi}_{\ell} \sim \ell \bar{g}_{\ell}
\label{phi-accum}
\end{equation}

\noindent From the gyrokinetic equation \ref{gyro-g} we have the nonlinear decorrelation time

\begin{equation}
\label{flux-phenom}
\tau_{NL}^{-1} \sim \ell^{-2}\varphi_{\ell} J_0(v/\ell)  \sim \ell^{-3/2}v^{-1/2}\varphi_{\ell}\cos(v/\ell - \pi/4)
\end{equation}

\noindent where we have again used the large argument approximation for $J_0$.  Averaging this expression, and then multiplying by $g_{\ell}^2$ we get

\begin{equation}
\frac{g_{\ell}^2}{\tau_{NL}} \sim \ell^{-3/2}v^{-1/2}\bar{\varphi}_{\ell}g_{\ell}^2
\end{equation}

\noindent Dividing by the background distribution function $F_0$, integrating over velocity and assuming a constant free energy flux gives

\begin{equation}
\ell^{-3/2}\bar{\varphi}_{\ell}\bar{g}_{\ell}^2 \sim \int v dv \frac{g_{\ell}^2}{\tau_{NL}F_0} \sim \bar{\varepsilon}
\end{equation}
 
\noindent Using the result \ref{phi-accum}, we have $\bar{g}_{\ell}^3\ell^{-1/2} \sim \bar{\varepsilon}$ so that we can write $\bar{g}_{\ell} \sim \bar{\varepsilon}^{1/3} \ell^{1/6}$ and $\bar{\varphi}_{\ell} \sim \bar{\varepsilon}^{1/3} \ell^{7/6}$.  The resulting spectra are $W_g(k) \propto k^{-4/3}$ and $E_{\varphi}(k) \propto k^{-10/3}$, where $E_{\varphi}$ is the $\varphi^2$ spectrum.

The inverse cascade range is analysed in the same manner.  We integrate the (phenomenological) gyrokinetic equation over velocity to get

\begin{align}
\frac{\bar{\varphi}_{\ell}}{\tau_{NL}} &\sim \ell^{-2}\bar{\varphi}_{\ell}\int v dv J_0^2(v/\ell) g_{\ell} \sim \ell^{-1}\bar{\varphi}_{\ell}\int dv \cos^2(v/\ell - \pi/4) g_{\ell} \nonumber\\
&\sim \ell^{-1/2} \bar{\varphi}_{\ell}\bar{g}_{\ell} \sim \ell^{-3/2} \bar{\varphi}_{\ell}^2
\end{align}

\noindent where we have again assumed a random walk accumulation to write $\int dv \cos^2(v/\ell - \pi/4) g_{\ell} \sim \ell^{1/2}\bar{g}_{\ell}$.  And assuming constancy of flux of the invariant $E$ in the inverse cascade range, we have $\bar{\varphi}_{\ell}^2/\tau_{NL} \equiv \varepsilon_E \sim \ell^{-3/2} \bar{\varphi}_{\ell}^3$.  Thus $\bar{\varphi}_{\ell} \sim \varepsilon_E^{1/3}\ell^{1/2}$ and $\bar{g}_{\ell} \sim \varepsilon_E^{1/3}\ell^{-1/2}$ and the spectra are $E(k) \sim E_{\varphi}(k) \sim k^{-2}$ and $W_g(k) \sim k^0$.

\subsection{The cascade through multiple scales}

Using the phenomenological scalings from this section, we may give a sketch of the fully developed cascade from CHM/Euler limit to the high-$k$ gyrokinetic limit.  As there are multiple special scales and invariant quantities, there are several possible scenarios.  For simplicity we consider just two scenarios given by figures \ref{large-inject-fig} and \ref{small-inject-fig}.  Equation \ref{invar-reln} states that the gyrokinetic invariant $W_g$ is composed of three separately conserved quantities in the CHM limit: $W_g = W_{\tilde{g}} + \lambda^2 E + Z$.  We use this relation to trace out the fate of $W_g$ in the long wavelength limit.


\subsubsection{Injection at large scales}
  
\begin{figure}
\begin{center}
\ifthenelse{\boolean{JpgFigs}}
{\includegraphics[width=\columnwidth]{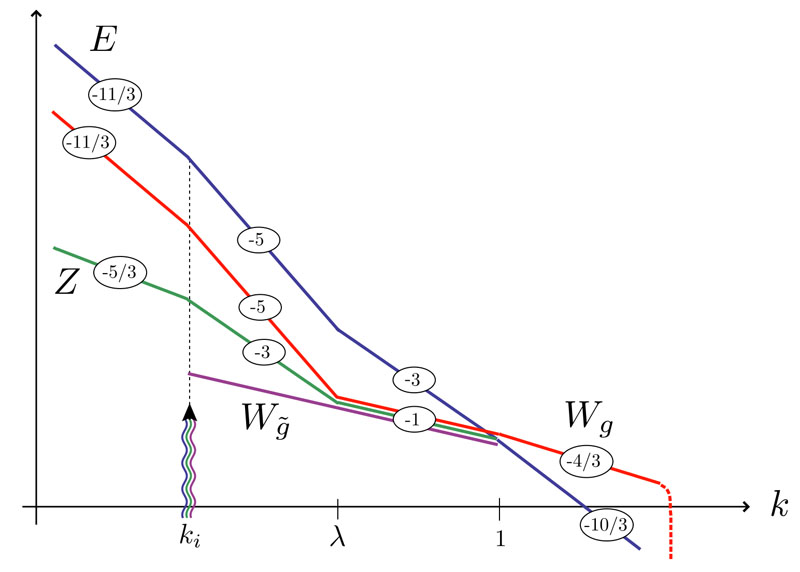}}
{\includegraphics[width=\columnwidth]{Figures/injection-large-scale.pdf}}
\end{center}
\caption{Scenario 1: Injection at large scales}
\label{large-inject-fig}
\end{figure}

In figure \ref{large-inject-fig} injection is at large scales, in the so-called potential limit of the CHM equation, where $k_i \ll \lambda$ and the invariant $E$ reduces to $E \approx \lambda^2 E_{\varphi}$.  In the stationary state, the injection of $E$ fuels the inverse cascade and the injection of $Z$ and $W_{\tilde{g}}$ feed into forward cascades which ultimately drive the cascade of $W_g$ at the Larmor radius scale $k \sim 1$.  In the so-called ``kinetic limit'' of the CHM equation, $\lambda \ll k \ll 1$, the contribution of $E$ to $W_g$ is negligible and $W_g = W_{\tilde{g}} + Z$.  In the potential limit, $k \ll \lambda$, the steep spectrum of $E$ is the dominate part of $W_g$, \ie $W_g \approx \lambda^2 E$.  This leads to the unexpected result that there must be a continual accumulation of the forward cascaded quantity $W_g$ at large scales, in the fully developed state.

\subsubsection{Injection at small scales}

\begin{figure}
\begin{center}
\ifthenelse{\boolean{JpgFigs}}
{\includegraphics[width=\columnwidth]{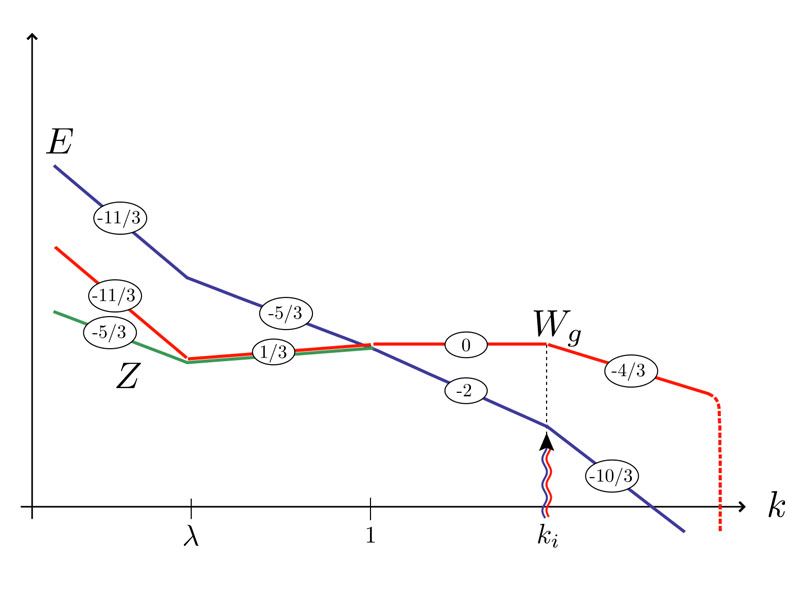}}
{\includegraphics[width=\columnwidth]{Figures/injection-small-scale.pdf}}
\end{center}
\caption{Scenario 2: Injection at small scales}
\label{small-inject-fig}
\end{figure}

In figure \ref{small-inject-fig} injection is at small scales, in the nonlinear phase-mixing regime, $k \gg 1$.  The free energy $W_g$ cascades forward to the dissipation scale whereas the second invariant $E$ cascades inversely.  Again, the identity of $W_g$ changes in the long wavelength limits.  In the so called kinetic limit of the CHM equation, $\lambda \ll k \ll 1$, we have $W_g \approx Z$ since there is no source of $W_{\tilde{g}}$ in this scenario.  In the potential limit $k \ll \lambda$, we have $W_g \approx \lambda^2E$ and the $W_g$ takes on the steeper spectrum of $E$.

\section{Statistical theory}\label{statistical-sec}

In this section we will analyse the gyrokinetic system, equations \ref{gyro-g} and \ref{qn-g} using familiar statistical methods from fluid turbulence \cite[we adopt somewhat the methods, terminology and notation of ][]{frisch}.  First we list the symmetries of gyrokinetic system and then formulate the statistical assumptions which follow.  Following this we derive the two exact third-order statistical relations which determine self-similarity in the inertial ranges for the forward and inverse cascades.

\subsection{Symmetries of the gyrokinetic system}

Statistical arguments in turbulence theory are anchored to the underlying symmetries of the dynamical equations of interest.  Although specific boundary conditions, initial conditions and forcing mechanisms break these symmetries at large scale, a guiding principle of fluid turbulence is that at sufficiently small scales, far from boundaries, a fully developed turbulence will emerge where the symmetries of the dynamical equations are restored in a statistical sense.  We will appeal to this principle in the arguments of this chapter.

The following table is a summary of symmetries of the gyrokinetic system:

\begin{subequations}
\begin{align}
&\mbox{\em Position-translation:} & ({\bf r}) \rightarrow ({\bf r} + {\boldsymbol \sigma})\\
&\mbox{\em Time-translation:} &(t) \rightarrow (t + \tau) \\
&\mbox{\em Constant potential shift:} &(g,\; \varphi) \rightarrow (g + (\alpha - 1)\psi F_0,\; \varphi + \psi) \label{shift-invariance}\\
&\mbox{\em Scaling transformations:} &(g,\; \varphi,\; {\bf r},\; v,\; t) \rightarrow (\gamma^{-1}g,\; \mu^2\gamma^{-1}\varphi,\; \mu {\bf r},\; \mu v,\; \gamma t) \label{scaling-invariance}\\
&\mbox{\em Rotations:} &({\bf r}) \rightarrow (A{\bf r});\;\;\; A \in SO(2) \\
&\mbox{\em Parity:} &(g,\; \varphi,\; {\bf r}) \rightarrow (-g,\; -\varphi,\; -{\bf r})
\end{align}
\end{subequations}

Translational invariance in time and position-space are obviously the basis for the assumptions of homogeneity and stationarity respectively.  The invariance of gyrokinetics to shift by an arbitrary constant potential is somewhat trivial since a constant potential produces no electric field and thus has no effect on dynamics.  (Note that for the transformation \ref{shift-invariance}, the shift in $g$ is Maxwellian so as to vanish under the collision operator.)  The scaling symmetries are described by the transformation \ref{scaling-invariance} and parameterised by two scaling factors, $\mu$ and $\gamma$.  This symmetry will be crucial in determining the scaling in the velocity-space increment.  In particular, for $\gamma = 1$, the scaling invariance \ref{scaling-invariance} implies a dual-scaling in velocity and position-space as described in section \ref{dual-self-sim-sec}.  Note that the scaling invariance applies to the collisionless limit.

\subsection{The ensemble average}

We will denote the ensemble average with an over-bar so that the correlation function of $g$ between two points in phase space, $({\bf r},\; v)$ and $({\bf r}^{\prime},\; v^{\prime})$, is defined

\begin{equation}
G =  \ensbl{ g({\bf R}, v_{}) g({\bf R}^{\prime}, v_{}^{\prime}) } = \ensbl{gg^{\prime}}
\end{equation}

\noindent Homogeneity in position-space (or equivalently, in gyro-centre space) follows from the translational invariance of the gyrokinetic equation and implies that $G$ is a function only of the increment $\ell = {\bf R}^{\prime} - {\bf R}$.  In contrast, velocity-space is fundamentally inhomogeneous as there is obviously no translational invariance in $v$-space.  (Velocity dependence in the Maxwellian and the gyro-average each break translational invariance.)  However, we expect a local homogeneity to apply to velocity-space for small enough velocity increment $\ell_v = v^{\prime} - v$.  In other words, the two-point statistical quantities such as structure and correlation functions should have a two-scale dependence in velocity-space, varying strongly with the increment $\ell_v$ but weakly (smoothly) with the centred velocity $\bar{v} = (v^{\prime} + v)/2$.

The remainder of this section will be devoted to deriving exact statistical relations in analogy to Kolmogorov's `four-fifths' law (or the `three-halves' law for two dimensional NS turbulence).  As there are two conserved quantities, there will be two third-order statistical relations.  We begin with the relation implied by the conservation of $W_g$ and finish with that for $E_{}$.

\subsection{third-order Kolmogorov relations}\label{third-order-result-W-sec}

Let's redefine $g$ to be normalised by an arbitrary mean-scale function of velocity-space, $\kappa(v_{})$.  This will allow us to develop a theory for arbitrary velocity-integrated moments of the invariant $g^2$.  (We will fix $\kappa$ at a specific point in the analysis, later.  This is done for the sake of clarity in presentation and we will demonstrate that the results are valid for general $\kappa$)  The definition of $g$ we now adopt is

\begin{equation}
g = (h - F_0\gyroavg{\varphi})/\kappa(v_{})\label{renormalized-g}
\end{equation}

\noindent The generalised free energy, originally defined by equation \ref{gen-free-energy-0}, is also redefined.  Using the ensemble average instead of explicit spatial averaging, we define

\begin{equation}
W_g = 2\pi\int_0^{\infty} v_{}dv_{}\frac{\ensbl{g^2}}{2} 
\label{gen-free-energy}
\end{equation}

\noindent which for $\kappa = F_0^{1/2}$ corresponds to the definition given by equation \ref{gen-free-energy-0} and we will see that this instance of $W_g$ corresponds to the free energy $W_{}$ in the large-$k$ limit.

From the gyrokinetic equation we derive the budget equation at scales $\ell$ and $\ell_v$

\begin{equation}
  \label{budget_real2}
\frac{\partial G}{\partial t} - \frac{1}{2}\bnabla_{\ell_{}}\cdot\Sthree= \ensbl{g\CollisionOpP{h^{\prime}}/\kappa^{\prime} + g^{\prime}\CollisionOp{h}/\kappa} + \ensbl{g\Source^{\prime}/\kappa^{\prime}} + \ensbl{g^{\prime}\Source/\kappa}
\end{equation}




\noindent where we have defined the third order structure function $\Sthree \equiv \ensbl{\delta{\bf v}_E\delta g^2}$ with $\delta g = g^{\prime} - g$ and $\delta{\bf v}_E = {\bf v}_E^{\prime} - {\bf v}_E$.  Also, we have used homogeneity and incompressibility of ${\bf v}_E$ to manipulate the nonlinear flux to obtain $-\ensbl{g^{\prime}{\bf v}_E\cdot\bnabla g} + \ensbl{g{\bf v}_E^{\prime}\cdot\bnabla^{\prime} g^{\prime}} = \bnabla_{\ell_{}}\cdot\Sthree/2$


Now if we take $\ellvec_{} = 0$, $\ell_v = 0$, the nonlinear term is zero by homogeneity.  The result (equivalent to equation \ref{gk-invariant-1}) is the `global budget' equation:

\begin{equation}
\frac{\partial\ensbl{g^2}}{\partial t} = 2\ensbl{g\CollisionOp{h}/\kappa} + 2\ensbl{g\Source}/\kappa
\label{budget_global_t}
\end{equation}

\noindent assuming stationarity, the left hand side of the equation is zero and we have the steady state balance between forcing and dissipation

\begin{equation}
\ensbl{g\Source}/\kappa = -\ensbl{g\CollisionOp{h}/\kappa} = \varepsilon(v_{})
\label{budget_global}
\end{equation}

\noindent where we have defined the kinetic free energy injection rate $\varepsilon(v_{})$.  To derive the `inertial range' law for $\Sthree$ we now assume that the forcing is restricted to a narrow range around a injection scale which we denote \ellF.  For simplicity we also assume that the only special velocity scale of the problem is the thermal velocity $v = 1$ and we therefore assume that the forcing acts on this scale in velocity-space.  For $\ell \ll \ellF$ and $\ell_v \ll 1$, the forcing term in the budget equation \ref{budget_real2} reduces to

\begin{equation}
\ensbl{g\Source^{\prime}}/\kappa^{\prime} + \ensbl{g^{\prime}\Source}/\kappa \approx 2\ensbl{g\Source}/\kappa = 2\varepsilon
\end{equation}

\noindent Now we take the limit of small collisionality.  That is, since the collision operator is proportional to a normalised collision frequency $\nu_{\star}$ (which is the collision frequency of the turbulent species divided by the diamagnetic frequency $v_{\mbox{\scriptsize{th}}}/L_n$) we are free to take this parameter as small as we like so that for fixed $\ell$ and $\ell_v$, the collisional term is negligible in the budget equation \ref{budget_real2}.  Thus by assuming small collisionality, stationarity $\ell_v \ll 1$ and $\ell \ll \ellF$ we have from equation \ref{budget_real2}

\begin{equation}
\bnabla_{\ell_{}}\cdot\Sthree = -4\varepsilon(v_{})
\label{four-fifths}
\end{equation}

\noindent where we have written the injection rate $\varepsilon$ as an explicit function of $v_{}$.  Using isotropy, it follows that

\begin{equation}
  \label{four-fifths-iso}
\Sthree = -2\hat{\ellvec_{}}(\varepsilon \ell + \mbox{const.})
\end{equation}

\noindent Note that the constant of integration is included here, which in general depends on $v$ and $\ell_v$.  In fluid turbulence, this constant is zero since the structure function in that case depends only on $\ell$, so must be zero when $\ell_{} = 0$.  However, with non-zero $\ell_v$ we must allow for possibility that $\Sthree$ tends to a constant as $\ell_{} \rightarrow 0$.  Indeed, we have not yet determined the $\ell_v$-dependence of $\Sthree$ (\ie $\varepsilon$ does not depend on $\ell_v$) and we will use the additional term to account for this dependence in section \ref{dual-self-sim-sec}.






\subsection{The $E_{}$ third-order Kolmogorov relation}

To derive the scale-by-scale budget equation for the invariant $E_{}$ we proceed analogously to the analysis leading to equation \ref{four-fifths-iso}.  However in this case, we are dealing with functions of real-space variable ${\bf r}$ and its increment $\ellvec_{r} = {\bf r}^{\prime} - {\bf r}$.  Also, velocity will appear as a dummy variable of integration, so it will simplify our notation to refer to a single variable $v$ in what follows--\ie ${\bf R} = {\bf r} - \rhovec(v)$, ${\bf R}^{\prime} = {\bf r}^{\prime} - \rhovec(v)$ and $g^{\prime} = g({\bf R}^{\prime}, v)$, \etc , from which it follows that $\ellvec_{} = \ellvec_{r}$ and we will refer to a single increment in position-space, $\ellvec$.  As our interest has shifted away from velocity dependence, let's also take the simplification $\kappa = 1$ in our definition of $g$, equation \ref{renormalized-g}.

We take the gyrokinetic equation \ref{gyro-g} for $g$ and $g^{\prime}$ and multiply by $\varphi^{\prime}$ and $\varphi$ respectively, then ensemble average.  Next we integrate over velocity $v_{}$.  The result is

\begin{multline}
(\alpha - \Gamma_0)\frac{\partial \Phi}{\partial t} - \bnabla_{\ell_{}}\cdot\SthreeE = \\ 2\pi \; \ensbl{\varphi^{\prime}\int v dv\angleavg{\CollisionOp{h}}} + 2\pi \; \ensbl{\varphi \int v dv \angleavgP{\CollisionOpP{h^{\prime}}}} + \ensbl{\varphi\SourceE^{\prime}} + \ensbl{\varphi^{\prime}\SourceE}
\label{budget-E}
\end{multline}

\noindent where we have defined $\Phi = \ensbl{\varphi \varphi^{\prime}}$ and $\SthreeE =  2\pi \int v dv \; \ensbl{\delta\gyroavg{\varphi} \delta {\bf v}_E\delta g }$\footnote{An alternate equivalent expression for \SthreeE is $\SthreeE =  \ensbl{\delta\varphi \; 2\pi \int v dv \angleavgRR{\delta {\bf v}_E\delta g }}$ where $\delta \varphi = \varphi^{\prime} - \varphi$ and the angle average with respect to ${\bf r}$ and ${\bf r}^{\prime}$ is defined $\angleavgRR{f({\bf R}^{\prime},{\bf R})} = (2\pi)^{-1}\int d\vartheta f({\bf r}^{\prime} - \rhovec(\vartheta), {\bf r} - \rhovec(\vartheta))$.} (with $\delta \gyroavg{\varphi} = \gyroavgP{\varphi^{\prime}} - \gyroavg{\varphi}$) which by homogeneity depends only on the increment $\ell_{}$.  Also by homogeneity, we have $\Gamma_0 \Phi = \Gamma_0^{\prime} \Phi = \Gamma_0^{(\ell_{})}\Phi$ where $\Gamma_0^{(\ell_{})}$ operates in $\ell_{}$-space.  Also, we have defined the $E$-forcing $\SourceE = 2\pi \int vdv \angleavg{\Source}$.  We can express the first term of equation \ref{budget-E} in terms of the rate of change of the invariant $E$ and structure functions of $\varphi$.  Using homogeneity we find the following relation

\begin{equation}
(\alpha - \Gamma_0)\Phi = 2E - \frac{1}{2}\left(\ensbl{\delta \varphi^2} - \ensbl{\delta \varphi \delta(\Gamma_0\varphi)}\right)
\end{equation}

\noindent where $\delta(\Gamma_0 \varphi) = \Gamma_0^{\prime}\varphi^{\prime} - \Gamma_0\varphi$.  We substitute this into equation \ref{budget-E}.  The structure function terms are zero in the stationary limit.  With the additional assumption of small collisionality, equation \ref{budget-E} becomes

\begin{equation}
\frac{\partial E}{\partial t} = \frac{1}{2}\bnabla_{\ell_{}}\cdot\SthreeE + \frac{1}{2}\ensbl{\varphi\SourceE^{\prime}} + \frac{1}{2}\ensbl{\varphi^{\prime}\SourceE}\label{budget-E-stationary}
\end{equation}

\noindent For $\ell = 0$ we have the global balance of $E$ which describes the continual accumulation due to forcing:

\begin{equation}
\frac{\partial E}{\partial t} = \ensbl{\varphi\SourceE}\label{global-budget-E} \equiv \varepsilonE
\end{equation}

\noindent where $\varepsilonE$ is the injection rate of $E$.  Now recall that the forcing is assumed to be restricted to a scale $\ellF$ as explained in the previous section on the forward cascade.  For the inverse cascade, we are considering an inertial range such that $\ell \gg \ellF$.  In this limit, the forcing term in equation \ref{budget-E-stationary} is zero and we may write the third order inertial range result for the inverse cascade of $E$:

\begin{equation}
\bnabla_{\ell_{}}\cdot \SthreeE = 2\varepsilonE
\label{four-fifths-w}
\end{equation}

\noindent Using isotropy, we may also write

\begin{equation}
\SthreeE = \varepsilonE\ellvec_{}
\label{four-fifths-w-iso}
\end{equation}

\subsection{The CHM/Euler limit}

\noindent In the CHM or Euler limit (the result is applicable to both Euler and CHM since the electron response does not enter in the nonlinearity), equation \ref{four-fifths-w} reduces to 

\begin{equation}
\bnabla_{\ell_{}}\cdot\ensbl{ \delta\varphi \; \delta{\bf v}_{E0}\; \delta(\nabla^2\varphi)} = -2\varepsilonE
\label{CHM-third-order}
\end{equation}


\noindent where we recall that ${\bf v}_{E0} = {\bf {\hat z}}\times\bnabla\varphi$ is the $E\times B$ velocity without gyro-average.  With some algebra, equation \ref{CHM-third-order} may be manipulated to find agreement with the result from \cite{boffetta} for CHM turbulence and the result from \cite{bernard} for two dimensional NS turbulence:

\begin{equation}
\bnabla_{\ell_{}}\cdot\ensbl{\delta {\bf u} |\delta {\bf u}|^2} = 4\varepsilonE
\end{equation}

\noindent where ${\bf u} \equiv -{\bf v}_{E0}$.  

The two gyrokinetic results of this section, equation \ref{four-fifths} and equation \ref{four-fifths-w}, hold for arbitrary spatial scales--\ie they hold above, below and at the Larmor radius scale.  (Note, however, that for equation \ref{four-fifths}, we have assumed the velocity scale satisfies $\ell_v \ll 1$.)  In the the following sections, to derive approximate power-law spectra for the conserved quantities, we will consider the limit $k \gg 1$ and $p \gg 1$.   This limit corresponds to the regime where non-linear phase mixing is dominant and gyrokinetics exhibits the phase-space cascade.

\section{The gyrokinetic turbulence cascade}

\label{free-energy-cascade-sec}

\subsection{Self-similarity hypothesis}

\label{dual-self-sim-sec}

For the forward cascade of free energy, scaling with respect to $\ell$ is described by the property of $\Sthree$ given by equation \ref{four-fifths-iso}.  As we have pointed out, self-similarity in $\ell_v$ is not predicted by the inertial range free energy-flux equation \ref{four-fifths-iso}.  Thus we will supplement equation \ref{four-fifths-iso} with an hypothesis as follows.  Given fixed scales $\ell$ and $\ell_v$, the difference $\delta g$ satisfies the {\it dual self-similarity hypothesis}

\begin{equation}
\delta g(\lambda \ell_{}, \lambda \ell_{v}) \doteq \lambda^{h_g}\delta g(\ell_{}, \ell_{v})
\label{dual-hypothesis}
\end{equation}

\noindent by which we mean that $\delta g$ exhibits this scaling statistically--\ie where it appears in an ensemble average.  For instance it implies $\Stwo(\lambda \ell, \lambda \ell_v) /\Stwo(\ell, \ell_v) = \lambda^{2h_g}$.  The hypothesis may be justified from two perspectives.  The first is the argument of decorrelation in velocity-space as has been described by \cite{schek-ppcf} and reiterated in section \ref{phenom-sec}.  A second point of view is based upon the fact that the collisionless gyrokinetic system is invariant to simultaneous scaling of ${\bf R}$ and $v$ by the same factor (this is the symmetry \ref{scaling-invariance} with $\gamma = 1$).  Thus, if we consider velocity increments $\ell_v \ll 1$ (that is, smaller than the thermal velocity, the characteristic scale of the Maxwellian), the statistics of the fluctuations should depend weakly on the centred velocity $\bar{v}$ and obey a dual scaling in $\ell$ and $\ell_v$--\ie equation \ref{dual-hypothesis}.  As we will prove later, the scaling for $\delta g$ implies a self-similarity for the gyro-averaged $E \times B$ velocity difference $\delta {\bf v}_E$, for small $\ell_{}$ and $\ell_v$:

\begin{equation}
\delta {\bf v}_E(\lambda \ell, \lambda \ell_v) \doteq \lambda^{\hE}\delta {\bf v}_E(\ell, \ell_v)
\label{stwoe-scaling}
\end{equation}

\noindent As a consequence of these scaling relations and equation \ref{four-fifths-iso} we see that the unknown constant in the expression for $\Sthree = \ensbl{\delta {\bf v}_E\delta g^2}$ must be linear in $\ell_v$ (if it is non-zero).  Thus we write

\begin{equation}
  \label{four-fifths-iso-hyp}
\Sthree = -2\hat{\ellvec_{}}(\varepsilon \ell + \varepsilon^{\prime} |\ell_v|)
\end{equation}

\noindent where $\varepsilon^{\prime}$ is an unknown constant.  The absolute value on $\ell_v$ is required, due to the fact that under the transformation $(\ell,\; \ell_v) \rightarrow (-\ell,\; -\ell_v)$ we must have $\Sthree \rightarrow -\Sthree$.  Finally, equation \ref{four-fifths-iso-hyp} implies that 

\begin{equation}
2h_g + \hE = 1
\label{hg-he}
\end{equation}

\subsection{The kinematics of gyrokinetic turbulence}

\label{spectral_density}

Before investigating spectral scaling, we must define the free energy spectral density.  In this section we introduce a way of characterising velocity scales using a zeroth-order Hankel transform.  The Hankel transform, being a two-dimensional Fourier transform with rotational symmetry, is also encountered in the theory of two-dimensional fluid turbulence.  The difference is that in gyrokinetics, dynamics are exactly independent of the angle of the perpendicular velocity whereas in isotropic fluid turbulence, only statistically averaged quantities are independent of orientation in space.  

We now introduce some of the notation and identities that will be needed.  Given a function $g(v)$, its Hankel transform is

\begin{equation}
\hat{g}(p) = \int_0^{\infty} vdvJ_0(pv)g(v) 
\end{equation}

\noindent Inversion of the transformation is easily achieved using the orthogonality of the zeroth-order Bessel functions, given by

\begin{equation}
\int_0^{\infty} J_0(p_1 x) J_0(p_2 x) x dx = \delta(p_1 - p_2)/p_1 
\end{equation}

\noindent Using this identity it is easy to see that the generalised Parseval's theorem for two functions $f$ and $g$ is

\begin{equation}
\int_0^{\infty} v dv f(v) g(v) = \int_0^{\infty} p dp \hat{f}(p)\hat{g}(p)  
\end{equation}



\noindent An integral involving three Bessel functions is encountered in deriving the mode coupling due to the nonlinearity.  This integral is

\begin{equation}
\int_0^{\infty} v dv J_0(p_1 v)J_0(p_2 v)J_0(p_3 v) = K(p_1,p_2, p_3) \label{three-bes-int}
\end{equation}

\noindent where if $p_1$, $p_2$ and $p_3$ form the sides of a triangle, then $K = 1/2\pi\Delta$ where $\Delta$ is the area of that triangle; if $p_1$, $p_2$ and $p_3$ do not form the sides of a triangle, then $K = 0$.  Applying this identity, we can obtain the mode coupling equation for Hankel-Fourier modes.  Neglecting collisions, the gyrokinetic equation, equation \ref{gyro-g}, in Hankel-Fourier space is

\begin{multline}
\frac{\partial \hat{g}({\bf k}, p)}{\partial t} = \\
 \frac{1}{2\pi}\int d^2{\bf k}^{\prime} \beta(k^{\prime})\; ({\bf k}-{\bf k}^{\prime})\cdot({\bf {\hat z}}\times{\bf k}^{\prime}) \int qdq \; K(k^{\prime}, p, q) \; \hat{g}({\bf k}^{\prime}, k^{\prime}) \hat{g}({\bf k}-{\bf k}^{\prime}, q)
\label{coupling}
\end{multline}

\subsection{The forward cascade of free energy $W_g$}

Returning to the problem at hand, the Hankel-Fourier transform of the distribution function $g$ is

\begin{equation}
\hat{g}({\bf k}, p) \equiv \frac{1}{2\pi}\int_{\field{R}}d^2{\bf R}\int_0^{\infty}v_{}dv_{}J_0(pv_{})\e^{-\dot{\imath}{\bf k}\cdot{\bf R}}g({\bf R},v_{})
\end{equation}

\noindent Now we transform and inverse-transform $g$ and use homogeneity to express the conserved quantity $\ensbl{g^2}$ in terms of the correlation function $G(\ellvec_{},v_{}^{\prime}, v_{})$:

\begin{eqnarray}
\ensbl{g^2}& = \int k dkpdp\ell_{} d\ell_{} v_{}^{\prime}dv_{}^{\prime}J_0(k\ell_{})J_0(pv_{}^{\prime})J_0(pv_{})G(\ell_{} ,v_{}^{\prime},v_{})
\label{spectral-manip}
\end{eqnarray}

\noindent where we have used the identities $\int J_0(ax)J_0(bx)xdx = \delta (b - a)/b$ and $\int d^2{\bf k} \e^{\dot{\imath}{\bf k}\cdot{\bf r}} = (2\pi)^2\delta ({\bf r})$.  Also, isotropy has been used to reduce the Fourier-transform in vector position increment $\ellvec$ to an Hankel transform in scalar increment $\ell$.  Now recall the mean (ensemble-averaged) generalised free energy is defined

\begin{equation}
W_g = 2\pi\int v_{}dv_{}\frac{\ensbl{g^2}}{2}
\end{equation}

\noindent From equation \ref{spectral-manip} it follows that 





\begin{equation}
W_g = \int dkdp\; W_g( k,p)
\end{equation}
 
\noindent where

\begin{equation}
W_g(k,p)  \equiv \pi pk\int \ell_{}d\ell_{}v_{}dv_{}v_{}^{\prime}dv_{}^{\prime}J_0(k\ell_{})J_0(pv_{}^{\prime})J_0(pv_{})G(\ell_{},v_{}^{\prime},v_{})
\label{sg-def}
\end{equation}

\noindent The spectral density can be expressed in terms of the transform function:  

\begin{eqnarray}
W_g(k, p)  = \pi\;pk\;\ensbl{\hat{g}(k,p)^2}
\end{eqnarray}

\noindent Notice that equation \ref{sg-def} is not invertible as with the case of the Weiner-Khinchin formula found in fluid turbulence--\ie the correlation function $G(\ell_{},v_{}^{\prime},v_{})$ cannot be recovered from $W_g(k,p)$.  Because of the lack of homogeneity in $v_{}$, we have integrated over two dimensions $v_{}$ and $v_{}^{\prime}$ to define scales in velocity-space via a single wavenumber p.

\subsection{Spectral scaling laws}

Now to establish the relationship between the scaling of the spectrum $W_g(k,p)$ and that of the structure function $\Stwo = \ensbl{\delta g^2}$ at small scales, we will first change velocity variables to the increment variable $\ell_v = v_{}^{\prime} - v_{}$ and the centred velocity $\bar{v} = (v_{} + v_{}^{\prime})/2$.

\begin{equation}
W_g(k,p) = \pi pk\int \ell_{} d\ell_{}(4\bar{v}^2 -  \ell_v^2)d\ell_v d\bar{v} J_0(k\ell_{})Q(p,\ell_v,\bar{v})G(\ell_{}, \ell_v, \bar{v})\label{sg-spectrum-def}
\end{equation}

\noindent where

\begin{equation}
Q(p, \ell_v, \bar{v}) = J_0[p(2\bar{v} + \ell_v )/2]J_0[p(2\bar{v} - \ell_v)/2]
\end{equation}

\subsection{Scales smaller than the Larmor radius: $k \gg 1$}

Until now, we have not made any assumption about the size of $k$ or $p$.  Now we now consider the phase-mixing range $k \gg 1$ and $p \gg 1$.  First, we make use the fact that the correlation function $G(\ell_v)$ must be peaked around small $\ell_v$ to approximate that the integral is dominated by $\ell_v \ll \bar{v}$.  We can also take $p\bar{v} \gg 1$ and use the large-argument approximation of the Bessel function.  In this limit, $Q(p, \ell_v, \bar{v})$ can be expressed

\begin{equation}
Q(p, \ell_v, \bar{v}) \approx \frac{1}{4\pi p \bar{v}}\left(\sin(2 p \bar{v}) + \cos(p \ell_v)\right)
\end{equation}

\noindent Substituting this expression into equation \ref{sg-spectrum-def} we see that the term proportional to $\sin(2 p \bar{v})$ oscillates rapidly and does not contribute in comparison, so we neglect this.  Using $G = (\ensbl{g^2} + \ensbl{(g^{\prime})^2} - \Stwo)/2$ we can express $W_g(k,p)$ in terms of $\Stwo$ for non-zero $k$ and $p$:

\begin{equation}
W_g(k,p) \approx -\frac{k}{2}  \int \ell_{} d\ell_{}\; \bar{v} d\bar{v} \; d\ell_v J_0(k\ell_{})\cos(p\ell_v)\Stwo
\label{sg-def-approx}
\end{equation}

\noindent Then self-similarity of $\delta g$ implies that $W_g(k, p)$ must satisfy the scaling

\begin{equation}
W_g(\lambda k, \lambda p) = \lambda^{\nu}W_g(k, p)
\label{sg-dual-scaling}
\end{equation}

\noindent with

\begin{equation}
\nu = -2 - 2h_g
\label{nu-hg}
\end{equation}

\noindent (Note that to arrive at this scaling, we have assumed that the bounds of the integration of $\ell_v$ in equation \ref{sg-def-approx} are not important since we are concerned with the small $\ell_v$ behaviour of the integral.)  To determine the scaling index $\nu$ we will now find its relationship to $\hE$, the scaling index of the gyro-averaged $E \times B$ velocity structure function $\delta {\bf v}_E$.  We will write $\StwoE = \ensbl{|\delta{\bf v}_E|^2}$ in terms of the spectral free energy density $W_g(k,p)$ and take the limit $\ell_v \ll \bar{v}$ as above.  We recall quasi-neutrality

\begin{equation}
\hat{\varphi} (k) = \beta(k)\int v_{}dv_{} J_0(k v_{}) \kappa (v_{})\hat{g}({\bf k}, v_{})\label{qn-g-k-2}
\end{equation}

\noindent where

\begin{equation}
\beta(k) = \frac{2\pi}{\alpha - \hat{\Gamma}_0(k^2)} 
\end{equation}

\noindent and we recall that $\hat{\Gamma}_0(k^2) = \e^{-k^2}I_0(k^2)$.  We momentarily assume the special case $\kappa(v_{}) = 1$.  We must stress that this assumption will not, in the end, affect the generality of the results.  After establishing the spectral scaling in $W_g$ for the case $\kappa = 1$, the scaling index of $\StwoE$ is fixed, and one may refer to the general result of equation \ref{four-fifths-iso-hyp} to establish the general scaling of $S_2$ and thus the general $W_g(k,p)$ spectral scaling laws.  The general-$\kappa$ scaling laws are found to be the same as those for to the case $\kappa = 1$.  Continuing, with $\kappa = 1$, we may write

\begin{equation}
\hat{\varphi} (k) = \beta(k) \hat{g}({\bf k}, k) \label{qn-g-k-p}
\end{equation}

\noindent which allows us to express $\StwoE = \ensbl{|\delta{\bf v}_E|^2}$ in the form

\begin{equation}
\StwoE  = \int dk H(k, \ell_{}, v_{}, v_{}^{\prime}) W_g(k, k)
\label{EB-struc-fcn}
\end{equation}

\noindent where 

\begin{equation}
H(k, \ell_{}, v_{}, v_{}^{\prime}) = -\frac{k \beta^2(k)}{\pi}\left[J_0^2(k v_{}^{\prime}) + J_0^2(k v_{}) - 2 J_0(k v_{}^{\prime})J_0(k v_{})J_0(k\ell_{})\right]
\end{equation}

\noindent Expanding this in the limit $k\bar{v} \gg 1$ and $\ell_v \ll \bar{v}$, as done for $Q$ above, only part of $H$ contributes significantly to the integral of equation \ref{EB-struc-fcn} and we may take

\begin{equation}
H(k, \ell_{}, \ell_v, \bar{v}) \approx \frac{2}{\alpha^2\bar{v}}\left[1 - \cos(k\ell_v)J_0(k\ell_{})\right]
\label{H-approx}
\end{equation}

\noindent Substituting this into equation \ref{EB-struc-fcn}, we have 

\begin{equation}
\StwoE  \approx \frac{2}{\alpha^2\bar{v}}\int dk \left[1 - \cos(k\ell_v)J_0(k\ell_{})\right] W_g(k, k)
\label{EB-struc-fcn-approx}
\end{equation}

\noindent From equation \ref{EB-struc-fcn-approx} we may determine the scaling index $\hE$ in terms of $\nu$, the scaling index of $W_g$:

\begin{equation}
2\hE = -1 - \nu
\label{he-nu}
\end{equation}

\noindent Now we may combine our results for the high-$k$ limit to obtain unique scaling indices for $\delta g$, $\delta{\bf v}_E$ and $W_g(k,p)$.  From equations \ref{hg-he}, \ref{nu-hg} and \ref{he-nu} we have

\begin{subequations}
\begin{align}
&h_g = 1/6 \\
&\hE = 2/3 \\
&\nu = -7/3
\end{align}
\end{subequations}

\noindent And, therefore, taking $p = k$ we have the power law

\begin{equation}
W_g(k, k) \propto k^{-7/3}
\label{sg-kk-scaling}
\end{equation}

\noindent which implies the large-$k$ limit of the second invariant $E_{}$ (which is simply the $\varphi^2$ spectrum) scales like

\begin{equation}
E_{}(k) \approx \alpha E_{\varphi}(k) \approx \frac{\alpha k}{4\pi} \ensbl{|\hat{\varphi}(k)|^2} \sim k^{-1} W_g(k,k) \propto k^{-10/3}
\end{equation}

\noindent This scaling for the electrostatic spectrum $E_{\varphi}(k)$ is in agreement with the previous predictions of \cite{schekochihin} (see section \ref{phenom-sec}).  

\subsection{The two-dimensional spectrum $W_g(k, p)$}\label{two-dim-spectrum-sec}

We have already seen that the two-dimensional spectrum $W_g(k, p)$ obeys the scaling law given by equation \ref{sg-dual-scaling}, with $\nu = -7/3$.  To obtain this scaling, we have assumed $k \gg 1$ and $p \gg 1$ but nothing about the relative sizes of $k$ and $p$.  In the subsidiary limits $p \ll k$ and $k \ll p$, we can show that the spectrum $W_g(k, p)$ has a power law separately in $k$ and $p$.  The results we obtain are

\begin{equation}
W_g(k, p) \propto 
\begin{cases}
k^{-2}p^{-1/3}\mbox{,\hspace{2mm} for \hspace{2mm}} k \gg p\\
p^{-2}k^{-1/3}\mbox{,\hspace{2mm} for \hspace{2mm}} k \ll p
\end{cases}
\label{skp-limit-scaling}
\end{equation}

We will present the argument for $p \ll k$ and omit the case $k \ll p$ which is analogous.  In the limit $p \ll k$, the scaling of the spectrum $W_g(k,p)$ is determined by the scaling of the $\Stwo$ in the limit that $\ell_{} \ll \ell_v$.  Thus we must determine the behaviour of the structure functions in this limit.  Since we now have a power law for $W_g(k,k)$, it is apparent from equations \ref{EB-struc-fcn-approx} and \ref{sg-kk-scaling} that $\StwoE = \ensbl{|\delta{\bf v}_E|^2}$ may be computed analytically as a function of $\ell_{}$ and $\ell_v$.  For general $\ell$ and $\ell_v$, the structure function $\StwoE$ is evaluated in terms of a hypergeometric function.  However, here we are only interested in the limits $\ell_{} \ll \ell_v$.  Thus if we take $W_g(k,k) = W_{g0}\;k^{-7/3}$ and expand the integral to first order in $\ell/\ell_v$, we obtain the result

\begin{equation}
\StwoE \approx C_1 \ell_v^{4/3} + C_2 \ell_v^{-2/3}\ell_{}^2
\label{s2e-lglv}
\end{equation}

\noindent where $C_1$ and $C_2$ are constants which depend on $W_{g0}$ and $\bar{v}$.  Next, we write $\Stwo \approx |\Sthree|/\sqrt{\StwoE}$ and obtain to first order

\begin{equation}
\Stwo \approx \frac{\varepsilon^{\prime}}{\sqrt{C_1}}\ell_v^{1/3}(1 + \frac{\varepsilon}{\varepsilon^{\prime}}\frac{\ell}{\ell_v})
\end{equation}

\noindent When we substitute this into the expression for $W_g(k,p)$, equation \ref{sg-def-approx}, the leading order term produces a delta function after integration over $\ell$ which will be zero for non-zero $k$.  Since we are assuming $k$ and $p$ are both much greater than one, this term is obviously zero.  Thus $W_g(k,p)$ is

\begin{align}
W_g(k,p) &\approx -\frac{k\varepsilon}{\sqrt{C_1}}  \int \ell_{} d\ell_{}\; \bar{v} d\bar{v} \; d\ell_v J_0(k\ell_{})\cos(p\ell_v)\;(\ell_v^{-2/3}\ell_{}) 
\label{sg-pllk-scaling}
\end{align}

\noindent It is easy to see that equation \ref{sg-pllk-scaling} implies the power law $W_g(k,p) \propto p^{-1/3}k^{-2}$.  Repeating this analysis for the limit $k \ll p$ completes the demonstration of equation \ref{skp-limit-scaling}.

\subsection{The one-dimensional spectra $W_g(k)$ and $W_g(p)$}

By integrating equation \ref{sg-spectrum-def} over $p$ we obtain

\begin{equation}
W_g(k) \equiv \int dp W_g(k,p) = -\pi \frac{k}{2} \int \ell_{}d\ell_{} v dv J_0(k\ell_{})\Stwo(\ell_{}, v, v)
\end{equation}

\noindent And integrating over $k$ we obtain

\begin{equation}
W_g(p) \equiv \int dk W_g(k,p) = -\frac{1}{2}\int \bar{v} d\bar{v} d\ell_{v}\cos(p\ell_{v})\Stwo(\ell = 0, \ell_v, \bar{v})
\end{equation}

\noindent The scaling of $\Stwo$ for zero gyro-position increment $\ell_{}$ can be obtained from equation \ref{s2e-lglv}.  The scaling of $\Stwo$ for zero velocity increment $\ell_v$ is obtained analogously.  We find the following power laws:

\begin{equation}
W_g(k) \propto k^{-4/3}
\label{sg-k-scaling}
\end{equation}

\noindent and

\begin{equation}
W_g(p) \propto p^{-4/3}
\label{sg-p-scaling}
\end{equation}

\noindent The power law for $W_g(k)$ has been predicted by \cite{schekochihin} while the power law for $W_g(p)$ is a new prediction.  As a consistency check one may integrate $W_g(k,p)$ directly, using the asymptotic forms from equation \ref{skp-limit-scaling}, to obtain roughly $W_g(p) \sim \int_0^p dk k^{-1/3}p^{-2} + \int_p^{\infty} dk k^{-2}p^{-1/3} \sim p^{-4/3}$.  This works likewise for $W_g(k)$.

\subsection{The inverse cascade of $E_{}$}

\label{inverse-cascade-sec}

We now turn to the scaling of the spectra for the high-$k$ inverse cascade range, $1 \ll k \ll k_i$.  We recall the result of equation \ref{four-fifths-w-iso}: $\SthreeE = \varepsilonE \ellvec_{}$ (which is valid for $\ell_v = 0$), where $\SthreeE =  2\pi \int v dv \; \ensbl{\delta\gyroavg{\varphi} \delta {\bf v}_E\delta g }$.  



We must take care in deducing the scaling index $h_g$ of the increment $\delta g$ from this expression for $\SthreeE$.  From the phenomenological arguments of section \ref{phenom-sec}, we expect that the velocity integration of $\delta g$ should, roughly speaking, introduce a factor of $\ell^{1/2}$.  Thus we cannot, for instance, argue that the scaling indices satisfy $h_g + \hE + \hPhi = 1$ from the linearity of $\SthreeE$--as was done to obtain equation \ref{hg-he}.

We will not use the random walk argument in this section.  However, we will make the additional assumption of local transfer in $k$-space (implicit in the phenomenology) to obtain an estimation of the effect of the velocity integration on scaling of $\SthreeE$.  In $k$-space, $\delta g$, $\delta {\bf v}_E$ and $\delta \gyroavg{\varphi}$ are expressed

\begin{align}
\delta g &= \frac{1}{2\pi}\int d^2{\bf k}_1\left[\e^{-\dot{\imath}{\bf k}_1\cdot{\bf R}^{\prime}} - \e^{-\dot{\imath}{\bf k}_1\cdot{\bf R}}\right] \hat{g}({\bf k}_1)\nonumber\\
\delta \gyroavg{\varphi} &= \frac{1}{2\pi}\int d^2{\bf k}_2\left[\e^{-\dot{\imath}{\bf k}_2\cdot{\bf R}^{\prime}} - \e^{-\dot{\imath}{\bf k}_2\cdot{\bf R}}\right] J_0(k_2 v)\hat{\varphi}({\bf k}_2)\nonumber\\
\delta {\bf v}_E &= \frac{1}{2\pi}\int d^2{\bf k}_3\left[\e^{-\dot{\imath}{\bf k}_3\cdot{\bf R}^{\prime}} - \e^{-\dot{\imath}{\bf k}_3\cdot{\bf R}}\right] J_0(k_3 v)(\hat{z}\times{\bf k}_3)\hat{\varphi}({\bf k}_3)
\end{align}

\noindent If we substitute these expressions this into the quantity $\ensbl{\delta \gyroavg{\varphi}\delta {\bf v}_E\delta g}$, locality in $k$-space implies that we may approximate $J_0(k_3 v) \approx J_0(k_1 v)$ and $J_0(k_2 v) \approx J_0(k_1 v)$ and transfer the two Bessel functions to the integral over ${\bf k}_1$.  This leads to the approximation

\begin{equation}
\SthreeE \approx \ensbl {\delta\varphi \delta{\bf v}_{E0} \; \int v dv \int d^2{\bf k}\left[\e^{-\dot{\imath}{\bf k}\cdot{\bf R}^{\prime}} - \e^{-\dot{\imath}{\bf k}\cdot{\bf R}}\right] J_0^2(k v) \hat{g}({\bf k}) } \label{S3E-approx}
\end{equation}

\noindent recalling that the $E\times B$ velocity (without gyro-average) is ${\bf v}_{E0} = \hat{z}\times\bnabla\varphi$.  The scaling of $\delta\varphi$ and $\delta {\bf v}_{E0}$ may be written in terms of the index $\nu$ by following the methods leading to the scaling index of $\StwoE$, $\hE$.  It is found that 

\begin{subequations}\label{extra-scalings}
\begin{align}
&\delta\varphi(\lambda \ell) \doteq \lambda^{-\nu/2} \delta\varphi(\ell)\\
&\delta{\bf v}_{E0}(\lambda \ell) \doteq \lambda^{-\nu/2-1} \delta{\bf v}_{E0}(\ell) 
\end{align}
\end{subequations}

\noindent Now we determine the scaling of the remaining velocity integral in equation \ref{S3E-approx}.  To do this we expand in the Hankel transform of $\delta g$ and perform the integration over velocity using equation \ref{three-bes-int}.  Then we square the quantity and ensemble average, applying homogeneity and isotropy.  We obtain 

\begin{multline}
\ensbl{\left(\int v dv \int d^2{\bf k}\left[\e^{-\dot{\imath}{\bf k}\cdot{\bf R}^{\prime}} - \e^{-\dot{\imath}{\bf k}\cdot{\bf R}}\right] J_0^2(k v) \hat{g}({\bf k}) \right)^2} \approx\\ 8\pi\int dk\; pdp\; K^2(k,k,p)[1 - J_0(k\ell)] W_g(k,p) \label{final-scaling}
\end{multline}

\noindent where the function $K$ is defined by equation \ref{three-bes-int}--it arises from the integral of three Bessel functions.  Also, we have made the approximation that $\ensbl{\hat{g}(p)\hat{g}(p^{\prime})} \approx 2\pi\ensbl{\hat{g}(p)\hat{g}(p^{\prime})}\delta(p-p^{\prime})$.  This approximation is analogous to the exact result in position-space due to homogeneity: $\ensbl{\hat{g}({\bf k}^{\prime})\hat{g}({\bf k})} = \delta({\bf k}^{\prime} + {\bf k})\ensbl{\hat{g}({\bf k}^{\prime})\hat{g}({\bf k})}$.  The Hankel transform velocity-space analogue may be proved\footnote{The identity $\delta(p^{\prime} - p) = \displaystyle{\lim_{a \rightarrow 0}}\; \frac{1}{2\pi}\int_{-1/a}^{1/a}\cos[\bar{v}(p^{\prime}-p)]d\bar{v}$ is useful here.} by taking the limit $k \gg 1$ and assuming that the correlation function $G(\ell_v)$ is highly peaked around $\ell_v = 0$ and varies smoothly in $\bar{v}$ (as we have been assuming in this chapter).

Combing the results of equations \ref{four-fifths-w-iso}, \ref{S3E-approx}, \ref{extra-scalings} and \ref{final-scaling} we have

\begin{equation}
\nu = -1
\end{equation}

\noindent which implies, using equation \ref{nu-hg}, that

\begin{equation}
h_g = -1/2
\end{equation}

\noindent leading to the spectra

\begin{subequations}
\begin{align}
& E_{}(k) \propto k^{-2} \\
&W_g(k,k) \propto k^{-1} \\
& W_g(k) \propto k^{0}
\end{align}
\end{subequations}

\subsection{Some comments on the two dimensional spectrum $W_g(k,p)$}

From the scaling relations obtained in this section we may write $\ensbl{\delta \gyroavg{\varphi}\delta {\bf v}_E\delta g} \propto \ell^{1/2}$.   However this result is specific to $\ell_v = 0$.  For $\ell_v \neq 0$, the self-similarity hypothesis \ref{dual-hypothesis} can only give $\ensbl{\delta \gyroavg{\varphi}\delta {\bf v}_E\delta g} \propto f(\ell,\ell_v)$, such that the function $f(\ell, \ell_v)$ obeys the scaling $f(\lambda \ell, \lambda \ell_v) = \lambda^{1/2}f(\ell, \ell_v)$.  Without more specific knowledge of the function $f$, it is not possible to determine power law spectra for $W_g(k,p)$ in the limits $k \gg p$ and $k \ll p$.  However, in the limit that $\ell = 0$ and $\ell_v \neq 0$ we must have $f \propto \ell_v^{1/2}$, implying that

\begin{equation}
W_g(p) \propto p^{0}
\end{equation}

\section{Conclusion}

\subsection{Summary}

In this chapter we have considered forced two dimensional gyrokinetics as a simple paradigm for kinetic plasma turbulence.  We have explored how the nonlinear phase-mixing mechanism gives rise to a phase-space cascade for scales much smaller than the Larmor radius.  We have investigated the relationship between the inertial range cascades in fluid theories (HM and two-dimensional Euler turbulence) and gyrokinetics, and found a simple relationship between the fluid invariants and the kinetic invariants, given by equation \ref{invar-reln}.  The gyrokinetic free energy invariant $W_g$ is cascaded to fine scales and is ultimately dissipated by collisions.  Concurrently, there is a gyrokinetic invariant $E$ which my in principle cascade inversely from very fine scales (\eg the electron Larmor radius) to long wavelength regimes of the CHM/Euler limits, much larger that the ion Larmor radius.  

In addition to a phenomenological derivation of the spectral scaling laws based on the previous work of \cite{schekochihin}, we have given derivations of exact third-order laws for the forward cascade range and for the inverse cascade range.  These derivations are based upon general considerations of symmetries of the gyrokinetic equation and an additional assumption of a two scale dependence in velocity-space.

Combining the exact third order results with a dual self-similarity hypothesis and the two-scale assumption in velocity-space, we are able to reproduce the phenomenological scaling laws for the both the forward cascade range and the inverse cascade range.  Since the  methods used are somewhat different, this can be seen as an independent confirmation of these scaling laws.

For the forward cascade range, the dual self-similarity hypothesis is also used to obtain novel predictions for the velocity-space spectrum.  To describe the velocity spectrum, we introduce the use of a zeroth order Hankel transformation.  This novel spectral treatment of perpendicular velocity-space may in general, be found useful for theoretical and numerical applications in gyrokinetics.  




\subsection{A note on forcing and universality}


In general, turbulence may be generated by linear instabilities which are induced by a mean-scale free energy gradient.  In this case, additional (linear) terms in the gyrokinetic equation must be included.  Although the present work has left the exact form of the forcing unspecified, we note that a simple phenomenological calculation shows that the additional terms do not invalidate the inertial range assumption at fine scales -- the details of these arguments are left for future work.  Thus it is possible that these scalings may be a universal feature of magnetised plasma turbulence.  

Encouragingly, the scaling for $E_{\varphi}(k)$ in the forward cascade range has already been independently observed in a fusion-relevant simulation by \cite{gorler}.  Also, the simulations found in \cite{tatsuno} demonstrate strong confirmation of the scalings predicted by \cite{schekochihin} and also demonstrates consistency with the velocity-space prediction for $W_g(p)$, given by equation \ref{sg-p-scaling}.  Numerical investigation of other novel predictions of this chapter, including the inverse cascade scaling laws and the detailed phase-space spectral predictions of section \ref{two-dim-spectrum-sec}, are planned.

\subsection{Future work}\label{future-work-sec}

There are some issues not addressed by this work which should be given high priority.  First, a more complete investigation of the gyrokinetic turbulence cascade would include physical content such as electromagnetic affects and the correct linear instability drives.  Electromagnetism introduces a fluctuating magnetic field into the nonlinearity which couples to the distribution function through Ampere's law and will enrich the problem substantially.

The present work requires significant extension to be applied to the realistic tokamak scenarios.  As a general point, it should be noted that the macroscopic properties of a fusion plasma are most strongly influenced by scales at the forcing range or larger.  This has meant that inertial range ideas are generally not a focus of modern magnetically confined fusion research.  However, some more immediate applications of this work to fusion include simulation benchmark and diagnostics, or the general use of the $v_{\perp}$ Hankel-transform spectral treatment in gyrokinetic linear and nonlinear problems.  

Inertial range understanding gyrokinetic turbulence may also enter in the problem of transport.  As turbulence modelling in fusion ultimately aims to produce accurate quantitative prediction of transport, the inertial range contribution to transport will likely need to be included as the field evolves.

Three dimensional dynamics must also be investigated.  First, the assumption that the parallel streaming term (see equation \ref{gyro_full}) may be neglected requires careful justification.  One instance where this term does play a role is in the Alfv\'{e}n wave cascade--which can be treated within gyrokinetic theory.  In this case, it is believed that a critical balance condition is satisfied where the generation of finer scales in $k_{\parallel}$ ensures that the parallel term remains important at all scales in the cascade forward.  On the other hand, a conventional assumption in tokamak turbulence is that one may take the size of $k_{\parallel}$ to be correspond to the distance along a field line from the stable side to the unstable side of the toroidal magnetic surface.  In that case, the nonlinearity will dominate over this term for sufficiently small (perpendicular) scales.

It is known that the parallel streaming term also gives rise to a linear phase-mixing, which generates fine scales in parallel velocity-space ($v_{\parallel}$) dependence of the distribution function.  Nonlinear simulations and theoretical work by \cite{watanabe} demonstrate how this parallel phase-mixing plays a role in the statistical steady state of the entropy cascade.  Their work describes a steady state where parallel phase mixing is treated as a passive scalar phenomenon and exhibits an inertial range with a power law spectrum.  However, this work is done employing a $v_{\perp}$-integrated version of the gyrokinetic equation, which suppresses perpendicular phase-mixing by assuming a fixed Maxwellian dependence in $v_{\perp}$-space.  Thus our work is complimentary to the theoretical work by \cite{watanabe}, although both works neglect inclusion of the third dimension in position-space.

In general, the present capabilities of numerical simulations allow a detailed investigation of (perpendicular) nonlinear phase mixing or (parallel) linear phase-mixing, but not both.  Thus a theoretical investigation of the full five-dimensional theory (that is, three position dimensions and two velocity dimensions) currently has the opportunity to pave the way to a more complete picture of the turbulent steady state in the gyrokinetic turbulence cascade.


\chapter{Conclusion}

\section{Highlights}

In this thesis we have seen the application of varying approaches to modeling and understanding the fully developed state of a turbulent magnetized plasma.  There has been a common motivation, in magnetically confined fusion; and there has been a common theme, in the multiple scales at play.  In chapters \ref{chap-2-sec}, \ref{secondary-chap} and \ref{phase-space-turbulence-chapter} we have seen this theme in the multiple scales approach to gyrokinetic transport theory, in disparate time-scales in the saturation of linear instabilities by secondaries, and in the treatment of velocity scales in the phase-space cascade with a two-scale assumption.

Chapter \ref{chap-2-sec} (and the appendix \ref{appendix-sec}) has introduced a synthesis of gyrokinetic and (neo)classical transport theories based on the method of multiple scales.  Chapter \ref{secondary-chap} has explored secondary instability theory as a simple model of an important nonlinear process in fully developed turbulence.  We have argued how this theory can be applied to mixing length phenomenology to shed light on important broader issues of tokamak physics.  Finally in chapter \ref{phase-space-turbulence-chapter} we have developed a framework for understanding the turbulent generation of phase-space structures and derived the inertial range scaling laws, following the methodology from neutral fluid turbulence.

\section{Final note}

Although each of the works in this thesis have produced novel results, they are intermediate steps in solving larger and more difficult problems -- and, naturally, there are important questions  and opportunities which remain.  The chapters of this thesis have already detailed some of the future work that is needed.  In the context of the combined work of this thesis, there are some general (and somewhat speculative) remarks that can be made.

Statistical theories of fully developed strong turbulence in the energy-containing range have been unable to capture the features that were subsequently observed in nonlinear simulations such as the levels of anisotropy and intensity and, ultimately, the thermal fluxes.  These analytical theories are based upon plausible assumptions about the nature of the fully developed state.  

For instance, a typical assumption is that, at each scale, the linear growth rate of unstable modes should be balanced against the nonlinear decorrelation time determined generically by the strength of the nonlinear term.  (Or more generally, one may assume that the nonlinear decorrelation time may be determined as a function of the linear growth rate spectrum -- see \cite{canuto-goldman-1}.)  However, for important cases in tokamak turbulence, it has turned out that this plausible assumption misses the mark.  For instance, although the linear theories of ETG and ITG turbulence are isomorphic, the saturated turbulent state in the energy-containing range are qualitatively very different, as discussed in chapter \ref{secondary-chap}.  

Secondary instability theory has shed light on this issue.  For the case of ETG turbulence it can be shown that, for sufficiently small wavenumbers, the magnitude of the nonlinearity required to destabilize a fast secondary becomes much larger than the level needed for a basic balance with the linear drive terms.  These weak secondaries have been used to successfully predict the shape of the saturated spectrum (see \cite{dorland3}) for an important part of the energy-containing range.  That is, it turns out that the shape of the ETG saturated spectrum is largely determined by the balance between secondary and primary growth rates.

As numerical simulations and focused analytical models build our intuition for the basic physical processes which govern the different turbulent ranges, there will be an opportunity to revisit the problem of a more complete and fundamental theoretical description.  Such a description must contain essential ingredients of the linear and nonlinear physics which leads to such things as self organization (or spectral condensation) and observed saturation amplitudes.  This theory would be a powerful tool in creating a deeper understanding of fusion confinement.  It could lead to a phenomenology of magnetized plasma turbulence with broad applicability and, in general, a mathematically and computationally simplified framework to enable efficient exploration of the vast parameter space for the operation of fusion devices.  Finally it may be hoped that this would be a path to the discovery of ways to tame turbulent plasma transport and realize the dream of small-sized and inexpensive fusion reactors.

\appendix
\chapter{Appendix: Gyrokinetics transport in toroidal geometry}\label{appendix-sec}

\section{Introduction}

The appendix is devoted to a lengthy derivation of transport and entropy balance for axi-symmetric toroidal geometry.  The procedure followed here is quite similar to that in chapter \ref{chap-2-sec}, so there is much repetition.  The appendix has not been produced in a form designed for publication in a journal (the work is still in progress), so it should be viewed as a collection of notes, though having several concrete results.

Others who have combined neoclassical and anomalous transport are \cite{shaing, balescu, sugama-transport-2}.  The work which is probably closest to ours is \cite{sugama-transport-2}.  They also derive the transport equations within the gyrokinetic ordering hierarchy, but employ an ensemble average to separate the mean-behavior, instead of the patch and time averages we use based on scale separations.  Also, the final form of heat transport we have provided is significantly manipulated as to compare closely with entropy balance, and so that the heating term is in a positive-definite form suitable for application to numerical simulations -- see \cite{barnes-thesis}.

\section{Equilibrium field geometry}
\label{geometry-sec}
\subsection{Axi-symmetry}
\label{axi-symm-sec}

We assume toroidal geometry and make use of axi-symmetry when needed.  However, we avoid explicitly using magnetic coordinates when possible.  We make the common choice of labeling magnetic surfaces by the flux variable $\psi$ which is related to the poloidal flux by $\psi = \psi_p/2\pi$.  The equilibrium magnetic field can then be expressed

\begin{equation}
{\bf B_0} = \nabla\psi\times\nabla\phi + I(\psi)\nabla\phi
\label{B0-eqn}
\end{equation}
\noindent where $\phi$ is the toroidal angle and $I(\psi)$ is a flux surface quantity which is proportional to the total poloidal current.

\subsection{Magnetic coordinates}
\label{mag-coord-sec}

\noindent There are many suitable choices for flux coordinates.  Our choice is the radial coordinate $\psi$, the poloidal angle coordinate $\theta$ and the toroidal angle $\phi$.  There is also some freedom in choosing the definition of $\theta$.  Here we define

\begin{equation}
\theta = \frac{I(\psi)}{q(\psi)}\int_0^{l_p}\frac{dl_p^\prime}{R|\nabla\psi|}
\end{equation}

\noindent where $l_p$ is the length along the flux surface as one integrates in the poloidal direction in a constant $\phi$ plane.  This choice of $\theta$ has the feature that field lines appear ``straight'' in the $\phi$-$\theta$ plane.  In particular, a field line satisfies the equation $\frac{d\phi}{d\theta} = q$, where $q$ is a constant on the flux surface called the safety factor.  (The safety factor is the ratio of the number of toroidal transits to the number of poloidal transits that a single field line makes as it traces out the flux surface.)  These coordinates provide another simple way to express ${\bf B_0}$

\begin{equation}
{\bf B_0} = \nabla\psi\times(\phi - q\theta)
\end{equation}

\noindent As a final note, the equilibrium field direction will be denoted by

\begin{equation}
{\bf b_0} = \frac{\bf B_0}{|{\bf B_0}|}
\end{equation}


\subsection{Annulus volume average}

\label{annulus-vol-sec}

The annulus volume average is defined over an interval $\Delta\psi$ containing the flux surface with label $\psi$.  In particular it is defined

\begin{equation}
\annavg{A} = \frac{\int_{\Delta V} d^3{\bf r}A}{\Delta V}
\end{equation}

\noindent where

\begin{equation}
\int_{\Delta V} d^3{\bf r}A = \int^{\psi + \Delta\psi/2}_{\psi - \Delta\psi/2}\int^{2\pi}_{0}\int^{2\pi}_{0}\frac{d\theta d\phi d\psi^{\prime}}{\nabla\psi^{\prime}\times\nabla\theta\cdot\nabla\phi}A(\psi^{\prime},\theta,\phi)
\end{equation}

\begin{figure}[htbp]
\begin{center}
\includegraphics[width=5in]{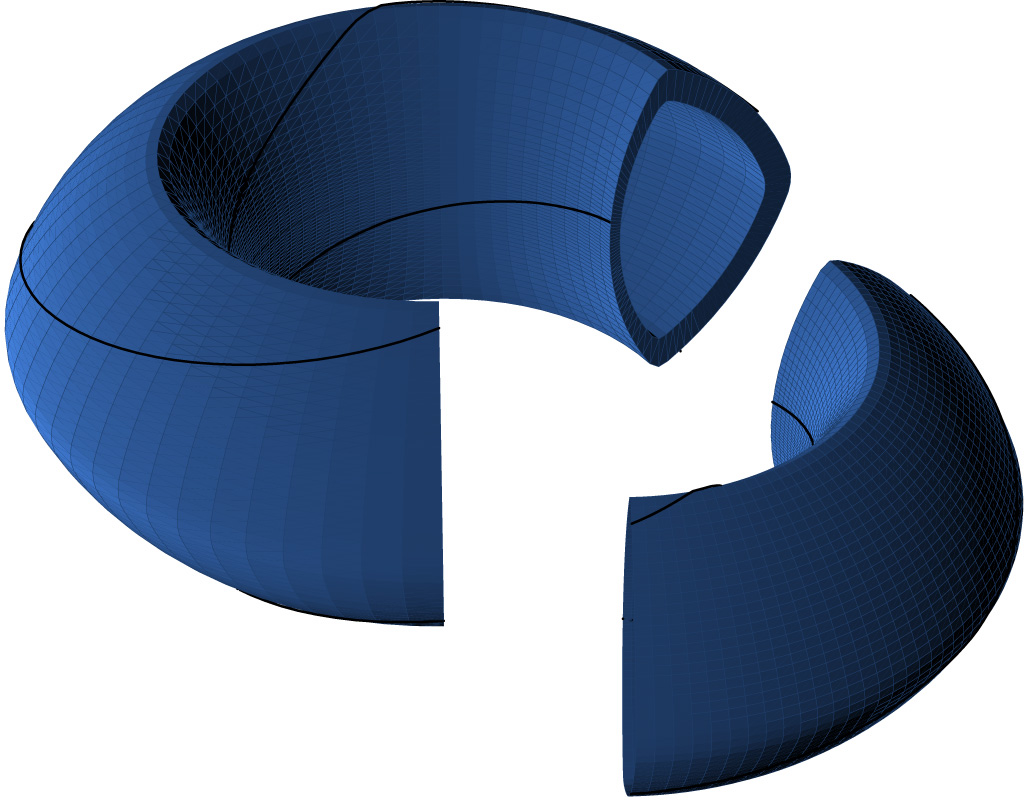}
\caption{The annulus volume.}
\label{annulus-volume-fig}
\end{center}
\end{figure}

\noindent The annulus volume average is inspired and closely related to the flux surface average from neoclassical transport theory.  In fact, when applied to equilibrium quantities, it is equivalent to the flux surface average since the shell can be considered to have infinitesimal thickness.  Indeed, the explicit form of the flux surface average in Ref.~\cite{hinton}, Eqn.~ 2.55, is identical to the annulus volume definition in the limit of infinitesimal volume $\Delta V$.  We will make use of this fact in adapting results from neoclassical theory, Refs.~\cite{hinton,bernstein}.  For the turbulent terms, the annulus has finite thickness and essentially combines a flux surface average with the patch volume average defined by Eqn.~\ref{patch-eqn}.  The annulus volume average of the divergence of a vector function ${\bf F}$ is evaluated in a slightly different manner to the flux surface average

\begin{eqnarray}
&\annavg{\nabla\cdot {\bf F}} = \nonumber \\
&\frac{1}{\Delta V}\int \nabla\cdot {\bf F} = \nonumber \\
&\frac{1}{\Delta V}\int d^{2}{\bf s}\cdot{\bf F} = \nonumber \\
&\frac{1}{\Delta V}\left[\int \frac{d^{2}s}{|\nabla\psi|}\nabla\psi\cdot{\bf F}\right]_{\psi - \Delta\psi/2}^{\psi + \Delta\psi/2} = \nonumber \\
&\frac{1}{\Delta V}\frac{d}{d\psi}\int \frac{d^{2}s d\psi}{|\nabla\psi|} \nabla\psi\cdot{\bf F} = \nonumber \\
&\frac{1}{\Delta V}\frac{d}{d\psi}(\Delta V\annavg{ {\bf F}\cdot\nabla\psi}) = \nonumber \\
&D^{(\Delta)}_{\psi}\annavg{ {\bf F}\cdot\nabla\psi}
\label{ann-div}
\end{eqnarray}

\noindent where we have introduced the operator $D^{(\Delta)}_{\psi}$ defined $D^{(\Delta)}_{\psi}A = \Delta V^{-1}d/d\psi (\Delta V\; A)$.  I will also make use of the analogous result for the flux surface average (Ref.~\cite{hinton}, Eqn.~ 2.58) which I copy here:

\begin{equation}
\fluxavg{\nabla\cdot {\bf F}} = \frac{d\psi}{d V}\frac{d}{d\psi}\frac{d V}{d\psi}\fluxavg{ {\bf F}\cdot\nabla\psi} = D_{\psi}\fluxavg{ {\bf F}\cdot\nabla\psi}
\label{flux-div}
\end{equation}

\noindent where $V$ is the total toroidal volume contained within the flux surface labeled $\psi$, and the we define $D_{\psi}A = d\psi/d V\;d/d\psi(d V/d\psi A)$.

\section{Proving $F_0$ is a flux surface quantity}

In Sec.~\ref{first-sec} we used the assumption that the equilibrium distribution function does not vary in the direction of the equilibrium magnetic field, ${\bf b}_0\cdot\nabla F_0 = 0$.  This can be proven for a system with closed flux surface geometry (e.g. axi-symmetric toroidal geometry).  First we recall the ${\cal O}(1)$ Fokker-Planck equation, Eqn.~\ref{order1}

\begin{multline}
{\bf v}\cdot\nabla F_0 + {\bf v}_\perp\cdot\nabla \delta f_1 +
\frac{q}{m}\left( -\nabla\varphi + {\bf v\times\delta
  B}\right)\cdot\frac{\partial F_0}{\partial {\bf v}}
-\Omega_0\left(\frac{\partial \delta f_1}{\partial \vartheta}\right)_{{\bf
    r}, v, v_\perp} \\
    = C(F_0, F_0)
\end{multline}

\noindent We use Boltzman's H-theorem to show that $F_0$ must be Maxwellian.  We multiply Eqn~(\ref{order1}) by $1 + \log F_0$, integrate over velocity space, but this time integrate over annulus volume
defined in Sec.~\ref{annulus-vol-sec}.  The troublesome term is evaluated as follows:

\begin{eqnarray}
\annavg{\int d^3{\bf v}\cdot\nabla (F_0 \ln F_0)} = \frac{1}{\Delta V}\int d^2 {\bf s}\cdot {\bf v}_{\perp} (F_0 \ln F_0) = 0
\end{eqnarray}

\noindent where we have used the divergence theorem to evaluate the volume integral as a surface integral and used the fact that the surface vector is perpendicular to the equilibrium magnetic field so that $d^2 {\bf s}\cdot {\bf v} = d^2 {\bf s}\cdot {\bf v}_{\perp}.$  This is obviously odd in $\vartheta$ since $F_0$ is independent of $\vartheta$.  So, the only term remaining from the ${\cal O}(1)$ equation is the collisional generation of entropy per unit volume in the annulus volume.  So we have

\begin{eqnarray}
\annavg{\int d^3{\bf v} C(F_0, F_0)\ln F_0} = 0
\end{eqnarray}

\noindent As before we have that $F_0$ is a Maxwellian.  Finally, we return to the ${\cal O}(1)$ equation and apply the patch average and average over $\vartheta$ leaving the desired result:

\begin{equation}
{\bf b_0}\cdot\nabla F_0 = 0
\end{equation}

\noindent Since we are considering a closed flux surface geometry (and ergodic field lines), the lack of variation along the magnetic field  implies that $F_0$ is constant on a fixed flux surface.  I.e. in space, $F_0$ is a function only of the radial coordinate $\psi$:

\begin{equation}
F_0 = F_0(\psi, v^2)
\end{equation}

\section{${\cal O}(\epsilon^2)$: Transport equations}
\label{transport-sec}

Here we calculate the particle transport and heat transport equations in
Gyrokinetics. To obtain the long time dependence of $n_0$ and $T_0$ we consider the
moment equations of the full un-gyro-averaged kinetic equation for  
any species ($\frac{df_s}{dt}= C_{sr}(f_s, f_r) + C_{ss}(f_s, f_s)$
where $C_{sr}(f_s, f_r)$ is collisions of species s on species r and
$C_{ss}(f_s, f_s)$ is like particle collisions).  This avoids having
to write out all the ${\cal O}({\epsilon}^2)$ terms from the expanded
equations.

\subsection{\bf Particle transport}

To obtain the time evolution of the density, we integrate the
Fokker-Planck equation over velocity and average over the annulus volume.

\begin{multline}
\int d^3{\bf v}\annavg{\left[\frac{\partial f_{s}}{\partial t} + {\bf v}\cdot\nabla f_s + \frac{q}{m}({\bf E} + {\bf v}\times{\bf B})\cdot \frac{\partial f_s}{\partial {\bf v}}\right]} \\
= \int d^3{\bf v}\annavg{ C(f,f)}
\end{multline}

\noindent The integral of the collisions over velocity will be zero to conserve
particles.  The $({\bf E} + {\bf v}\times{\bf B})\cdot \frac{\partial
  f_s}{\partial {\bf v}}$ terms will be zero because they are a
perfect divergence in velocity space.  What is left is the continuity equation.

\begin{equation}
\int d^3{\bf v}\annavg{ 
\left[\frac{\partial f_{s}}{\partial t} + \nabla\cdot ({\bf v}f_s)\right]} = 0
\label{particle_FP-appendix}
\end{equation}

\noindent The flux term is evaluated using Eqn.~\ref{ann-div}

\begin{equation}
\annavg{{\bf v}\cdot\nabla f_s} = D^{(\Delta)}_{\psi}\annavg{\nabla\psi\cdot{\bf v}f_s}
\end{equation}

\noindent We can transform this expression by considering the following

\begin{align}
&\int d^3{\bf v}\annavg{ 
R{\bf v}\cdot {\hat \phi}\left[ {\bf v} \times {\bf B}_0 \cdot
  \frac{\partial f}{\partial v} \right] }
= - \int d^3{\bf v}\annavg{ 
{\hat \phi}\cdot {\bf v}\times{\bf B}_0 fR} \nonumber \\
& = \int d^3{\bf v}\annavg{ 
{\bf v}\cdot \nabla \psi f}
\label{v_grad-appendix}
\end{align}

\noindent where we first integrated by parts, and then used ${\bf B}_0 = \nabla \psi\times \nabla \phi + \nabla \psi F(\psi)$.  At this point, we can again substitute the Fokker-Plank equation in for ${\bf v} \times {\bf B}_0 \cdot \frac{\partial f}{\partial v}$.  Thus, we can write

\begin{eqnarray}
&\int d^3{\bf v}\annavg{\nabla\cdot ({\bf v}f_s)} = D^{(\Delta)}_{\psi} \int d^3{\bf v}\annavg{ 
({\bf v}\cdot \nabla \psi f)} = \nonumber \\
& D^{(\Delta)}_{\psi}\int d^3{\bf v}\annavg{\frac{mR}{q}{\bf v}\cdot {\hat \phi}\left[ \frac{q}{m}{\bf v} \times {\bf B}_0 \cdot \frac{\partial f}{\partial v} \right]} = \nonumber \\
&-D^{(\Delta)}_{\psi} \int d^3{\bf v}\annavg{ \frac{mR}{q}{\bf v}\cdot {\hat \phi}\left[\frac{\partial f_{s}}{\partial t} + {\bf v}\cdot\nabla f_s + \frac{q}{m}({\bf E} + {\bf v}\times\delta{\bf B})\cdot \frac{\partial f_s}{\partial {\bf v}} - C(f,f)\right] }
\label{particle_start}
\end{eqnarray}

\noindent We can rewrite Eqn~(\ref{particle_FP-appendix}) as

\begin{eqnarray}
&\int d^3{\bf v}\annavg{\frac{\partial f_{s}}{\partial t}}= \nonumber \\
&D^{(\Delta)}_{\psi}\int d^3{\bf v}\annavg{\frac{mR{\bf v}\cdot {\hat \phi}}{q}\left[\frac{\partial f_{s}}{\partial t} + {\bf v}\cdot\nabla f_s + \frac{q}{m}({\bf E} + {\bf v}\times\delta{\bf B})\cdot \frac{\partial f_s}{\partial {\bf v}} - {C(f,f)}\right]}
\label{density_moment-appendix}
\end{eqnarray}

\noindent At this point, we would use our solution for the distribution
function given by Eqn~(\ref{fform}) and substitute it into
Eqn~(\ref{density_moment-appendix}).  Most terms are zero or negligible by averaging and ordering arguments.  We leave the work for Sec.~\ref{particle-appendix}.  The result is

\begin{eqnarray}
&\annavg{\frac{\partial n_{0s}}{\partial t}} = \nonumber \\
&D^{(\Delta)}_{\psi}\int d^3{\bf v}\annavg{
R{\hat \phi}\cdot\left[F_m \frac{\partial {\bf A_0}}{\partial t} + \nabla \chi h_{gk}
+ \frac{m{\bf v}}{q}\;C(\rhovec \cdot\nabla F_m - h_{nc}, F_m)\right]}
\label{density-transport}
\end{eqnarray}

\noindent where $F_m$ is the Maxwellian.  The first term is the Ware pinch.  The second term is accounts for the turbulent or anomalous transport driven by the fluctuating fields and the associated drift velocities.  The collisional term contains classical and neoclassical particle transport.

\subsection{Electron particle transport}

The above equation and analysis in Sec.~\ref{particle-appendix} is
independent of species.  The density transport for electrons is
simplified by integrating the electron gyrokinetic and neoclassical
equations over velocity.  The result obtained is that $h_{gk}$ and
$h_{nc}$ are both odd $v_\parallel$. The resultant electron density
transport equation reads 

\begin{equation}
\annavg{\frac{\partial n_{0e}}{\partial t}} = 
 D^{(\Delta)}_{\psi}\int d^3{\bf v}\annavg{
R{\hat \phi}\cdot\left[F_{me} \frac{\partial {\bf A_0}}{\partial t} + \nabla \varphi h_{gke}\right]}
\label{electrondensity-transport}
\end{equation}

\noindent Note if we assume the Boltzmann response for electrons, the anomalous term (the second term) will integrate away which implies there will be no anomalous particle transport of electrons for ITG-driven turbulence insofar as the Boltzmann response holds.

\subsection{\bf Heat transport}

To calculate heat transport, we multiply the Fokker-Planck equation by $\frac{1}{2}m{\bf v}^2$, integrate over velocity, and average.  Following the same methodology as with the particle transport, 

\begin{multline}
\int d^3{\bf v}\annavg{\frac{mv^2}{2} 
\left[\frac{\partial f}{\partial t} + {\bf v}\cdot\nabla f + 
\frac{q}{m}({\bf E} + {\bf v}\times{\bf B})\cdot \frac{\partial
  f}{\partial {\bf v}}\right]} \\
= \int d^3{\bf v}\annavg{ \frac{mv^2}{2}C(f,f)}
\label{heating_moment_start}
\end{multline}

\noindent The second term in Eqn~(\ref{heating_moment_start}) can be rewritten in the same manner as
Eqn~(\ref{v_grad-appendix}).   

\begin{align}
&\int d^3{\bf v}\frac{mv^2}{2}\annavg{{\bf v}\cdot\nabla f} = D^{(\Delta)}_{\psi} \int d^3{\bf
  v}\frac{mv^2}{2}\annavg{{\bf v}\cdot\nabla\psi f} \nonumber \\
&= D^{(\Delta)}_{\psi}\int d^3{\bf v}\annavg{\frac{mR}{2}
{\bf v}\cdot {\hat \phi}\frac{\partial}{\partial {\bf v}}\cdot ({\bf v}\times{\bf B}_0f_s v^2} \nonumber \\
&= \frac{m}{2}D^{(\Delta)}_{\psi}\int d^3{\bf v}
\annavg{\frac{mv^2R}{q}{\bf v}\cdot {\hat \phi}\left[\frac{q}{m}({\bf v}\times{\bf B}_0)\cdot
  \frac{\partial f}{\partial {\bf v}}\right]}
\end{align}

\noindent The last line was obtained using $({\bf v}\times{\bf b}_0)\cdot
\frac{\partial v^2}{\partial {\bf v}} = 0$.  Reworking of
Eqn~(\ref{heating_moment_start}) yields 

\begin{align}
\int d^3{\bf v}\annavg{\frac{mv^2}{2} 
\frac{\partial f}{\partial t}} = 
& -\frac{m}{2}D^{(\Delta)}_{\psi} \int d^3{\bf v}\annavg{
  \frac{mv^2R}{q}{\bf v}\cdot {\hat 
    \phi} \left[\frac{q}{m}({\bf v}\times{\bf B}_0)\cdot \frac{\partial
      f}{\partial {\bf v}}\right]} \nonumber \\
& + \int d^3{\bf v}\annavg{\left[ q{\bf E}\cdot{\bf v}f + \frac{mv^2}{2}C(f,f)\right]}
\label{heating_moment-appendix}
\end{align}

\noindent Like the particle transport analysis, using Eqn~(\ref{fform}) in
Eqn~(\ref{heating_moment-appendix}) is left for Sec.~(\ref{heat-appendix}).
The result we obtain is

\begin{align}
&\annavg{\frac{3}{2}\frac{\partial (n_{0s}T_{0s})}{\partial t}} -
  D^{(\Delta)}_{\psi}\int d^3{\bf v}\annavg{\frac{mR}{2}F_mv^2{\bf \hat{\phi}}\cdot\frac{\partial {\bf
	A_0}}{\partial t} }
\nonumber \\ & -D^{(\Delta)}_{\psi}\int d^3{\bf v}\annavg{\left[ 
\frac{mv^2R}{2}(\nabla \chi \cdot {\bf \hat{\phi}})h_{gk} 
+ \frac{m^2v^2R}{2q}{\bf v}\cdot {\hat \phi}C(h_{nc} - \rhovec \cdot \nabla F_m, F_m)\right]}
\nonumber \\ & - \int d^3{\bf v}\annavg{ 
q \left [\frac{\partial \chi}{\partial t}h_{gk} - v_{\parallel}{\bf
  b_0}\cdot\frac{\partial {\bf A_0}}{\partial t}h_{nc} + {\bf
  v}\cdot\frac{\partial {\bf A_0}}{\partial t}(\rhovec\cdot\nabla F_m)
  \right]} \nonumber \\
& = n_{0s} \nu_{{\cal E}}^{sr}(T_r - T_s)\label{pre-heat-transport} 
\end{align}

\noindent The first term will give us the temperature evolution.  The second
through fourth terms are heat convected by the particle transport.
The $\frac{\partial \chi}{\partial t} h_{gk}$ term is gyrokinetic
heating.  The second and third terms on line three are heating from
the Ware pinch.  The last term is heat exchange between species.  

\subsection{Electron temperature transport}

Because $h_{gke}$ and $h_{nce}$ both are odd in $v_\parallel$, the above heat transport equation is again simplified for electrons.  

\begin{align}
&\annavg{\frac{3}{2}\frac{\partial (n_{0e}T_{0e})}{\partial t}} - D^{(\Delta)}_{\psi}\int d^3{\bf v}\annavg{\frac{m_eR}{2}F_{me}v^2{\bf \hat{\phi}}\cdot\frac{\partial {\bf A_0}}{\partial t}}
\nonumber \\ 
& -D^{(\Delta)}_{\psi}\int d^3{\bf v}\annavg{\frac{m_ev^2R}{2}(\nabla \varphi \cdot {\bf \hat{\phi}})h_{gke}} \nonumber \\
& - \int d^3{\bf v}\annavg{q \left [\frac{\partial \varphi}{\partial t}h_{gke} - {\bf v}\cdot\frac{\partial {\bf A_0}}{\partial t}(\rhovec\cdot\nabla F_{me})\right]} = n_{0e} \nu_{{\cal E}}^{ei}(T_i - T_e)\label{electron-pre-heat-transport} 
\end{align}

\section{Closure of the moment equations}
\label{closure-sec}

In this section we describe the necessary steps to obtain closure and complete the transport time step to obtain the equilibrium evolution.  At this point we have the equations for turbulence (the electron and ion gyro-kinetic equations), and the equations for transport (heat and density).  The prescription for solving this system is to evolve the gyro-kinetic distribution function over many {\em turbulent} time steps and use the averaged solution as input for the transport equations.  In other words the turbulence should evolve to generate an average steady-state turbulent transport (if, however, the transport grew unbounded and did not reach this steady-state, the gyro-kinetic approximation would become invalid altogether).  In principle one would then proceed to evolve the density and pressure (temperature) profiles a {\em transport} timestep to obtain the new equilibrium profiles.  However, this step is not quite complete because we have left to solve for the time evolution of the equilibrium field geometry.  We now show how the evolution of the equilibrium geometry is obtained to complete closure.

\subsection{Flux surface motion}

The partial time derivatives such as $\partial n_0/\partial t$ have been thus far implicitly taken at fixed ${\bf r}$.  However, since the temperature and density are flux surface quantities and the flux surfaces themselves move on the turbulent time scale, it can be useful to know how density and temperature of each species evolves on a fixed flux surface.  The following argument can be found in Ref.~\cite{bernstein}.  For a fixed point in space, the motion of flux surfaces must satisfy a continuity equation  of the following form:

\begin{equation}
\frac{\partial \psi}{\partial t} + {\bf v}_{\psi}\cdot\nabla\psi = 0
\end{equation}

\noindent where ${\bf v}_{\psi}$ is the flux surface velocity.  Since motion parallel to surfaces is not physically meaningful we can write without loss of generality that

\begin{equation}
{\bf v}_{\psi} = - \frac{\partial \psi}{\partial t}\frac{\nabla \psi}{|\nabla \psi|^2}
\end{equation}

\noindent Thus we arrive at the following identity

\begin{equation}
\left(\frac{\partial}{\partial t}\right)_{\bf r} = \left(\frac{\partial}{\partial t}\right)_{\psi} - {\bf v}_{\psi}\cdot\nabla
\end{equation}

\noindent Combining this result with Eqn.~\ref{flux-div} we have

\begin{equation}
\fluxavg{ \left(\frac{\partial F}{\partial t}\right)_{\bf r}} = \left(\frac{\partial}{\partial t}\right)_{\psi}\fluxavg{F} + D_{\psi}\fluxavg{F\frac{\partial\psi}{\partial t}}\label{flux-motion}
\end{equation}

\noindent An expression for $\frac{\partial \psi}{\partial t}$ follows directly from taking the inner produce of $\nabla\psi$ with Faraday's Law for the equilibrium field $-\partial{\bf B_0}/\partial t = \nabla\times{\bf E_{0i}}$ (see also \cite{hinton})

\begin{equation}
\frac{\partial \psi}{\partial t} = -R{\bf \hat{\phi}}\cdot{\bf E_{0i}}\label{psi_rate}
\end{equation}

\noindent where we define ${\bf E_{0i}} = -\partial{\bf A_0}/\partial t$, the equilibrium inductive electric field.  We now use Eqn.~\ref{flux-motion} and Eqn.~\ref{psi_rate} to rewrite the transport equations.   Using the fact that density and temperature are flux surface quantities we write

\begin{equation}
\annavg{ \frac{\partial n_{0s}}{\partial t} } = \fluxavg{ \frac{\partial n_{0s}}{\partial t} } = \frac{d n_{0s}}{d t} - D_{\psi}\fluxavg{n_{0s}R{\bf \hat{\phi}}\cdot{\bf E_{0i}}}
\end{equation}

\noindent and

\begin{equation}
\frac{3}{2}\annavg{ \frac{\partial n_{0s}T_{0s}}{\partial t}} = \frac{3}{2}\fluxavg{ \frac{\partial n_{0s}T_{0s}}{\partial t}} = \frac{3}{2}\frac{d n_{0s}T_{0s}}{d t} - D_{\psi}\fluxavg{\frac{3}{2}n_{0s}T_{0s}R{\bf \hat{\phi}}\cdot{\bf E_{0i}}}
\end{equation}

\noindent Substituting these expressions into the density (Eqn.~\ref{density-transport}) and heat (Eqn.~\ref{heat-transport}) transport equations  gives cancellation with the ware pinch terms:

\begin{equation}
\frac{d n_{0s}}{d t} 
= D_{\psi}\int d^{3}{\bf v} \fluxavg{R{\hat \phi}\cdot\left[\nabla \chi h_{gk} + \frac{m{\bf v}}{q}\;C(\rhovec \cdot\nabla F_m - h_{nc}, F_m)\right]}
\label{density-transport2}
\end{equation}

\noindent and

\begin{align}
&\frac{3}{2}\frac{d n_{0s} T_{0s}}{d t} =
\nonumber \\ & D_{\psi}\int d^{3}{\bf v} \fluxavg{\frac{mv^2R}{2}{\bf \hat{\phi}}\cdot\left[\nabla \chi h_{gk} 
- \frac{m}{2q}{\bf v}C(h_{nc} - \rhovec \cdot \nabla F_m, F_m)\right]}
\nonumber \\ & - \int d^{3}{\bf v} \fluxavg{\frac{h_{gk}T}{F_m}R{\hat \phi}\cdot\nabla \chi \frac{\partial F_m}{\partial \psi}}
- \int d^{3}{\bf v} \fluxavg{\frac{h_{gk}T}{F_0}C(h_{gk})}
\nonumber \\ & + \int d^{3}{\bf v}\fluxavg{\frac{T}{F_m}\left(\frac{mR}{q}\frac{\partial F_m}{\partial \psi}{\bf \hat{\phi}\cdot b_0}v_{\parallel}C(h_{nc}) - h_{nc}C(h_{nc})\right)}
\nonumber \\ & - \fluxavg{T\frac{\partial \psi}{\partial t}\frac{\partial n_{0s}}{\partial \psi}}
\nonumber \\ & + n_{0s} \nu_{{\cal E}}^{sr}(T_r - T_s).
\label{heat-transport2} 
\end{align}

\noindent Notice that we have adopted the flux surface average in place of the annulus volume average for all terms in the transport equations (including the turbulent terms).  In fact, the formal smoothing procedure we have been employing thus far is more than needed.  The toroidal symmetry of our system of equations ensures that the flux surface average will be sufficient as an effective ensemble average.  In other words subsequent smoothing over intermediate time and space ($\Delta \psi$) intervals are redundant and we may compactly express the final transport equations only in terms of the flux surface average from neoclassical theory.  Consider now the form of the equilibrium magnetic field given bye Eqn.~\ref{B0-eqn} in Sec.~\ref{axi-symm-sec}:

\begin{equation}
{\bf B_0} = \nabla\psi\times\nabla\phi + I(\psi)\nabla\phi\nonumber
\end{equation}

\noindent The equilibrium field is determined when one knows the spatial dependence of the poloidal flux function $\psi(R,Z)$ and also the toroidal field function $I(\psi)$.  It is a natural and common choice to therefore solve for these quantities.  This is essentially done using Faraday's Law and the Grad Shafranov Equation.  The equations obtained, however, depend on the flux volume averaged quantity $\annavg{{\bf B_0\cdot E_{0i}}}$ which must be obtained by solving the neoclassical equations Eqn.~\ref{gyro-slow} and Eqn.~\ref{gyro-slow-e}.

\subsection{Evolution of the toroidal field function $I(\psi)$}

Our stated aim is to now solve for the time evolution of functions $I(\psi)$ and $\psi(R, Z)$.  We follow Bernstein \cite{bernstein} and take the toroidal component of Faraday's Law to obtain

\begin{equation}
\frac{\partial I}{\partial t}R^{-2} = \nabla\cdot(\nabla\phi\times{\bf E_{0i}})
\end{equation}

\noindent Applying the flux surface average to this equation then yields

\begin{equation}
\fluxavg{\frac{\partial I}{\partial t}R^{-2}} = D_{\psi}\fluxavg{({\bf E_{0i}}\cdot{\bf B_0} - \frac{\partial\psi}{\partial t}R^{-2} I)}
\end{equation}

\noindent Finally, the last term will drop out upon using Eqn.~\ref{flux-motion}

\begin{equation}
\left(\frac{\partial}{\partial t}\right)_{\psi}\fluxavg{\frac{I}{R^2}} = D_{\psi}\fluxavg{{\bf B_0}\cdot{\bf E_{0i}}}\label{i-rate}
\end{equation}

\noindent Which can be written in an alternate form by using the chain rule on the left hand side and another application of Eqn.~\ref{flux-motion} 

\begin{equation}
\left(\frac{\partial I}{\partial t}\right)_{\psi} = \frac{I}{\fluxavg{R^{-2}}}D_{\psi}\fluxavg{R^{-2}\frac{\partial\psi}{\partial t}} + \frac{1}{\fluxavg{R^{-2}}}D_{\psi}\fluxavg{{\bf B_0}\cdot{\bf E_{0i}}}
\label{i-rate2}
\end{equation}

\noindent This equation gives a way to solve for $I(\psi)$ at the next time step.  Combined with the Grad Shafranov equation and the heat transport equation (Eqn.~\ref{heat-transport2}) one can calculate $\psi(R,Z)$ at the next time step.  The Grad Shafranov equation is

\begin{equation}
R^2\nabla\cdot(R^{-2}\nabla\psi) = -I\frac{\partial I}{\partial\psi} - \mu_0 R^2\frac{\partial P}{\partial \psi}\label{grad-shaf}
\end{equation}

\noindent where 

\begin{equation}
P_0 = \displaystyle\sum_s n_{0s}T_{0s}
\end{equation}

\noindent is the total equilibrium pressure.  However, we are not quite done because we have not yet shown how one obtains the quantity $\fluxavg{{\bf B_0}\cdot{\bf E_{0i}}}$.

\subsection{The neoclassical equation and obtaining $\fluxavg{{\bf B_0}\cdot{\bf E_{0i}}}$}

\noindent The quantity $\fluxavg{{\bf B_0}\cdot{\bf E_{0i}}}$ is not known {\it a priori}.  It depends on the inductive equilibrium field which is dependent on the evolution of the equilibrium magnetic field for which we are seeking a solution.  The quantity $\fluxavg{{\bf B_0}\cdot{\bf E_{0i}}}$ can be determined, however, by solving the neoclassical equation.  One uses the solution to calculate the flux volume averaged parallel current which is known by Ampere's law in terms of spatial derivatives of the equilibrium magnetic current.  We shall see that this gives an expression for $\fluxavg{{\bf B_0}\cdot{\bf E_{0i}}}$ in terms of the solution to the neoclassical equation and known quantities.  We recall the neoclasical equation for ions (electrons are treated analogously)

\begin{equation}
v_{\parallel}{\bf b}_0\cdot \nabla h_{nc} - C(h_{nc}) = -{\bf v}_D \cdot\nabla F_m + q\frac{F_m}{T_0 }v_{\parallel}{\bf b_0}\cdot{\bf E_{0i}}
\end{equation}

\noindent The approach (see \cite{hinton} and \cite{bernstein}) is to transform $h_{nc}$ so as to obtain an equation involving $\fluxavg{{\bf B_0}\cdot{\bf E_{0i}}}$ and not the full ${\bf B_0}\cdot{\bf E_{0i}}$.  The transformation we will use is

\begin{equation}
h_{nc} = \tilde{h}_{nc} + \frac{qF_m}{T_0}\int^{\theta}\frac{d\theta^{\prime}}{{\bf b_0}\cdot\nabla\theta^{\prime}}\left[ {\bf b_0}\cdot{\bf E_{0i}} - B_0\frac{\fluxavg{{\bf B_0}\cdot{\bf E_{0i}}}}{\fluxavg{B_0^2}}\right]
\end{equation}

\noindent where $\theta$ is the poloidal magnetic coordinate defined in Sec.~\ref{mag-coord-sec}.  Note that the second term in the integral is determined by the requirement that $h_{nc}$ is periodic in $\theta$.  Note also that the integral term is proportional to the maxwellian of the species.  If one examens the transport equations it is apparent that this maxwellian part will yield zero since it in collisional terms involving other maxwellians.  Therefore, we will be able to use $\tilde{h}_{nc}$ in place of $h_{nc}$ wherever it appears.  Due to axi-symmetry, we can write the first term of the neoclassical equation as $v_{\parallel}{\bf b_0}\cdot\nabla\theta\frac{\partial h_{nc}}{\partial\theta}$.  The resulting equation for $\tilde{h}_{nc}$ is

\begin{equation}
v_{\parallel}{\bf b_0}\cdot\nabla\tilde{h}_{nc} + C(\tilde{h}_{nc}) = -{\bf v}_D\cdot\nabla F_m + v_{\parallel}\frac{qF_m}{T_0}B_0\frac{\fluxavg{{\bf B_0}\cdot{\bf E_{0i}}}}{\fluxavg{B_0^2}}
\end{equation}

\noindent Due to linearity, we may further split up the solution into two pieces for the two driving terms.

\begin{equation}
\tilde{h}_{nc} = g_1\fluxavg{{\bf B_0}\cdot{\bf E_{0i}}} + g_2
\end{equation}

\noindent where $g_1$ and $g_2$ separately satisfy the equations

\begin{eqnarray}
v_{\parallel}{\bf b_0}\cdot\nabla g_1 - C(g_1) = v_{\parallel}\frac{qF_m}{T_0}\frac{B_0}{\fluxavg{B_0^2}} \nonumber
\\v_{\parallel}{\bf b_0}\cdot\nabla g_2 - C(g_2) = -{\bf v}_D\cdot\nabla F_m
\end{eqnarray}

\noindent These equations must be solved to obtain $\tilde{h}_{nc}$.  The flux volume averaged parallel current can be expressed in terms of these solutions.

\begin{equation}
\fluxavg{J_{\parallel}} = \fluxavg{ \int d^3{\bf v}\;v_{\parallel}(q\tilde{h}_{nci} - e\tilde{h}_{nce})}
\end{equation}

\noindent this is then equated to an expression for $\fluxavg{J_{\parallel}}$ obtained from the Grad-Shafranov Equation and the parallel component of Ampere's Law

\begin{equation}
\fluxavg{J_{\parallel}} = \fluxavg{\frac{\partial I}{\partial \psi}B_0 + \frac{I}{B_0}\frac{\partial P}{\partial \psi} }
\end{equation}

\noindent The result is an expression for $\fluxavg{{\bf B_0}\cdot{\bf E_{0i}}}$

\begin{equation}
\fluxavg{{\bf B_0}\cdot{\bf E_{0i}}} = \frac{\fluxavg{\frac{\partial I}{\partial \psi}B_0 + \frac{I}{B_0}\frac{\partial P}{\partial \psi} } - \fluxavg{ \int d^3{\bf v}\;v_{\parallel}(qg_{2i} - eg_{2e})}}{\fluxavg{ \int d^3{\bf v}\;v_{\parallel}(qg_{1i} - eg_{1e})}}
\label{e-parallel-b}
\end{equation}

\subsection{Closure}

The equations needed for closure of transport are Eqn.~\ref{density-transport2}, Eqn.~\ref{heat-transport2}, Eqn.~\ref{i-rate}, Eqn.~\ref{grad-shaf} and Eqn.~\ref{e-parallel-b}.  The density transport equation can be integrated a time step independently.  The others, however are mutually dependent and must be solved simultaneously and consistently to obtain $P(\psi)$, $I(\psi)$ and $\psi(R,Z)$ at the next time step.  The heat transport equation (Eqn.~\ref{heat-transport2}) depends on $\fluxavg{\partial\psi/\partial t}$ and the evolution equation for $I(\psi)$ depends on $\fluxavg{R^{-2}\partial\psi/\partial t}$.  The value of $\partial\psi/\partial t$ obtained by evolving $\psi(R,Z,t)$ using the Grad Shafranov equation must therefore be checked for consistency with the value used in Eqn.~\ref{i-rate} and Eqn.~\ref{e-parallel-b}.

\section{Entropy balance}
\label{entropy-sec}

\noindent The purpose of this section is to calculate the entropy production of
the plasma.  Standard theory predicts that the time change of the
local entropy will be due to either convection or collisions.  

\begin{equation}
\annavg{\frac{\partial S}{\partial t}} = D^{(\Delta)}_{\psi} \int d^{3}{\bf
  v} \annavg{\nabla\psi\cdot\left({\bf v}f \ln f\right)}  -
  \int d^{3}{\bf v} \annavg{C(f, f) \ln f}
\label{entropy_start}
\end{equation}

\noindent By manipulating the left hand side of Eqn~(\ref{entropy_start}), we
can insert the density and temperature evolution equations we have
derived in Sec.~\ref{transport-sec}.  

\begin{align}
&\annavg{\frac{dS}{dt}} \approx - \int d^3{\bf
  v}\annavg{\frac{\partial (F_m ln F_m)}{\partial t}} \nonumber \\
&=  -\annavg{\frac{\partial n_0}{\partial t}}\left[
  \ln n_0 + \frac{3}{2} \ln (\frac{m}{2 \pi}) - \frac{3}{2} \ln T + 1 \right]
 + \frac{3}{2T} \annavg{\frac{\partial (n_0T_0)}{\partial t}}
\label{entropy-f}
\end{align}

\noindent Our goal will be to show that the terms we have calculated for the
density and temperature transport indeed produce only convection or
entropy production through collisions.  In order to achieve this
result, we must manipulate Eqn~(\ref{pre-heat-transport}), the
temperature transport equation.  The details
of the manipulation can be found in Sec.~(\ref{heat-manip-sec}).

\begin{align}
&\annavg{\frac{3}{2}\frac{\partial (n_{0s} T_{0s})}{\partial t}} = D^{(\Delta)}_{\psi}\int d^3{\bf v}\annavg{\frac{mR}{2}F_mv^2{\bf \hat{\phi}}\cdot\frac{\partial {\bf A_0}}{\partial t}} \nonumber \\ 
& + D^{(\Delta)}_{\psi} \int d^3{\bf v} \annavg{ \frac{mv^2R}{2}(\nabla \chi \cdot {\bf \hat{\phi}})h_{gk}} \nonumber \\ 
& - D^{(\Delta)}_{\psi} \int d^3{\bf v} \annavg{\frac{m^2v^2R}{2q}{\bf v}\cdot {\hat \phi}C(h_{nc} - \rhovec \cdot \nabla F_m, F_m)} \nonumber \\ 
& - \int d^3{\bf v}\annavg{ \frac{h_{gk}T}{F_m}R{\hat \phi}\cdot\nabla \chi \frac{\partial F_m}{\partial \psi} - \frac{h_{gk}T}{F_0}\gyroavg{C(h_{gk})}} \nonumber \\ 
& + \int d^3{\bf v}\annavg{\frac{T}{F_m}\left(\frac{mR}{q}\frac{\partial F_m}{\partial \psi}{\bf \hat{\phi}\cdot b_0}v_{\parallel}C(h_{nc}) - h_{nc}C(h_{nc})\right)} \nonumber \\ 
& - \int d^3{\bf v}\annavg{RT{\bf \hat{\phi}}\cdot\frac{\partial {\bf A_0}}{\partial t}\frac{\partial F_m}{\partial \psi}} \nonumber \\ 
& + n_{0s} \nu_{{\cal E}}^{sr}(T_r - T_s).
\label{heat-transport} 
\end{align}

\noindent We can now substitute in from Eqns~(\ref{density-transport}) and
(\ref{heat-transport}) into Eqn~(\ref{entropy-f}). 

\begin{align}
&\annavg{\frac{\partial S}{\partial t}} = \nonumber \\
&\int d^{3}{\bf v}\left( \ln n_0 + \frac{3}{2}\ln(\frac{m}{2\pi}) - \frac{3}{2}\ln T + 1\right) \nonumber \\  
&D^{(\Delta)}_{\psi}\annavg{ R{\hat \phi}\cdot(\nabla \chi h_{gk} + \frac{\partial {\bf A_0}}{\partial t}F_m + \frac{m{\bf v}}{q}\;C(\rhovec \cdot\nabla F_m - h_{nc}, F_m))} \nonumber \\
&+ \frac{1}{T}\left[ D^{(\Delta)}_{\psi}\int d^{3}{\bf v}\annavg{
\frac{mv^2R}{2}{\bf \hat{\phi}}\cdot\left (\nabla \chi h_{gk} + \frac{\partial {\bf A_0}}{\partial t}F_m 
- \frac{m{\bf v}}{q}C(h_{nc} - \rhovec \cdot \nabla F_m, F_m)\right )}\right]\nonumber \\ 
&+ \int d^{3}{\bf v} \annavg{ \frac{\partial \ln F_m}{\partial \psi}R{\hat \phi}\cdot\left(
 - \nabla \chi h_{gk}  - \frac{\partial {\bf
     A_0}}{\partial t}F_m 
+ \frac{mv_\parallel {\bf b_0}}{q} C(h_{nc})\right)} \nonumber \\
&+ \int d^3{\bf v}\annavg{-\frac{h_{nc}}{F_0} \left<C(h_{nc}, F_m)\right> -
  \frac{h_{gk}}{F_0}\left <C(h_{gk},F_m)\right >} + n_{0s} \nu_{{\cal
      E}}^{sr}\frac{(T_r - T_s)}{T_s} \label{entropy-mid} 
\end{align} 

\noindent At the moment, the last line of Eqn~(\ref{entropy-mid}) contains terms
that are neither convection or collisional.  What is left to do is pull out the $D^{(\Delta)}_{\psi}$ to act on the entire second line, put the $\frac{1}{T}$ inside the $D^{(\Delta)}_{\psi}$ of the second line, and evaluate $\partial F_m/\partial \psi$ on the third line.  We find


\begin{align}
&\annavg{\frac{\partial S}{\partial t}} = \nonumber \\
&-D^{(\Delta)}_{\psi} \int d^3{\bf v} (\ln F_m + 1) \nonumber \\
&\annavg{R{\hat \phi}\cdot(\nabla \chi h_{gk} + \frac{\partial {\bf A_0}}{\partial t}F_m + \frac{m{\bf v}}{q}\;C(\rhovec \cdot\nabla F_m - h_{nc}, F_m))} \nonumber \\
&- \int d^3{\bf v}\annavg{\frac{\delta f_1}{F_m} \left<C(\delta f_1, F_m)\right>} + n_{0s} \nu_{{\cal E}}^{sr}\frac{(T_r - T_s)}{T_s} \nonumber \\
\end{align}

\noindent where we take $\delta f_1 = -\rhovec\cdot\nabla F_m + h_{nc} + h_{gk}$, noting that here the Boltzmann part does not contribute to collisional entropy production.  This form is exactly what we expect from Eqn.~\ref{entropy_start}.  The last two terms can be easily identified as the perturbative expansion of $fC(f,f)$ with the final term giving the inter-species collisional entropy exchange.

\section{Particle transport calculation details}
\label{particle-appendix}

This section will demonstrate how to obtain
Eqn~(\ref{density-transport}).  We begin with
Eqn~(\ref{density_moment-appendix}).

\begin{multline}
\int d^3{\bf v}\annavg{\frac{\partial f_{s}}{\partial t}} = \\
D^{(\Delta)}_{\psi}\int d^3{\bf v}\annavg{\frac{mR{\bf v}\cdot {\hat \phi}}{q}
\left[\frac{\partial f_{s}}{\partial t} + {\bf v}\cdot\nabla f_s + \frac{q}{m}({\bf E} + {\bf v}\times\delta{\bf B})\cdot \frac{\partial f_s}{\partial {\bf v}} - {C(f,f)}\right]}
\label{density_moment2}
\end{multline}

\noindent Our goal will be to use the solution
for the distribution function given by Eqn~(\ref{fform}) and evaluate
the terms in Eqn~(\ref{density_moment2}).  Plugging the solution for
$f$ into the left hand side of Eqn~(\ref{density_moment2}), we obtain up to
${\cal O}({\epsilon^2})$ 

\begin{equation}
\int d^3{\bf v}\annavg{ 
\frac{\partial f_{s}}{\partial t}}= \annavg{\frac{\partial n_{0s}}{\partial t}} +
\int d^3{\bf v}\annavg{ \frac{\partial
  h}{\partial t}} +
\int d^3{\bf v}\annavg{ \frac{\partial \delta f_2}{\partial t}}
\label{density_time-appendix}
\end{equation}

\noindent We now have in our equation $\frac{\partial n_{0s}}{\partial t}$, which is the time
evolution of the density.  The remaining terms in Eqn
\ref{density_time-appendix} will drop out upon time averaging. The first term on the right hand side of Eqn.~\ref{density_moment2} is treated
analogously to Eqn.~\ref{density_time-appendix} but the order is one smaller
because of the preceding factors.  Thus, it contributes
nothing. The next term to calculate in Eqn.~\ref{density_moment-appendix} is also zero as we will see.

\begin{equation}
\int d^3{\bf v}\annavg{ 
\frac{mR}{q}{\bf v}\cdot {\hat \phi}({\bf v}\cdot\nabla)(F_0 + h + \delta f_2)}
\label{convective-term-density-moment}
\end{equation}

\noindent The $\delta f_2$ can be rewritten as a perfect divergence to the order we are calculating and will drop in ordering after averaging it over space.    The $F_0$ term is odd in velocity space and will integrate away.  The gyrokinetic part of the $h$ term will space averages to zero, and all that is left is a neoclassical term.  We rewrite this by defining a symmetric pressure tensor ${\bf P}$.

\begin{equation}
{\bf P} = \int d^3{\bf v}\;\;{\bf vv}h_{nc} = \int d^3{\bf v} \left[v_{\parallel}^2{\bf \hat{b}}_0{\bf \hat{b}}_0 + \frac{v_{\perp}^2}{2}({\bf I} - {\bf \hat{b}}_0{\bf \hat{b}}_0)\right]h_{nc} + {\cal O}(\epsilon^2)
\end{equation}

\noindent so that we have

\begin{equation}
\frac{m}{q}\int d^3{\bf v} \annavg{\hat{\phi}R\cdot(\nabla\cdot{\bf P})} = \frac{m}{q}\int d^3{\bf v} \annavg{ \nabla\cdot({\bf P}\cdot\hat{\phi}R) - {\bf P}:{\bf L}}
\end{equation}

\noindent where ${\bf L}$ is the antisymmetric tensor ${\bf \nabla(\hat{\phi}R)} = {\bf \hat{\phi}}{\bf\hat{R}} - {\bf\hat {R}}{\bf\hat{\phi}}$. (Note: ${\bf\hat{R}}$ is the radial unit vector in cylindrical coordinates.)  The second term is identically zero being the double dot product of a symmetric and an antisymmetric tensor.  The first term is a perfect divergence and we again use the divergence theorem to evaluate it.  Referring to the above expression for ${\bf P}$ we see that ${\bf P}\cdot{\bf\hat{\phi}}$ is everywhere parallel to the flux surface (perpendicular to the surface element vector) so the divergence theorem integral yields zero.  The field terms in Eqn.~\ref{density_moment2} can be written as 

\begin{equation}
\int d^3{\bf v}\annavg{ 
\frac{mR}{q}{\bf v}\cdot {\hat \phi}
\frac{q}{m}({\bf E} + {\bf v}\times\delta{\bf B})\cdot
\frac{\partial}{\partial {\bf v}} (F_0 + h)}
\end{equation}

\noindent When considering $F_0$ terms, the $-\rhovec\cdot\nabla F_m$ term is shown to be too small in order since the preceding perturbed fields have a small patch average.  Since $\frac{\partial F_m}{\partial {\bf v}} \propto {\bf v} F_m$ we worry only about the electric field for the $F_m$ and the $\delta f_1$ Boltzman correction terms and write

\begin{align}
\int d^3{\bf v}\annavg{ \frac{mR}{q}{\bf v}\cdot {\hat \phi}
\frac{q}{m}{\bf E}\cdot \frac{\partial F_m}{\partial {\bf v}}[1-\frac{q\varphi}{T}]} \nonumber \\
= \int d^3{\bf v}\annavg{ R(\nabla\varphi + \frac{\partial {\bf A}}{\partial t})\cdot {\hat \phi}F_m} \nonumber \\
= \int d^3{\bf v}\annavg{F_mR{\hat \phi}\cdot \frac{\partial {\bf A_0}}{\partial t}}
\label{density_field_eq}
\end{align}

\noindent Were we have used that $R{\bf \hat{\phi}}\cdot\nabla\varphi$ integrates to zero over the annulus.  What is left will time average to be negligible.  We now write ${\bf E} + {\bf v}\times\delta{\bf B}$ as $-\nabla\chi - {\bf v}\cdot\nabla{\bf\delta A}$ and calculate the h
  term by integrating by parts in velocity, so that
\begin{align}
&\int d^3{\bf v}\annavg{ 
R{\bf v}\cdot {\hat \phi}(-\nabla\chi - {\bf v}\cdot\nabla{\bf\delta A})\cdot \frac{\partial}{\partial {\bf v}}h_{gk}} = 
  \int d^3{\bf v}\annavg{ 
R{\bf \hat{\phi}}\cdot\nabla\chi 
h_{gk}}
\label{density_field-appendix}
\end{align}

\noindent where the ${\bf v}\cdot\nabla{\bf A}$ term is removed in two parts.
First the $v_{\parallel}{\bf b_0}\cdot\nabla{\bf\delta A}$ is written
as a total divergence with negligible error and vanishes due to
divergence theorem.  Second, the ${\bf v_{\perp}}\cdot\nabla{\bf\delta
  A}$ is removed by employing Eqn.~\ref{deexi} after integration by
parts.  (Note that what we are interested in is the gyrokinetic part
of $h$--the term with $h_{nc}$ does not enter at this order,
because the patch average of $\nabla\chi$ and ${\bf v}\cdot\nabla{\bf A}$
are small.)  The final term in Eqn.~\ref{density_moment2} simply
involves plugging in values for the distribution function. 
\begin{equation}
\int d^3{\bf v}\annavg{ 
\frac{mR}{q}{\bf v}\cdot {\hat \phi}C(\rhovec \cdot
\nabla F_m - h_{nc}, F_m)}
\label{density_collision-appendix}
\end{equation}

\noindent Recall that we denote $F_m$ as the Maxwellian so as to distinguish it from the
modified Maxwellian $F_0$ defined in Eqn.~\ref{modmax}.  Collecting
terms, and recalling the $D^{(\Delta)}_{\psi}$ in front of the integral inside Eqns~(\ref{density_field_eq}), (\ref{density_field-appendix}), and (\ref{density_collision-appendix}) from Eqn~(\ref{density_moment2}) , we have the
transport equation for particles. 

\begin{multline}
\annavg{\frac{\partial n_{0s}}{\partial t}}= \\
D^{(\Delta)}_{\psi}\int d^3{\bf v}\annavg{R{\hat \phi}\cdot\left[F_m \frac{\partial {\bf A_0}}{\partial t} + \nabla \chi h_{gk} + \frac{m{\bf v}}{q}\;C(\rhovec \cdot\nabla F_m - h_{nc}, F_m)\right]}
\end{multline}

\section{Heat transport calculation details}
\label{heat-appendix}

In this section, our goal will is to substitute our solution for the distribution function into Eqn~(\ref{heating_moment-appendix}).  

\begin{align}
\int d^3{\bf v}\annavg{\frac{mv^2}{2} 
\frac{\partial f}{\partial t}} = 
& -\frac{m}{2}D^{(\Delta)}_{\psi} \int d^3{\bf v}\annavg{
  \frac{mv^2R}{q}{\bf v}\cdot {\hat 
    \phi} \left[\frac{q}{m}({\bf v}\times{\bf B}_0)\cdot \frac{\partial
      f}{\partial {\bf v}}\right]} \nonumber \\
& + \int d^3{\bf v}\annavg{\left[ q{\bf E}\cdot{\bf v}f + \frac{mv^2}{2}C(f,f)\right]}
\label{heating_moment2}
\end{align}

\noindent We begin with the left hand side of Eqn~(\ref{heating_moment2}) will yield 

\begin{equation}
\annavg{\frac{3}{2}\frac{\partial(n_{0s}T_{0s})}{\partial t}}
+ \int d^3{\bf v}\annavg{ 
\frac{1}{2}m{\bf v }^{2}
\frac{\partial}{\partial t}(h + \delta f_2) }
\label{heating_time-appendix}
\end{equation}

\noindent The other terms in the time derivative will be negligible after time average.  The first term on the right hand side of Eqn~(\ref{heating_moment2}) will be treated very similar to the particle
transport starting with Eqn~(\ref{particle_start}).  

\begin{align}
&\frac{m}{2}D^{(\Delta)}_{\psi}\int d^3{\bf v}\annavg{v^2\frac{mR}{q}{\bf v}\cdot {\hat \phi}\left[\frac{q}{m}({\bf v}\times{\bf B}_0)\cdot \frac{\partial f_s}{\partial {\bf v}}\right]} \nonumber\\
&= -D^{(\Delta)}_{\psi}\int d^3{\bf v} \nonumber \\
&\annavg{v^2\frac{m^2R}{2q}{\bf v}\cdot {\hat \phi}\left[\frac{\partial f_{s}}{\partial t} + {\bf v}\cdot\nabla f_s + \frac{q}{m}({\bf E} + {\bf v}\times\delta{\bf B})\cdot \frac{\partial  f_s}{\partial {\bf v}} - C(f,f)\right]}
\label{first-term-heating}
\end{align}

\noindent The significant difference between this equation and the particle transport term is
obviously the $v^2$ inside the integral.  Fortunately, all the terms that were
negligible in the density transport will still be negligible as we
will see.  The time derivative will time average away.  We can remove
all parts of the next term  except the neoclassical part of $h$
following the same arguments used in particle transport.  What is left
we manipulate using tensor notation.  We define a symmetric tensor
${\bf R}$ ( the ``energy-weighted stress tensor'', \cite{hinton}). 

\begin{equation}
{\cal {\bf R}} = \int d^3{\bf v}\;\;(mv^2/2){\bf vv}h_{nc} = \int d^3{\bf v} (mv^2/2)\left[v_{\parallel}^2{\bf \hat{b}}_0{\bf \hat{b}}_0 + v_{\perp}^2({\bf I} - {\bf \hat{b}}_0{\bf \hat{b}}_0)\right]h_{nc} \nonumber
\end{equation}

\noindent The argument is finished exactly as before.  The field terms in Eqn~(\ref{first-term-heating}) will produce transport.  

\begin{equation}
-\int d^3{\bf v}\annavg{
\frac{mv^2}{2} \frac{mR}{q}{\bf v}\cdot {\hat \phi}
\left[\frac{q}{m}({\bf E} + {\bf v}\times{\bf \delta B})\cdot \frac{\partial}{\partial {\bf v}}\right](F_0 + h_{gk})}
\end{equation}

\noindent First we calculate the $F_0$ part using the methods from particle transport.

\begin{equation}
-\int d^3{\bf v}\annavg{\frac{mv^2R}{2}F_m{\bf
 \hat{\phi}}\cdot\frac{\partial {\bf A_0}}{\partial t}}
\label{stuff66a}
\end{equation}

\noindent What is left is

\begin{equation}
-\int d^3{\bf v}\annavg{
\frac{mv^2}{2} \frac{mR}{q}{\bf v}\cdot {\hat \phi}
\left[\frac{q}{m}(-\nabla\chi - {\bf v}\cdot\nabla{\bf \delta A})\cdot \frac{\partial}{\partial {\bf v}}\right]h_{gk}}
\label{stuff66}
\end{equation}

\noindent For the $h_{gk}$ part in Eqn(~\ref{stuff66}), the integration by parts
is more complex than with the analogous term in the particle transport
equation.  This is due to the $v^2$ factor.  The ${\bf v}\cdot\nabla{\bf \delta A}$
term is removed as with the density transport case of
Eqn~(\ref{density_field-appendix}).  The result is 

\begin{eqnarray}
-D^{(\Delta)}_{\psi}\int d^3{\bf v}\annavg{\frac{mv^2R}{2}(\nabla \chi \cdot {\bf \hat{\phi}})h_{gk} + mR{\bf v} \cdot {\bf \hat{\phi}}({\bf v} \cdot \nabla\varphi)h_{gk}}
\label{heating_convection-appendix}
\end{eqnarray}

\noindent We will later see that the second term of Eqn~(\ref{heating_convection-appendix})
will cancel with part of the electric field in
Eqn~(\ref{heating_moment2}).  The collisional terms of Eqn~(\ref{first-term-heating})
work similarly to density transport.

\begin{eqnarray}
D^{(\Delta)}_{\psi}\int d^3{\bf v}\annavg{
\frac{m^2v^2R}{2q}{\bf v}\cdot {\hat \phi}C(h_{nc} - \rhovec \cdot \nabla F_m, F_m)}
\label{heating_vdot-appendix}
\end{eqnarray}

\noindent This covers all of the terms from Eqn~(\ref{first-term-heating}).  The next step is
to return to Eqn~(\ref{heating_moment2}) and calculate the electric
field term.  The scalar potential
part of the electric field is an order larger than the $\frac{\partial
  A}{\partial t}$ term, but as we will see most of it will cancel.
This will result in the part of the scalar potential that does not
average away acting on the same order as $\frac{\partial A}{\partial
  t}$.

\begin{align}
&\int d^3{\bf v}\annavg{ q {\bf v}\cdot\nabla\varphi f} =  
\int d^3{\bf v}\annavg{ q\left[-  \frac{\partial\varphi}{\partial t} f  + 
( \frac{\partial\varphi}{\partial t} + {\bf v}\cdot\nabla\varphi) f\right]}\nonumber \\
&=  \int d^3{\bf v}\annavg{q\left[- \frac{\partial\varphi}{\partial t}  f - 
( \frac{\partial f}{\partial t} + {\bf v}\cdot\nabla f) \varphi +  \frac{\partial(\varphi f)}{\partial t} + \nabla(\varphi f)\cdot {\bf v}\right]} \nonumber \\
&=  q\int d^3{\bf v}\annavg{ - \frac{\partial\varphi}{\partial t} f -  \frac{q}{m}\frac{\partial}{\partial {\bf v}}\cdot \left[( {\bf E} + \frac{{\bf v}\times {\bf B}}{c})\varphi f \right] } \nonumber \\
& + q\int d^3{\bf v}\annavg{ C(f,f) \varphi +  \frac{\partial(\varphi f)}{\partial t} + ({\bf v}\cdot \nabla)(\varphi f)} \nonumber \\
&=  \int d^3{\bf v}\annavg{ 
q\left[-  \frac{\partial\varphi}{\partial t} f +  \frac{\partial(\varphi  f)}{\partial t}  
+ ({\bf v}\cdot \nabla)(\varphi f) \right]}
\label{mess}
\end{align}

\noindent where we have integrated by parts in time and space between lines one 
and two, between lines two and three we use $\frac{df}{dt}=C(f,f)$ and, 
finally, we use gauss's law in velocity space and that all collisions conserve particles 
to obtain the final line. The second to last term that we obtain can be time
averaged to zero.  We find that the very last term will cancel
the last term of Eqn~(\ref{heating_convection-appendix}) when we plug in $h$ for f
in the equation.  The $F_0$ part of the final term is removed by $\vartheta$ integration and patch averaging to higher order.  The remaining part of this final term is manipulated as follows (note that $h_{nc}$ does not enter due to the small patch average of $\varphi$):

\begin{align}
&\int d^3{\bf v}\annavg{ 
q({\bf v}\cdot \nabla)(\varphi h_{gk}) }
=D^{(\Delta)}_{\psi}\int d^3{\bf v}\annavg{ 
q {\bf v}\cdot\nabla\psi(\varphi h_{gk}) }
\nonumber \\
&= D^{(\Delta)}_{\psi}\int d^3{\bf v}\annavg{ 
qR {\bf v}\cdot{\bf \hat{\phi}}\varphi \left[{\bf v\times
    B_0}\cdot\frac{\partial h_{gk}}{\partial {\bf v}}\right]} \nonumber\\ 
&= -D^{(\Delta)}_{\psi}\int d^3{\bf v}\annavg{ 
qR {\bf v}\cdot{\bf \hat{\phi}}\varphi B_0 \left({\frac{\partial h_{gk}}{\partial \vartheta}}\right)_r} \nonumber \\
&= -D^{(\Delta)}_{\psi}\int d^3{\bf v}\annavg{ 
qB_0R {\bf v}\cdot{\bf \hat{\phi}}\varphi (\frac{{\bf v}\cdot \nabla}{\Omega}h_{gk} + \left({\frac{\partial
    h_{gk}}{\partial \vartheta}}\right)_R)} \nonumber\\
&= D^{(\Delta)}_{\psi} \int d^3{\bf v}\annavg{ 
mR {\bf v}\cdot{\bf \hat{\phi}}({\bf v} \cdot \nabla\varphi)h_{gk}}
\end{align}

\noindent This indeed cancels with the second term in
Eqn~(\ref{heating_convection-appendix}).  Between lines 1 and 2 we used
Eqn~(\ref{v_grad-appendix}).  Between lines 4 and 5 we integrated by parts in
space to move the gradient onto the $\varphi$.  All that is left for us
to calculate from the scalar potential term in ${\bf E}\cdot {\bf v}$ is 

\begin{equation}
\frac{m}{2} \int d^3{\bf v}\annavg{ \;
v^2\frac{q}{m}{\bf E}\cdot \frac{\partial \delta f_1}{\partial {\bf v}}} = 
 -\int d^3{\bf v}\annavg{   \frac{\partial}{\partial t}\left[\varphi - 
\frac{v_{\parallel}A_{\parallel}}{c} - \frac{{\bf v_{\perp}}\cdot {\bf 
A_{\perp}}}{c}\right] q\delta f_1 }
\label{pre-heating}
\end{equation}

\noindent We now substitute our solution for $\delta f_1$.  The adiabatic/Boltzmann term gives

\begin{equation}
\annavg{
\int d^3{\bf v} 
 q \frac{\partial}{\partial t}[\varphi - 
\frac{v_{\parallel}A_{\parallel}}{c} - \frac{{\bf v_{\perp}}\cdot {\bf 
A_{\perp}}}{c}] (-q\frac{\varphi}{T_{0}}F_{m})} = \frac{d}{dt}
\annavg{[\frac{n_0}{T_0}(\frac{\varphi^2}{2})]}
\label{Boltz_Ev}
\end{equation}

\noindent since the $A_{\parallel}$ term is odd in $v_{\parallel}$ and the $A_{\perp}$ 
gyro-averages to zero (or equivalently is odd in $v_{\perp}$).  This
term will time average away, leading to the result that at this order
the ${\bf E}\cdot{\bf v}\frac {q\varphi}{T}F_0$ term produces no
heating.  We collect what is left of the heating terms from $\delta f_1$ of
Eqn~(\ref{pre-heating}):

\begin{equation}
- \int d^3{\bf v}\annavg{ 
q \left [\frac{\partial \chi}{\partial t}h_{gk} - v_{\parallel}{\bf
    b_0}\cdot\frac{\partial {\bf A_0}}{\partial t}h_{nc} + {\bf
    v}\cdot\frac{\partial {\bf A_0}}{\partial t}(\rhovec\cdot\nabla
  F_m) \right] }
\label{Ev_heating}
\end{equation}

\noindent The final term we need to calculate is the collisional energy exchange terms between 
species are now included -- note like particle collisions do not produce a
loss of energy and thus do not appear (this is easy to prove).  The interspecies collisional energy exchange is standard since to this order it is between Maxwellian species.  Specifically we have

\begin{equation}
\int d^3 {\bf v}\annavg{\frac{1}{2}mv^{2}C_{sr}(f_s, f_r)} = n_{0s} \nu_{{\cal E}}^{sr}(T_r - T_s)
\label{heating_collisions-appendix}
\end{equation}

\noindent where $\nu_{{\cal E}}^{sr}$ is found in \cite{helander}.  It is $\sqrt{m_e/m_i}$ smaller than the ion collision rate which is itself $\sqrt{m_e/m_i}$ smaller than the electron collision rate.  We now collect all the results given by Eqns~(\ref{stuff66a}), (\ref{heating_convection-appendix}), (\ref{heating_vdot-appendix}), (\ref{Ev_heating}), and (\ref{heating_collisions-appendix}).  

\begin{align}
&\annavg{\frac{3}{2}\frac{\partial (n_{0s}T_{0s})}{\partial t}} - D^{(\Delta)}_{\psi}\int d^3{\bf v}\annavg{\frac{mR}{2}F_mv^2{\bf \hat{\phi}}\cdot\frac{\partial {\bf A_0}}{\partial t} } \nonumber \\ 
& -D^{(\Delta)}_{\psi}\int d^3{\bf v}\annavg{\left[ \frac{mv^2R}{2}(\nabla \chi \cdot {\bf \hat{\phi}})h_{gk} + \frac{m^2v^2R}{2q}{\bf v}\cdot {\hat \phi}C(h_{nc} - \rhovec \cdot \nabla F_m, F_m)\right]}
\nonumber \\ 
& - \int d^3{\bf v}\annavg{ q \left [\frac{\partial \chi}{\partial t}h_{gk} - v_{\parallel}{\bf b_0}\cdot\frac{\partial {\bf A_0}}{\partial t}h_{nc} + {\bf v}\cdot\frac{\partial {\bf A_0}}{\partial t}(\rhovec\cdot\nabla F_m) \right]} \nonumber \\
& = n_{0s} \nu_{{\cal E}}^{sr}(T_r - T_s) \label{pre-heat-transport2}
\end{align}

\subsection{Heat transport manipulation}
\label{heat-manip-sec}

The purpose of this subsection is to manipulate
Eqn~(\ref{pre-heat-transport2}) into a form that will readily apply to
Sec.~\ref{entropy-sec}.  We can rewrite the bracketed terms of Eqn~(\ref{pre-heat-transport2})
by using the gyro-kinetic equations (\ref{gyro-slow}) and
(\ref{the-gyro-equation}).  A change of variables trick will convert our integral
into a gyro average with negligible error so that we may write  

\begin{align}
&\int d^3{\bf v}\annavg{ q \frac{\partial \chi}{\partial t} h_{gk}} = \int d^3{\bf v}\annavg{ q \frac{\partial \gyroavg{\chi}}{\partial t} h_{gk}} \nonumber \\ 
& = \int d^3{\bf v}\annavg{ \frac{T}{F_0} \left [ \frac{1}{2}\frac{\partial h^2_{gk}}{\partial t} + \frac{1}{2}v_{\parallel}{\bf b_0}\cdot\nabla h_{gk}^2 + \frac{1}{2}{\bf v}_{\chi}\cdot \nabla h^2_{gk} + h_{gk}{\bf v}_{\chi}\cdot \nabla F_0 \right ] }\nonumber \\
& + \int d^3{\bf v}\annavg{ \frac{T}{F_0} \left [\frac{1}{2}{\bf v}_D \cdot\nabla h^2_{gk} - h_{gk}\gyroavg{C(h_{gk})} \right ] }
\end {align}

\noindent We have multiplied the gyrokinetic equation \ref{the-gyro-equation} by $\frac{Th}{F_0}$ to obtain the right hand side of the equation.  Most of this is zero.  The first term time averages.  The second term averages to zero over the annulus.  (One uses that ${\bf b_0}\cdot\nabla v_{\parallel} = {\bf v_\perp}\cdot({\bf b_0}\cdot\nabla{\bf b_0})$ then gyro-averages.)  Then the annulus average is done employing
magnetic coordinates.)  The third term annulus averages to higher order also using $\nabla\cdot{\bf v}_{\chi} = 0$ and applying the divergence theorem.  The fifth term is argued away by order by writing
it as a full divergence with negligible error ($\nabla\cdot{\bf v}_D$ is non-zero but small).  Two terms remain: 

\begin{equation}
\int d^3{\bf v}\annavg{ 
q \frac{\partial \left <\chi\right >}{\partial t} h_{gk}}
= \int d^3{\bf v}\annavg{ 
\frac{h_{gk}T}{F_0} \left [{\bf v}_{\chi}\cdot \nabla F_0
- \left <C(h_{gk},F_m)\right > \right ]}
\label{heating_electric-appendix}
\end {equation}

\noindent We use axi-symmetry and write ${\bf B_0} = \nabla\psi\times\nabla\phi + F(\psi)\nabla\phi$.  (Note: $F(\psi) \equiv R{\bf \hat{\phi}}\cdot{\bf B_0}$ is not to be confused with $F_0$.)  Then we integrate over the annulus to obtain

\begin{equation}
\int d^3{\bf v}\annavg{h_{gk}T {\bf v}_{\chi}\cdot
\nabla \ln F_0} = - \int d^3{\bf v} \annavg{R{\hat
  \phi}\cdot\nabla \chi h_{gk}T \frac{\partial\ln F_m}{\partial \psi}}
\end{equation}

\noindent where we have removed the Boltzman part of $F_0$ by averaging over $\vartheta$
and integrating by parts in space to obtain a ${\bf b_0}\cdot\nabla$
integral which annulus averages to higher order.  Next we use the
neoclassical equation (Eqn.~\ref{gyro-slow}) to write 

\begin{align}
&-\int d^3{\bf v}\annavg{ \; qv_{\parallel}{\bf
      b_0}\cdot\frac{\partial {\bf A_0}}{\partial t}h_{nc}} \nonumber
  \\ &= \int d^3{\bf v}\annavg{\frac{Th_{nc}}{F_0}\left(\frac{\partial h_{nc}}{\partial t} +
  v_{\parallel}{\bf b}_0\cdot \nabla h_{nc} + {\bf v_D}\cdot\nabla F_0
  - \left<C(h_{nc})\right> \right)} \nonumber \\ & =
  \int d^3{\bf v}\annavg{\frac{Th_{nc}}{F_0}\left({\bf v_D}\cdot\nabla F_0 -
  \left<C(h_{nc})\right>\right) }
\end{align}

\noindent  where we have eliminated the first two terms on the second line by time average and annulus average respectively.  Note that we only need to keep the $F_m$ part of $F_0$ here since patch averaging/gyro-averaging cleans up the rest.  We need to rewrite the first term on the final line to show it is mostly collisional.  We again use axi-symmetry to obtain

\begin{align}
&\int d^3{\bf v}\annavg{ \; Th_{nc}{\bf v_D}\cdot\nabla\ln F_m} \nonumber\\ &= -\int d^3{\bf v}\annavg{ \frac{mT}{q}h_{nc}R{\bf \hat{\phi}}\cdot[\nabla\cdot({\bf b_0b_0}v^2_{\parallel})]\frac{\partial \ln F_m}{\partial \psi}}\nonumber\\ &= \int d^3{\bf v}\annavg{ \frac{mv^2_{\parallel}TR}{q}\frac{\partial \ln F_m}{\partial \psi}{\bf \hat{\phi}\cdot b_0}{\bf b_0}\cdot\nabla h_{nc}}\label{partie}
\end{align}

\noindent where we have integrated by parts between line two and three (refer to
the earlier arguments in the particle transport derivation with
pressure tensor ${\bf P}$).  Next, we again refer to the
neoclassical equation (\ref{gyro-slow}) to write 

\begin{equation}
v_{\parallel}{\bf b_0}\cdot\nabla h_{nc} = \left<C(h_{nc})\right> - {\bf v_D}\cdot\nabla F_0 - q\frac{F_0}{T}v_{\parallel}{\bf b_0}\cdot\frac{\partial {\bf A_0}}{\partial t}
\end{equation}

\noindent The middle ${\bf v_D}$ term will integrate to zero in $v_{\parallel}$
upon substitution into equation \ref{partie}.  We will soon see that
the last term will cancel.    Let us collect the recent results: 

\begin{equation}
\int d^3{\bf v}\annavg{ q \frac{\partial \chi}{\partial t} h_{gk}} = \int d^3{\bf v}\annavg{ \frac{h_{gk}T}{F_m} \left [-R{\hat \phi}\cdot\nabla \chi \frac{\partial F_m}{\partial \psi} - \left <C(h_{gk},F_m)\right > \right ]}
\end{equation}

\noindent and

\begin{align}
&-\int d^3{\bf v}\annavg{ \; qv_{\parallel}{\bf b_0}\cdot\frac{\partial {\bf A_0}}{\partial t}h_{nc}} = \nonumber \\
&\int d^3{\bf v}\annavg{\frac{T}{F_m}\left(\frac{mR}{q}\frac{\partial F_m}{\partial \psi}{\bf \hat{\phi}\cdot b_0}v_{\parallel}[\left<C(h_{nc})\right> - \frac{qF_m}{T}v_{\parallel}{\bf b_0}\cdot\frac{\partial {\bf A_0}}{\partial t}] - h_{nc}\left<C(h_{nc})\right>\right)}
\end{align}

\noindent Finally we need to rewrite the last term on the left hand side of equation \ref{pre-heat-transport}.  The result is

\begin{multline}
\int d^3{\bf v}\annavg{q{\bf v}\cdot\frac{\partial {\bf A_0}}{\partial t}\rhovec\cdot\nabla F_m} = \\
\int d^3{\bf v}\annavg{ R\left[ -T{\bf \hat{\phi}}\cdot\frac{\partial {\bf A_0}}{\partial t}\frac{\partial F_m}{\partial \psi} + m{\bf b_0}\cdot\frac{\partial {\bf A_0}}{\partial t}{\bf b_0\cdot\hat{\phi}}v^2_{\parallel}\frac{\partial F_m}{\partial \psi} \right]}
\end{multline}

\noindent We substitute these last three results into our heat transport equation (\ref{pre-heat-transport}) to obtain

\begin{align}
&\annavg{\frac{3}{2}\frac{\partial (n_{0s} T_{0s})}{\partial t}} = D^{(\Delta)}_{\psi}\int d^3{\bf v}\annavg{\frac{mR}{2}F_mv^2{\bf \hat{\phi}}\cdot\frac{\partial {\bf A_0}}{\partial t}} \nonumber \\ 
& + D^{(\Delta)}_{\psi} \int d^3{\bf v} \annavg{ \frac{mv^2R}{2}(\nabla \chi \cdot {\bf \hat{\phi}})h_{gk}} \nonumber \\ 
& - D^{(\Delta)}_{\psi} \int d^3{\bf v} \annavg{\frac{m^2v^2R}{2q}{\bf v}\cdot {\hat \phi}C(h_{nc} - \rhovec \cdot \nabla F_m, F_m)} \nonumber \\ 
& - \int d^3{\bf v}\annavg{ \frac{h_{gk}T}{F_m}R{\hat \phi}\cdot\nabla \chi \frac{\partial F_m}{\partial \psi} - \frac{h_{gk}T}{F_0}\gyroavg{C(h_{gk})}} \nonumber \\ 
& + \int d^3{\bf v}\annavg{\frac{T}{F_m}\left(\frac{mR}{q}\frac{\partial F_m}{\partial \psi}{\bf \hat{\phi}\cdot b_0}v_{\parallel}C(h_{nc}) - h_{nc}C(h_{nc})\right)} \nonumber \\ 
& - \int d^3{\bf v}\annavg{RT{\bf \hat{\phi}}\cdot\frac{\partial {\bf A_0}}{\partial t}\frac{\partial F_m}{\partial \psi}} \nonumber \\ 
& + n_{0s} \nu_{{\cal E}}^{sr}(T_r - T_s).
\label{heat-transport-final} 
\end{align}


\bibliographystyle{uclathes}
\addcontentsline{toc}{section}{References}
\bibliography{thesis_bib}    

\end {document}